\documentclass[structabstract]{aa}                                           
\usepackage{natbib}
\bibpunct{(}{)}{;}{a}{}{,} 

\usepackage{graphicx}
\usepackage{txfonts}
\begin{document}
%
\def\etal{et al.~\/}
\def\eg{{\it e.g.~\/}}
\def\etc{{\it etc.~\/}}
\def\ie{{\it i.e.~\/}}
\def\cf{{\it c.f.~\/}}  

\def\micron{{$\mu$m}}

\def\fout{{$S_{out}$~\/}}
\def\fin{{$S_{in}$~\/}}
\def\deltaf{{$\Delta S$~\/}}
\def\dfin{{$\Delta S/S_{in}$~\/}}
\def\dfout{{$|\Delta S|/S_{out}$~\/}}
\def\absdfin{{$\left\vert \Delta S/S_{in} \right\vert$~\/}}
\def\absdf{{$\left\vert \Delta S \right\vert$~\/}}

\def\f24{{S$_{24\mu}$m}}
\def\f100{{S$_{100\mu}$m}}
\def\f250{{S$_{250\mu}$m}}
\def\f350{{S$_{350\mu}$m}}
\def\f500{{S$_{500\mu}$m}}

\def\LIR{{L$_{IR}$}}
\def\FIR{{L$_{FIR}$}}
\def\L24{{L$_{24\mu}$m}}
\def\zz{{\it z~\/}} 

\def\hst{{\it HST~\/}}
\def\herschel{{\it Herschel}}
\def\spitzer{{\it Spitzer}}
\def\goodsh{GOODS-{\it Herschel~\/}}
\def\spire{{SPIRE~\/}}
\def\pacs{{PACS~\/}}
\def\goods{{GOODS~\/}}


\def\cirba{65\%}   
\def\cirbb{94\%}   
\def\cirbc{74\%}   
\def\cirbd{37\%}   
\def\cirbe{8\%}     

\title{GOODS-\herschel :  identification of the individual galaxies responsible for the 80-290\micron\ cosmic infrared background}

\author{R.~Leiton \inst{1,2}
\and D.~Elbaz \inst{1}
\and K.~Okumura\inst{1}
\and H.~S. Hwang \inst{3}
\and G.~Magdis \inst{4}
\and B.~Magnelli \inst{5}
\and I.~Valtchanov\inst{6}
\and M.~Dickinson\inst{7}
\and M.~B\'ethermin \inst{1,8}
\and C.~Schreiber\inst{1}
\and V.~Charmandaris\inst{9,10}
\and H.~Dole\inst{11}
\and S.~Juneau\inst{1}
\and D.~Le Borgne\inst{12,13}
\and M.~Pannella \inst{1,12}
\and A.~Pope \inst{14}
\and P.~Popesso \inst{5}
}
\institute{Laboratoire AIM-Paris-Saclay, CEA/DSM/Irfu - CNRS - Universit\'e Paris Diderot, CEA-Saclay, pt courrier 131, F-91191 Gif-sur-Yvette, France\\ %
\email{roger.leiton@cea.fr, roger.leiton@astro-udec.cl} 
\and Astronomy Department, Universidad de Concepci\'on, Concepci\'on, Chile. %
\and School of Physics, Korea Institute for Advanced Study, 85 Hoegiro, Dongdaemun-gu, Seoul 130-722, Republic of Korea %
\and Department of Physics, University of Oxford, Denys Wilkinson Building, Keble Road, Oxford OX1 3RH, UK %
\and Max-Planck-Institut f\"ur Extraterrestrische Physik (MPE), Postfach 1312, 85741, Garching, Germany %
\and Herschel Science Centre, European Space Astronomy Centre, Villanueva de la Ca\~nada, 28691 Madrid, Spain %
\and National Optical Astronomy Observatory, 950 North Cherry Avenue, Tucson, AZ 85719, USA %
\and European Southern Observatory, Karl-Schwarzschild-Strasse 2, D-85748 Garching, Germany %
\and Department of Physics, University of Crete, GR-71003 Heraklion, Greece %
\and Institute for Astronomy, National Observatory of Athens, GR-15236 Penteli, Greece %
\and Univ Paris Sud, Institut d'Astrophysique Spatiale (UMR 8617), B\^at. 121, 91405 Orsay, France; CNRS, Orsay 91405, France %
\and Institut d’Astrophysique de Paris, UMR 7095, CNRS, 98bis boulevard Arago, F-75005 Paris, France %
\and Sorbonne Universit\'es, UPMC Univ. Paris 06, UMR 7095, Institut d'Astrophysique de Paris, F-75005 Paris, France %
\and Department of Astronomy, University of Massachusetts, Amherst, MA 01003, USA %
}

   \date{Received ; accepted }

 
  \abstract
   {}
 {We propose a new method of pushing \herschel\ to its faintest detection limits using universal trends in the redshift evolution of the far infrared over 24\,$\mu$m colours in the well-sampled GOODS-North field. An extension to other fields with less multi-wavelength information is presented. This method is applied here to raise the contribution of individually detected \textit{Herschel} sources to the cosmic infrared background (CIRB) by a factor 5 close to its peak at 250\,$\mu$m and more than 3 in the 350 and 500\,$\mu$m bands.}  
   {We produce realistic mock \herschel\ images of the deep PACS and SPIRE images of the GOODS--\textit{North} field from the \goodsh Key Program and use them to quantify the confusion noise at the position of individual sources, \ie, estimate a ``local confusion noise". Two methods are used to identify sources with reliable photometric accuracy extracted using 24\,\micron\ prior positions. The \textit{clean index ($CI$)}, previously defined but validated here with simulations, which measures the presence of bright 24\,\micron\ neighbours and the \textit{photometric accuracy index ($PAI$)} directly extracted from the mock \textit{Herschel} images.}
   {Both methods converge to comparable depths and fractions of the CIRB resolved into sources individually detected with \textit{Herschel}. After correction for completeness, thanks to our mock \textit{Herschel} images, individually detected sources make up as much as 54\% and 60\% of the CIRB in the PACS bands down to 1.1 mJy at 100\,\micron\ and 2.2 mJy at 160\,\micron\  and 55, 33, and 13\% of the CIRB in the SPIRE bands down to 2.5, 5, and 9 mJy at 250\,\micron, 350\,\micron, and 500\,\micron, respectively.  The latter depths improve the detection limits of \herschel\ by factors of 5 at 250\,\micron, and 3 at 350\,\micron\ and 500\,\micron\ as compared to the standard confusion limit. Interestingly, the dominant contributors to the CIRB in all \herschel\ bands appear to be distant siblings of the Milky Way ($z$$\sim$0.96 for $\lambda$$<$300\,$\mu$m) with a stellar mass of $M_{\star}$$\sim$9$\times$10$^{10}$\,M$_{\odot}$.}
{}

  \authorrunning{R. Leiton et al.}
\titlerunning{\textit{Herschel} identifies individual galaxies responsible for the 80-290\micron\ CIRB}
   \maketitle
%

\section{Introduction}
\label{SEC:intro}

Primeval galaxies in the process of forming their first generation of stars were predicted to generate a diffuse extragalactic background light (EBL)
in the optical and near-infrared (NIR) \citep{partidge67}, later on extended to the far-IR \citep{low68,kaufman76}. 
\cite{puget96} reported the first measurement of the cosmic IR background (CIRB), an isotropic IR signal appearing in the all-sky maps of the FIRAS experiment onboard COBE with an integrated intensity of  $\sim$30$\pm$5~nWm$^{-2}$sr$^{-1}$ between 100 and 1000\,\micron. 
The CIRB represents about the half of the EBL \citep{dole06,lagache05}. Although it sums up only $\sim$2.3\% of the intensity of the cosmic microwave background (CMB), it is the strongest emission due to galaxies so that it represents the net energetic budget of galaxies.
The cosmic UV-optical (COB) background and CIRB are expected to be generated by the energy produced through stellar nucleosynthesis and, for a minor fraction, by black-hole accretion and emission from the diffuse intergalactic medium \citep{hauser01}. The contribution of accretion in the form of active galactic nuclei emission is expected to represent a fraction of $\sim$15\% of the CIRB \citep{jauzac11}.
Since the discovery of the CIRB, a major issue in observational cosmology has been to identify the individual sources responsible for this diffuse light which contains the net energetic budget of nucleosynthesis integrated over cosmic time, hence the detailed information on how and when stars formed in galaxies.

Several factors make the correct identification
of sources a challenge when
multi-wavelength data of extragalactic objects 
are combined. The resolution $\theta$ of a telescope having a collecting
mirror of a diameter $D$ is limited by diffraction,
\ie $\theta \sim \lambda/D$, a problem particularly difficult
 to treat for far-infrared (FIR) and submillimetre instrumentation
 ($\lambda \geq $ 100\,\micron). It is possible to
 attack this problem by constructing larger collecting
 surfaces for the telescopes (as well as instruments
 with smaller detection elements) until the desired angular
 resolution is reached. However, this increases substantially the associated costs that scale with the size of the telescope mirror to the 2-2.5 power \citep{vanbelle04}, as well as rocket limitations that prevent the launch of big telescopes in space.
Poor spatial resolution plays a major role in  
\textit{confusion noise}, which can be defined as the uncertainty in
the measurement of the flux of a point source owing to the
variance of the underlying background induced by extremely faint sources
or by the contamination of nearby point sources.
The origin of the background may be
 instrumental (caused by the thermal radiation of the telescope itself),
intrinsically diffuse (caused by the interstellar medium, \eg galactic cirrus,
see \citealt{helou90}) or may be the result of the accumulated light of faint objects (the extragalactic background light).

The launch of the \herschel\ satellite \citep{pilbratt10} has opened a new window on the FIR Universe by allowing the detection of individual galaxies down to uniquely faint limits with deep extragalactic surveys (for a comprehensive review see \citealt{lutz14}). Thanks to its 100-500\,$\mu$m coverage, \herschel\ provides the best insight into the sources responsible for the peak emission of the CIRB. Although it is the largest telescope ever sent into space, its 3.5m-class observatory does suffer from the so-called confusion limit, and the depth of these extragalactic surveys is essentially limited by the fluctuations of the signal due to a sea of unresolved sources, together with the blending from bright neighbours \citep{berta10,oliver10,nguyen10}.
Confusion is inherent to IR data,  particularly in the 250--500\,$\mu$m SPIRE bands, where the confusion limit is as high as $\sim$20 mJy \citep{oliver10,nguyen10}, but in this work we attempt to develop a method to go beyond the nominal confusion limit. The confusion noise results from a combination of effects 
 including the beam size, the instrumental noise, 
 astrometric uncertainties, and the emission of faint, undetected sources. 
Isolating the effect of each of these contributions to the confusion is a challenging task
(\eg  degeneracy between the position of the sources and the effect of the
uncertainty in individual pixels produced by the instrumental noise, pixel sensitivity, and 
local fluctuations due to faint sources).

From this point, this paper is divided into the following sections.
Section~\ref{SEC:method} explains the new method to incrementally correct
the observed \herschel\ colours from shorter to longer wavelengths;
\S\ref{SEC:dataset} presents the ancillary dataset supporting this work;
\S\ref{SEC:inc} analyses the role of  clean galaxies, blending and clustering
effects in the evolution of mid-IR (MIR) to FIR colours as a function of redshift;
\S\ref{SEC:colour} presents the relations that best fit the colour-redshift
relations, as well as the iterative corrections applied to correct blending effects;
\S\ref{SEC:imagemake} describes how the realistic mock images were made;
\S\ref{SEC:confusion} provides an analysis of the confusion noise in the extracted sources from the simulated maps.
Finally, \S\ref{SEC:cirb} presents new values for the resolved CIRB and the typical redshifts and stellar masses of the galaxies responsible for its origin, and \S\ref{SEC:conclusions} summarises the main results of this work.

\section{The method}
\label{SEC:method}
This work relies on the combination of the information of \spitzer\ and \herschel\ from the shortest to the largest wavelengths thanks to the existence of universal mid-to-far infrared colours of galaxies as a function of redshift. The success of this propagation of the information from the shortest to the longest wavelengths relies on the recently found evidence of universality of the IR spectral energy distribution (SED) of galaxies \citep{elbaz11}, itself reflecting more general scaling laws now often referred to as a ``main sequence of star-forming galaxies". Recent evidence suggests that most galaxies grow in stellar mass at a steady-state rather than from episodic, merger-driven starbursts, and new scaling laws have emerged from this scenario of galaxy formation.

First, a tight correlation between the specific SFR ($sSFR = SFR/M_{\star}$) and the stellar mass $M_{\star}$ at different redshifts : $z$ $\sim$0 \citep{brinchmann04,salim07}, $z$$\sim$1 \citep{elbaz07,noeske07}, $z$$\sim$2 \citep{daddi07a,pannella09}, $z$$\sim$3 \citep{magdis10,elbaz11}, $z$$\sim$4 \citep{daddi09}. In this scaling law, most galaxies fall on a \textit{main sequence} (MS) with steady star formation, while a small fraction of outliers experience starburst (SB) episodes.

Second, there is a correlation between the total IR luminosity $L_{IR}$ and the luminosity at 8\micron, $L_{8\mu\mathrm{m}}$. The $L_{IR}$ is mostly due to the FIR emission of big grains of dust, while the 8\micron\ emission is dominated by polycyclic aromatic hydrocarbon (PAH) molecules. Again, most galaxies fall on a main sequence with a median $IR8 = L_{IR}$/$L_{8\mu\mathrm{m}}$  of $\sim$4.9. Outliers to this sequence, \ie  $L_{IR}$/$L_{8\mu\mathrm{m}}$ $>>$4.9, are associated with mergers that show compact star-forming regions, where PAH molecules are more efficiently destroyed hence decreasing in $L_{8\mu\mathrm{m}}$ \citep{elbaz11}.

The third scaling law separating a dominant mode of star formation from a minor mode of starbursts is provided by the extension of the Schmidt-Kennicutt law \citep{schmidt59, kennicutt98}, the tight relation between projected densities of SFR and gas, to high redshifts \citep{daddi10,genzel10}. 


As a result, the mode in which galaxies form their stars
seems to follow a set of simple rules. The infrared SED of
distant galaxies is on average very similar to the one of local galaxies and these simple
scaling laws that have been found can be used to extrapolate total IR luminosities
from single photometric measurements.
In this work, considering the potential of extrapolating the IR fluxes from
shorter wavelengths and taking advantage of the unique multi-wavelength
data existing in the GOODS-\textit{North} field, we present a method of digging
below the standard global confusion limit (\ie the statistical limit that does
not account for the actual positions of individual sources) and reach a
\textit{local confusion limit}, accounting for these positions, to resolve most of the cosmic IR background even in the SPIRE bands that are most affected by confusion.

We achieve this by building realistic simulated images of a cosmological deep field that is well-sampled in redshift and has a very complete multi-wavelength dataset. The robustness of the source extraction and the accuracy of the photometry are studied in the simulations, aiming to find a criterion that determines the reliability of fluxes by 
\begin{itemize}
\item quantifying the impact of confusion on the uncertainty in the determination of \herschel\ flux densities
\item understanding the relative impact of the different sources of noise
\item improving the selection of prior sets of target positions to improve the search for faint galaxies in the process of catalogue building. 
\end{itemize}
The resulting criteria are designed to be applicable to other cosmological fields that may lack a complete and well-controlled redshift, and multi-wavelength ancillary datasets.

Consequently, we proceed according to the following steps
\begin{itemize} 
\item We exploit the fact that prior sources, seen at shorter 
wavelengths with higher spatial resolution (\eg 24\,\micron\ sources), 
are located at positions that are tightly correlated with the presence of FIR sources.
\item Thus, instead of defining a global confusion noise (produced by a floor of faint sources
percolating over the full image, and to limit detections above 3$\sigma$ of this global noise)
we use our simulations and the information at the shorter bands of each given \herschel\ band
to attribute to each source a photometric accuracy index. This index measures a ``local confusion noise", \ie the impact of confusion at the actual position of individual sources, by estimating the difference between the input and measured flux densities of every individual source used to produce the mock \textit{Herschel} image.
\item We use this index to isolate a subsample of sources that are the least affected by confusion from neighbouring sources.
\item We define this revised detection limit for a subset of the sources as the \textit{local confusion noise}. 
\item We are then able to dig into the heavily confused deep FIR images, improve the \herschel\ detection limits, resolve a greater fraction of the cosmic IR background (CIRB), and determine what fraction of the CIRB can be resolved into individual sources. Since sources that are rejected for being affected by random projections of sources in the line of sight are not biased, the properties of the sources with clean photometry are expected to be representative of the whole sample.
\end{itemize}

\section{The GOODS-\textit{North} dataset}
\label{SEC:dataset}
The full $10'\times15'$ Great Observatories Origins Deep Survey (GOODS) North field has been imaged with PACS \citep{poglitsch10} at 100 and 160\,$\mu$m, and SPIRE \citep{griffin10} at 250, 350, and 500\,$\mu$m as part of the \goodsh {\it Key Project} \citep{elbaz11}. By construction, the SPIRE images properly  encompass the GOODS-N field but also cover a wider area that is included in the FITS images of the present data release but not in the catalogues due to the lack of ancillary data in the external parts, in particular deep 24\,$\mu$m catalogues that are required to perform prior \textit{Herschel} source extraction. The total observing time in GOODS-N is 124 hours and 31.1 hours for PACS and SPIRE, respectively. 

The mosaics used in the present paper are the same ones as those described in \cite{elbaz11}. In short, the data were reduced within the HIPE environment in the \textit{Herschel} Common Science System (HCSS) with the official PACS Photometer pipeline \citep{wieprecht09}. Maps were created from the timelines of each AOR via the HCSS photProject algorithm, which is equivalent to the drizzle method (Fruchter \& Hook 2002). Each individual AOR map was projected with the same world coordinate system (WCS). Since the SPIRE bolometers do not fill the field of view, scanning at a particular angle is necessary in order to provide relatively homogeneous coverage, hence the observations were executed as cross-linked scan maps. Each single SPIRE cross-linked map covers $22^\prime \times 19^\prime$ with 3 repetitions for a total field of view of 900 arcmin$^2$.

The pixel sizes are 1.2~arcsec and 2.4~arcsec for the PACS maps at 100$\,\mu$m and 160$\,\mu$m maps and 3.6, 4.8, and 7.2~arcsec for the SPIRE maps at 250, 350, and 500\,$\mu$m, respectively. These significantly oversample the PSF full width at half-maximum (FWHM) by a factors of 5, corresponding to a sampling finer than Nyquist by a factor 2 (see Table~\ref{TABLE:fwhm} for a summary of the FWHM of  \spitzer\ and \herschel.)

The present SPIRE data from the GOODS--\herschel\ project in GOODS-North are the deepest data ever obtained at 200--600\,$\mu$m, together with those similarly deep data obtained by the HerMES\footnote{Herschel Multi-tiered Extragalactic Survey} consortium \citep{oliver10} in GOODS-South. We note that deeper data exist in the GOODS-South field for PACS for the central $8'\times8'$ region from GOODS--\herschel and extended over the whole GOODS-South field in the PEP\footnote{PACS Evolutionary Probe (PEP) Guaranteed-time Survey}-GOODS--\herschel\ combination \citep{magnelli13}. The 3--$\sigma$ depth for the \pacs data in GOODS-North are 1.1~mJy (100\,\micron), 2.7~mJy (160\,\micron) as compared to 0.6~mJy (100\,\micron), 2~mJy (160\,\micron) after combining the GOODS-\textit{Herschel} data with the PEP ones \citep{magnelli13}\footnote{As explained in Section 5 of \cite{magnelli13}, although the 3-$\sigma$ detection limit is 1.3 mJy at 160\,$\mu$m, sources below 2 mJy are affected by confusion, hence reliable individual detections are limited to a depth of 2 mJy rather than 1.3 mJy.}. Due to the confusion limit, the data at 160\,\micron\ are mildly deeper, while the 100\,$\mu$m data in the central region of GOODS-South are nearly twice deeper. However this deeper level is obtained at the centre only of the field and at the expense of a varying coverage on the outskirts of the field. Since the present paper presents a method to deal with confusion, and that this effect only starts to play a dominant role with \textit{Herschel} above 160\,$\mu$m, the following results would be comparable if done in one or the other field. Forthcoming papers will apply this method to all the CANDELS fields, \ie the two GOODS fields, as well as the central part of COSMOS and the UDS field in the framework of the Astrodeep FP7 European project. 

  \begin{table}
    \centering
    \begin{tabular}{c c c c c c}
\hline
\multicolumn{6}{c}{\textit{Spitzer}} \\
\hline
      $\lambda$ ($\mu$m) & 3.6& 4.5& 24& 70 &\\
      FWHM & $1.6"$ & $1.6"$ & $5.7"$& 18$"$ &\\
     \hline  
\multicolumn{6}{c}{\textit{Herschel}} \\
\hline
      $\lambda$ ($\mu$m) & 100& 160& 250& 350& 500 \\
      FWHM & $6.7"$& 11$"$& 18.1$"$& 24.9$"$ & 36.6$"$ \\
     \hline  
     \end{tabular}
       \caption{Full Width Half Maximum (FWHM) of the Point Spread Function (PSF) of the \textit{Spitzer} IRAC and MIPS bands (upper part), and \textit{Herschel} PACS and SPIRE bands (lower part).}
  \label{TABLE:fwhm}
  \end{table}

\subsection{Source extraction method}
\label{SEC:srcext}
Source extraction on the real \textit{Herschel} images, as well as on the mock images that will be later on described was performed, as described in Section 4.1 of Magnelli et al. (2013, see also \citealp{magnelli09}). A MIPS-24\,$\mu$m catalogue ($S_{24\mu\mathrm{m}}$ $\geq$\,20\,$\mu$Jy) was built from a list of sources coming from a blind SExtractor \citep{bertin96} source extraction on an IRAC mosaic made from the weighted average of the 3.6\,$\mu$m and 4.5\,$\mu$m images, using a `Mexhat' convolution kernel to enhance deblending in crowded regions. Due to the pixel sizes of the IRAC camera, the spatial resolution of both IRAC bands is given by the same FWHM$\sim$1.6\arcsec.

As a result of this choice, the position uncertainty of the \textit{Herschel} sources is the one of the IRAC priors but the accuracy of the separation of neighbouring sources is controlled by the actual FWHM of the PSF at the \textit{Herschel} wavelengths (listed in Table~\ref{TABLE:fwhm}). If we had a perfect knowledge of the pointing of the satellite, and in the absence of instrumental noise, sources could be deblended with a nearly infinite accuracy, which we actually tested. But in the presence of these two sources of uncertainty, even accurate prior positions are limited by the fact that we are not certain of the position in the \textit{Herschel} maps and due to the instrumental noise, which introduces an extra uncertainty on the centroid of the PSF that is fitted. The resulting flux uncertainties are therefore very difficult to quantify, requiring the most accurate mock \textit{Herschel} mosaics to reproduce these sources of uncertainties to quantify precisely the difference between input and measured flux densities for sources distributed with the actual source positions in the field. This is the key bottleneck of \textit{Herschel} extragalactic surveys that we are addressing here. Note that stacking on a larger number of source positions does reduce the effect of blending, since only the central sources are repeated, while neighbouring sources are diluted in the stacks, but that this is at the expense of studying large galaxy populations with no insurance on the actual fraction of sources that are making the signal. We are here interested in digging deeper into the confusion regime but following individual sources that can later on be studied.

We used MIPS-24\,\micron\ priors from \cite{magnelli13} imposing a minimum flux density of 20\,$\mu Jy$ (3-$\sigma$ limit, while for PACS-160\,\micron\ and SPIRE-250\,\micron\ we restricted the 24\,\micron\ priors to fluxes larger than 30\,$\mu$Jy  (5-$\sigma$ limit) (reducing the number of priors by about 35\%). For SPIRE-350 and 500\,\micron\, we used the 24\,\micron\ prior positions for sources with a S/N $\leq$ 2 detected at 250\,\micron. These criteria were adopted from Monte Carlo simulations to avoid using too many priors that would result in subdividing flux densities artificially, while producing residual maps (after PSF subtracting the sources brighter than the detection limit) with no obvious sources remaining.

It is important to avoid using a too large number of prior positions since this would lead to an ``over-deblending issue'', namely the flux density of a single \textit{Herschel} source may be distributed between several priors with high uncertainty. Therefore, in the event where two 24\,$\mu$m sources would turn out to be closer than 1.5 pixel (\ie FWHM/3.5 or Nyquist/1.5), we only kept the brightest one for the source extraction. This choice does not impact the source extraction itself -- it is only used to avoid overestimating the noise in the measurement by trying to deblend too nearby sources -- but it means that we have favoured one out of the two candidates as counterpart. The implications of this choice will be the matter of a forthcoming paper where we show that in a minor fraction of the cases, this approach leads to the missing of distant SPIRE sources that are wrongly attributed to more nearby galaxies (Shu et al. 2015, in prep.)

The MIPS-24\,\micron\ data probe fainter IR galaxies thanks to a less intense $k$-correction resulting in a flatter redshift evolution of the minimum $L_{IR}$ that can be detected. The \spitzer--MIPS 24\,\micron\ image reach deeper confusion levels (20~$\mu$Jy for 3$\sigma$) up to \zz $\sim$ 3 than \herschel\ data in the \goods fields and most \herschel\ sources have a 24\,\micron\ counterpart \citep{elbaz11}. A minor number of \herschel\ objects (21 galaxies, $<$1\%) have no 24\,\micron\ counterparts. Those sources have been identified as MIPS-dropouts, and show strong silicate features in absorption at 9.7 and 18\micron\ falling in the MIPS-24\,\micron\ band when the galaxies are at $z\sim$1.3 and $z\sim$0.4, respectively \citep{magdis11}. This means that source extraction based on 24\,\micron\ prior positions can recover most of the sources in the \herschel\ images. It is expected that above a redshift of $z \sim$3, the SPIRE bands will be more sensitive to distant sources at the depth reached here in \goodsh than MIPS-24\,\micron. As we will show in Section~\ref{SEC:nomissedsrc}, this will concern only a small fraction of the sources that will not play a dominant role in the confusion limit.

\subsection{Flux uncertainties}
\label{SEC:uncertainties}

Two methods have been classically used in previous studies (\eg \citealt{magnelli09,magnelli13,elbaz11}) to determine measurement uncertainties on the \textit{Herschel} flux densities. The first one consists in measuring the noise level in a ``residual map" at the position of the source. To produce such residual map, a first source extraction is performed to determine the 3-$\sigma$ detection threshold, then sources brighter than 3-$\sigma$ are identified, and only those ones are kept in a second iteration of the source extraction code on the original images. This leads to the production of a residual map where only 3-$\sigma$ sources have been subtracted, thus avoiding subtracting highly uncertain sources that would produce noise artificially. Keeping this residual map as a noise map, a third source extraction is then performed using all prior positions as in the first iteration but keeping the previous residual map as a noise map. This noise measurement is our ``local residual noise level".

A second method consists in injecting a small number of sources with a given flux density in the real \textit{Herschel} mosaic, extracting the sources using their all the prior positions, and including the ones of the extra sample, for which positions have been blurred in the input prior list by a random uncertainty mimicking the position uncertainty from the satellite. Reproducing this step many times for each flux density, we obtained an uncertainty on the measurement that we call ``global simulated noise level". This noise level can be compared to the local residual noise level for any of sources of similar flux densities. When this comparison was done, it was found that the two noise levels were equivalent for PACS, while for SPIRE the local residual noise level was underestimating the global simulated noise level by a factor 3/5, \ie sources expected to be measured at the 5-$\sigma$ level, were in fact detected at 3-$\sigma$. The inconvenient of this measurement is that it is not local, it is an average measurement over several positions on the map.

In the present paper, we will introduce a better determination of the noise level that will be defined from the use of realistic \textit{Herschel} mosaics. First, these mosaics will be used to demonstrate that the previous two noise levels can be made consistent. We will show -- in Figures~\ref{FIG:snrcompPACS},\ref{FIG:snrcompSPIRE} discussed in Section~\ref{SEC:noise} -- that the factor 3/5 is indeed reproduced in the realistic mock mosaics. Second, we will be able to determine a ``local simulated noise level" by determining the noise that affects the measurement of a source at its real position but on the mock mosaic. This local simulated noise level will be determined by a ``photometric accuracy index" (PAI) that will compare the measured and input flux densities of sources in our simulated images, where all sources are injected rather than a small sample randomly injected at positions that do not reproduce the actual local confusion noise level of sources.

\subsection{Redshifts and stellar masses}
\label{SEC:redshifts}
Spectroscopic redshifts come from  \cite{barger08} and Stern et al. (in preparation).

Photometric redshifts were obtained using EAZY\footnote{Publicly available at http://code.google.com/p/EAZY-photoz} adopting a linear combination of the seven standard EAZY galaxy templates as described in \cite{pannella14}. When compared to the spectroscopic sample, the photometric redshifts present a relative accuracy of 
$\Delta$$z$=($z_{\rm phot}$-$z_{\rm spec}$)$/$(1+$z_{\rm spec}$)$\leq$3\,\% 
with less than 3\,\% catastrophic outliers (\ie, objects with $\Delta$$z$$>$0.2).

Stellar masses were determined with FAST\footnote{Publicly available at http://astro.berkeley.edu/$\sim$mariska/FAST.html} \citep{kriek09} on the U to 4.5\,$\mu$m PSF-matched aperture photometry from \cite{pannella14}, using \cite{bruzual03} delayed exponentially declining star-formation histories, which mimicks the trend observed with the redshift evolution of the SFR-M$_{\star}$ main sequence of star-forming galaxies  (see \citealt{pannella14} for more details). We use solar metallicities, a Salpeter initial mass function, and the \cite{calzetti00} reddening law with $A_V$ up to 4 magnitudes.

\section{Universal mid-to-far infrared colours of star-forming galaxies as a function of redshift}
\label{SEC:inc}
In order to incrementally take advantage of the better spatial resolution of shorter wavelengths from 24 to 500\,\micron, we start by comparing the flux density of individually detected \herschel\ sources to their 24\,\micron\ emission as a function of redshift. The existence of an IR main sequence regime followed by most galaxies, as opposed to a minor population of galaxies in a starburst mode, suggests that most distant galaxies do not depart by large factors from a typical FIR/24\,\micron\ ratio, at any given redshift, for each one of the \herschel\ bands. We therefore extend here the IR main sequence concept based on the $IR8 = L_{IR}/L_{8\mu\mathrm{m}}$ colour index \citep{elbaz11} to the more general relations $L_{Herschel}/L_{24\mu\mathrm{m}/(1+z)}$, where $L_{Herschel}$ is the luminosity in each one of the \herschel\ bands, \ie $L_{100\mu\mathrm{m}/(1+z)}$, $L_{160\mu\mathrm{m}/(1+z)}$, ...

Our goals are to \textit{(i)} find the best recipe to construct as realistic mock \herschel\ images as possible that we will use to test the robustness of source extraction against confusion, \textit{(ii)} determine a methodology to benefit from the shorter wavelength \herschel\ bands to improve the source extraction in the longer bands on real images, mainly using PACS to deblend SPIRE, \textit{(iii)} estimate the completeness of this source extraction approach, and correct the contribution of faint \herschel\ sources to the cosmic IR background for incompleteness.

 \begin{figure*}
 \centering
      \includegraphics[width=7.25cm]{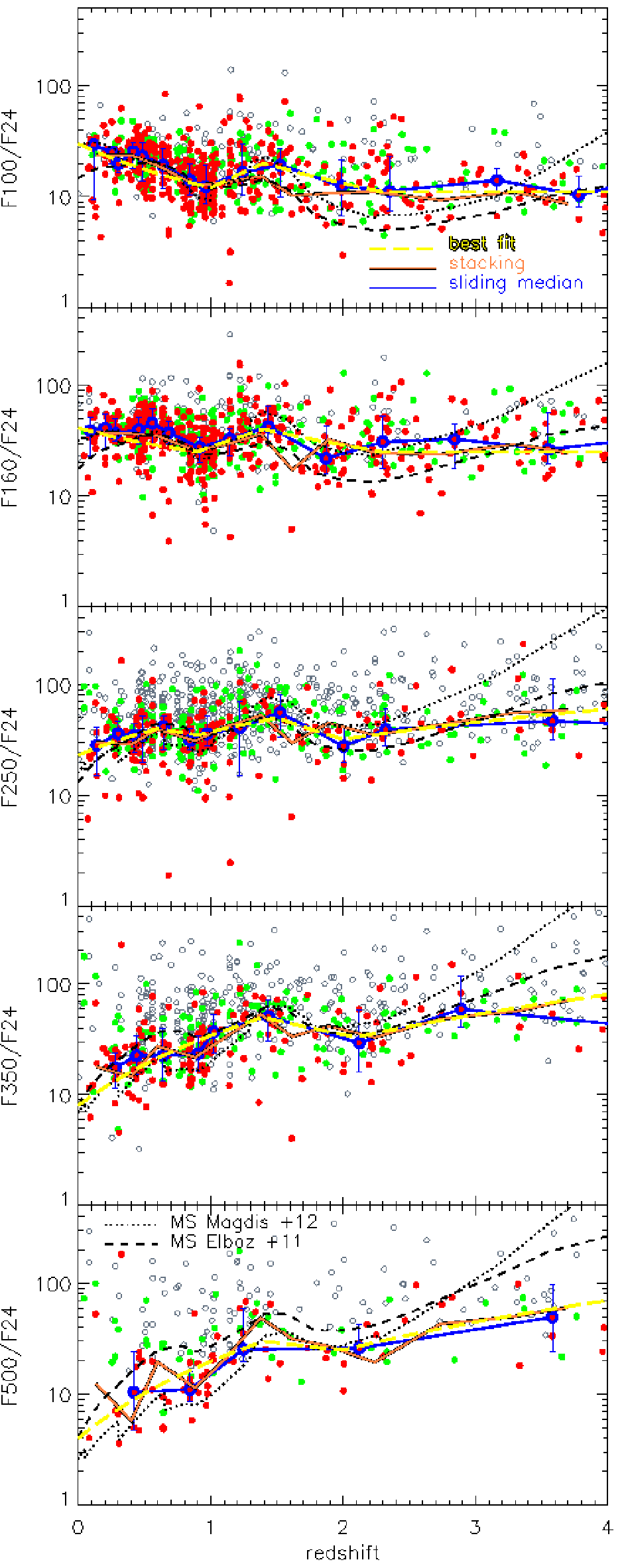}     
      \includegraphics[width=7.25cm]{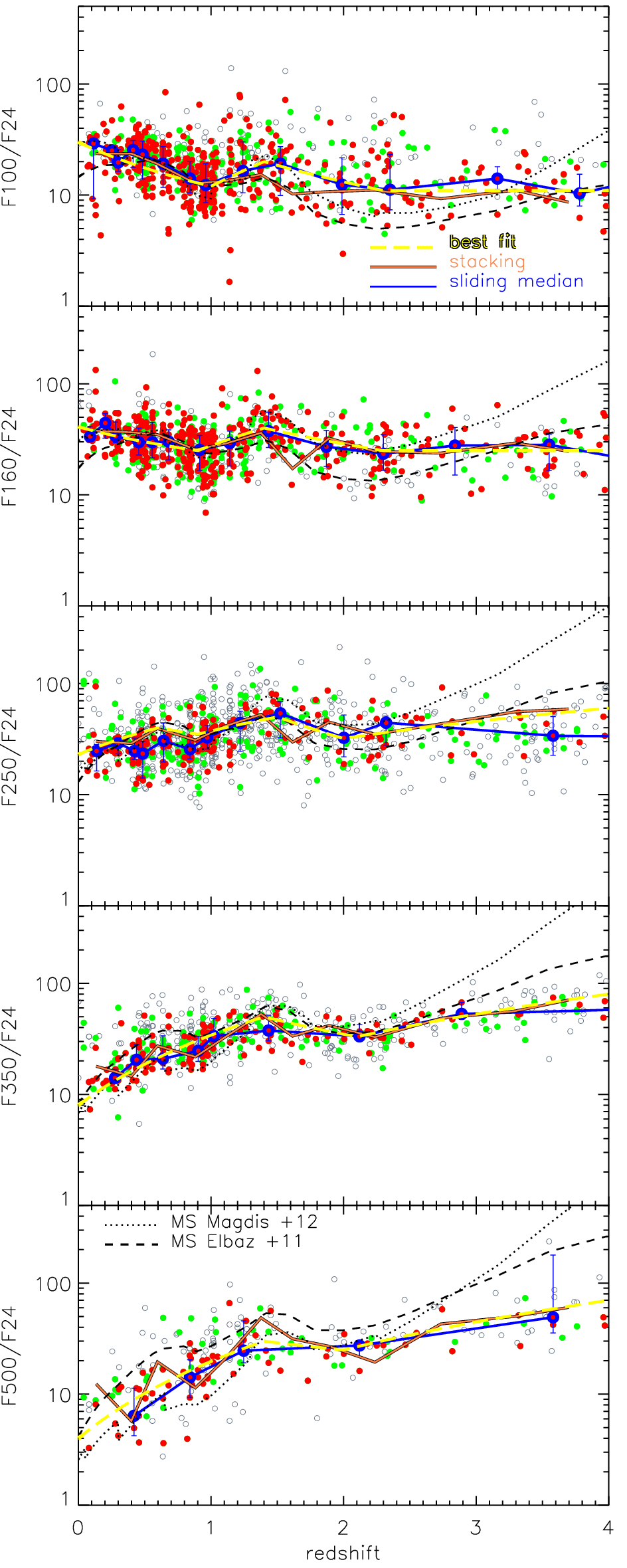}    
      \caption{Redshift evolution of the mid-to-far infrared colours of \textit{Herschel} galaxies. Each plot shows the ratio of the observed \herschel\ over MIPS-24\,$\mu$m flux densities as a function of redshift. Figures on the \textbf{left column} show the colours from the observed \textit{Herschel} measurements. Figures on the \textbf{right column} show the same dataset but after having corrected the 160 to 500\,$\mu$m over 24\,$\mu$m ratios of the individual galaxies by the offset observed in the 100 to 350\,$\mu$m shorter wavelength bands.
The points indicate the positions of individually detected sources separated in 3 populations, depending on the presence of bright 24\,$\mu$m neighbours (\ie with a $S_{24\mu\mathrm{m}}$ greater than half of the central source) in their close environment (at a distance of 12\arcsec for PACS and 20\arcsec for SPIRE). This number is the ``clean index", CI. Filled red dots: galaxies with no bright neighbours (CI=0); filled green dots:  galaxies with at most one bright neighbour (CI$\leq$1); open circles: remaining 3-$\sigma$ sources (CI$\geq$2). The sliding median of the CI=0 galaxies is shown with a blue line and blue circles with central red dots. The trends for the mean stacking at the positions of the 24\,$\mu$m prior sources brighter than 50$\mu$Jy are illustrated with orange lines. The yellow line corresponds to the best-fit relations described in Eq.~\ref{EQ:bf100} to~\ref{EQ:bf500} of \S\ref{SEC:colour} that fit the sliding median. The colours obtained for the IR spectral energy distributions of the library of \cite{magdis12} (black dotted line) and the main sequence SED of \cite{elbaz11} are shown for comparison.)
 }
     \label{FIG:bestfit}
  \end{figure*}
\subsection{Clean galaxies}
\label{SEC:cleangals}

The goal of this section is to determine the existence of universal trends in terms of FIR/MIR colours of galaxies as a function of redshift. In order to minimise the effects of bright neighbours on galaxies, which would artificially make such relations noisier, we have chosen to perform these measurements on a sub-population of ``clean" galaxies. Clean galaxies are defined by exploiting the a priori knowledge of the underlying galaxy population on the projected density map, assuming that the local spatial distribution of the 24\,\micron\ sources also traces the local spatial distribution of sources at longer wavelengths. Here we define a \textit{Clean Index} (CI) as the number of bright 24\,\micron\ neighbour sources ($S_{neighbour}$ $>$0.5$S_{central}$) that are closer to the central source, $S_{central}$, than 12\arcsec\ for PACS and 20\arcsec\ for SPIRE. It was found by \cite{hwang10} that sources with 2 or more 
``bright'' 24\,\micron\ neighbours systematically exhibit non-physical jumps in the FIR SED, resulting from their \herschel\ fluxes, while sources with 0 or 1 neighbour did not. In its original definition, the CI was computed within a distance of 20\arcsec\ by reference to the SPIRE 250\,\micron\ band. This distance that corresponds to 1.1$\times FWHM_{250\mu\mathrm{m}}$ being too conservative for PACS, we have translated it to the PACS 160\,\micron\ band, \ie 12\arcsec\ (equal to 1.1$\times FWHM_{160\mu\mathrm{m}}$).

\subsection{Redshift evolution of the far/mid-infrared colour of galaxies}
\label{SEC:clean}

Here we present a new empirical method based on the correlation of FIR/24\,\micron\ colours of individual galaxies as a function of redshift
 to extrapolate the \herschel\ fluxes to be used to produce realistic simulations. 
Libraries of spectral energy distributions are widely used to fit the FIR emission of galaxies and extrapolate their bolometric IR luminosity \citep{chary01,dale02,dacunha08}. However, uncertainties in our knowledge about the physics of galaxy evolution and 
not well-constrained measurements in the long wavelengths -- due to detection and confusion limits-- lead to large uncertainties 
when using such libraries, established for local galaxies, to extrapolate FIR fluxes of distant galaxies. 
After having used SED libraries, and found them unsatisfactory to predict the FIR fluxes based on 24\,$\mu$m as compared to the actual \herschel\ measurements 
(in particular the \spire data for which extrapolations are more extreme in terms of the wavelength gap), we 
 adopt here a more phenomenological approach to predict \herschel\ fluxes by fitting the observed evolution of the FIR/24\,\micron\,ratio as a function of redshift, and accounting for the departure of individual galaxies from this typical trend based on the closest shortest \herschel\ band.
The resulting trends will afterwards be compared to standard SED libraries. 
 
We found that the observed FIR/24\,\micron\ colours show well-defined trends for each \herschel\ band as a function of redshift.
Part of the dispersion of these colours comes from the effect of source blending. To minimise this effect in the determination of the typical colour of individual galaxies, we have split our sources in three categories, depending on the presence of bright neighbours at 24\,\micron, within a distance depending on the \herschel\ band considered. This technique has been used to identify ``clean'' candidate sources \citep{hwang10,elbaz10,elbaz11} as sources with the least probability to be affected by the confusion with a bright neighbour, where ``bright" designates sources with a flux density of at least half that of the central source. 
The evolution of the observed $S_{Herschel}$/$S_{24\mu\mathrm{m}}$ flux density ratios as a function of redshift are presented on the left panel of Figure~\ref{FIG:bestfit}. In Figure~\ref{FIG:bestfit}, sources defined as ``clean" because they have zero or one bright 24\,\micron\ neighbour are shown with red and green dots, respectively. We find that ``clean" \herschel\ sources follow a well-defined evolution of the FIR/24\,\micron\ colour with redshift with an average dispersion of only 0.23$\pm$0.07 dex. The error bars in Figure~\ref{FIG:bestfit} represent the median absolute deviation, MAD, \ie the interval in $log_{10}$ containing 68\,\% of galaxies around the median \textit{Herschel}$/$24\,$\mu$m colour. The average of this dispersion over the full redshift range is 0.23$\pm$0.07 dex, with values that vary with wavelength : MAD= [0.17, 0.16, 0.18, 0.21, 0.31] dex at [100, 160, 250, 350, 500]\,$\mu$m. Note that the MAD would have been larger without splitting sources in redshift, \ie MAD around the median colour of all redshifts = [0.20, 0.18, 0.20, 0.33, 0.42]. 

The sliding median of the sources without bright neighbours is shown with a blue line joining blue dots with red centres. In order to obtain the purest relations, we used here the filled red dots, \ie the galaxies fulfilling the condition of CI=0 (that is to say sources having no bright neighbour), to define the best-fitting relations and the dispersion around these relations that are shown with the solid blue line in Figure~\ref{FIG:bestfit}.

The green points with at most one bright neighbour are distributed symmetrically around this trend and we have tested that their sliding median does not present any major departure from the one seen with the cleanest sources, \ie no bright neighbour. While the technique allowed us to minimise the impact of blending, we are limited to the sources that can be detected by \textit{Herschel}. Therefore, in order to check whether this trend remains reliable to fainter levels, we have compared it to stacked measurements.

\subsection{Comparison with stacking measurements}
\label{SEC:stacks}
We have stacked postage stamp images on a selection of 24\,$\mu$m prior positions in order to determine whether the colour trends seen for detected sources were also representative of a wider selection or more typical of the brightest sources. For each of the sources in our galaxy sample, we produced a cutout, centred on the nearest pixel to the sky position of the source, in the PACS and SPIRE images of 60$\times$60 pixels, which corresponds to an angular scale of about 10 times the image beam. The cutouts were then stacked to create mean images in selected bins of redshift and stellar mass. The resulting measurements were corrected for flux boosting by clustering using the recipe described in the Appendix B2 in \citet{schreiber14}.
\begin{figure}
\centering
\includegraphics[width=8cm]{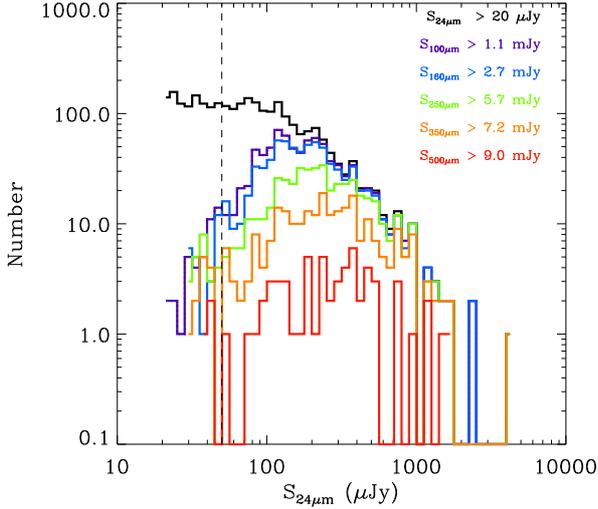}
\caption{Histogram showing the number of 24\,$\mu$m sources as a function of flux density in bins of 0.05 dex (black curve). The coloured lines show the number of 24\,$\mu$m sources in the same flux density bins that are detected by \textit{Herschel} at 100 to 500\,$\mu$m.}
\label{FIG:histo24}
\end{figure}

We stacked only prior 24\,$\mu$m positions of sources brighter than 50\,$\mu$Jy. The reason for this choice comes from the number of independent beams that can be used within 150 square arc minutes with the large SPIRE PSF FWHM. Within this area there are about 400, 870, and 1650 beams at 500, 350, and 250\,$\mu$m, respectively. Using all 24\,$\mu$m sources brighter than 20\,$\mu$Jy, meaning about 2400 sources, would imply that we would be stacking more positions than there are independent beams in the area. Obviously, we would not gain any information by doing this. Instead we decided to stop at 50\,$\mu$Jy in order to reduce the number of priors to nearly 1000. We note that the \textit{Herschel} detections rapidly drop for 24\,$\mu$m priors below 200\,$\mu$Jy. Between 200 and 50\,$\mu$Jy, they decrease by a factor 10 for PACS and 100 for SPIRE at the \cite{elbaz11} detection limits of 1.1 and 2.7 mJy at 100 and 160\,$\mu$m 5.7, 7.2, and 9 mJy at 250, 350, and 500\,$\mu$m (see Figure~\ref{FIG:histo24}).

Remarkably, the redshift evolution of the FIR/24\,\micron\ colours measured using stacking (orange line) follows closely the one of the sliding median (blue line and dots), showing that the detected sources do not present different SEDs than fainter undetected sources probed by stacking down to the limit probed here.

The narrow dispersion of the clean samples allowed us to define relations that best fit the median colours of individual sources as a function of redshift for each \herschel\ band (see \S \ref{SEC:colour} for more details). These best-fitting relations are shown with a yellow line in Figure~\ref{FIG:bestfit}. The nodes of the yellow lines were chosen arbitrarily to reproduce the sliding median with a minimum of relations.

\subsection{Blending and clustering effects}
\label{SEC:clustering}
Sources located at prior positions close to two or more bright 24\,\micron\ neighbours (black open circles in \ref{FIG:bestfit}) show 
FIR/24\,\micron\ ratios more dispersed and systematically higher than the best-fitting relations over the whole redshift range. This suggests that their colours are strongly affected by blending with nearby bright sources, hence are not characteristic of the colours of individual galaxies. 
 
In general, the non-clean sources become confused just by the presence of a large projected density of sources at their position (by chance associations rather than physical association in most cases) and there is no reason for them to have a different nature than the clean ones. 

A possible caveat though, would be the possibility that clean sources favour sources that are physically less clustered than the general population. Indeed, it is known that two effects are playing a role in the confusion of \textit{Herschel} mosaics: a projected random association and a physical association linked to clustering. Clustering can boost flux densities by several percents as described in the Table B.3 of \cite{schreiber14}. The clean index would preferentially reject those sources and, possibly, bias the colours of the resulting galaxies if clustered sources were found to be, for example, warmer than non clustered ones. In order to test this effect, we have kept the same number of 24\,$\mu$m sources and redistributed them randomly in the field. Physical clustering is therefore present only in the real source distribution and absent in the random one. We do find that the fraction of clean sources rises when distributing randomly the 24\,$\mu$m sources. In the real catalogue, 38.8\,\% of the sources are ``clean", \ie present at most one bright neighbour with $>$50\,\% of its 24\,$\mu$m flux density within 20\arcsec, while in comparison this fraction rises to 44.6\,\% in the randomly distributed sample. Concequently, we would find 7\,\% more clean sources in the absence of clustering. In other words, our clean index has led us to reject 7\,\% of the sources because they were physically associated with bright neighbour due to real physical associations. We do not know whether such sources are presenting different far-IR colours, but their small fraction suggests that their presence in the sample would not have strongly modified the colour trends that are used here.

\section{An incremental method to predict flux densities in the \herschel\ bands}
\label{SEC:colour}
%
%
%
We have shown in the previous section that the clean population is representative of the whole population,
as well as of the detected sources since they follow the same trend as the staked sources. This also implies that 
we can use the colours of clean detections to constrain 
a proto-typical IR SED of galaxies or at least to extrapolate
from 24\,\micron\ the flux densities of all prior sources in the 
\herschel\ bands used here  to make the mock images.
For comparison, in Figure~\ref{FIG:bestfit} we also show the main sequence SED from \cite{magdis12} 
(black dotted line), where the correct SED is chosen as a function of redshift,
and also the main sequence SED from \cite{elbaz11} (black dashed line).
These SEDs globally agree with the colour trends that we find here.
Noticeable differences appear at $z$ $>$3, where the SEDs are not constrained but sources at such high redshifts make a negligible contribution to the confusion limit. We note though that these large differences are however useful for the identification of candidate $z$ $>$3 sources.

In the following, we will show that the dispersion of ``clean" sources around the sliding median of the FIR/MIR relations is physical and not due to measurement uncertainties. To demonstrate this, we will determine the offset of sources with respect to the median trend at a given wavelength, \eg 100\,$\mu$m, then correct for this offset the following larger wavelength, \ie 160\,$\mu$m. If the dispersion is physical and dominated by FIR/MIR differences rather than changes in dust temperature, then the dispersion at 160\,$\mu$m will be reduced by the same amount. We will make such corrections iteratively up to 500\,$\mu$m. After having shown that this method does reduce by a substantial amount the dispersion of galaxies around the median trends, we will use these median trends and the offsets measured at ``shorter wavelengths" to reversely predict the observed ``longer wavelengths" flux densities of sources starting from 24\,$\mu$m to 500\,$\mu$m. This method will allow us to attribute realistic observed \textit{Herschel} flux densities to all 24\,$\mu$m priors and we will then produce mock \textit{Herschel} images from these catalogues that we will compare to the real images to verify that they do contain the same amounts of confusion noise due to undetected sources.

\begin{figure}
    \includegraphics[width=9cm]{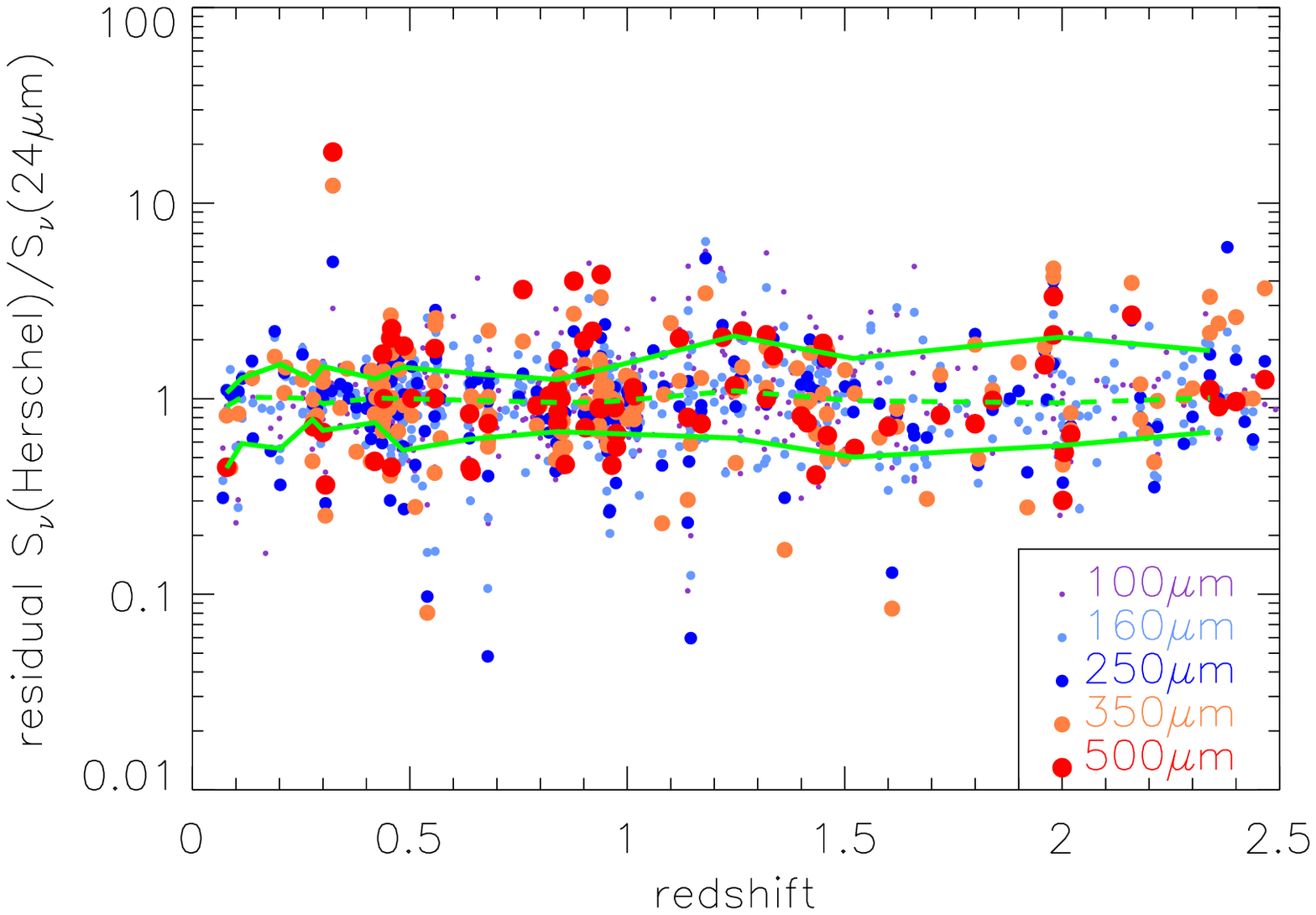}
    \includegraphics[width=9cm]{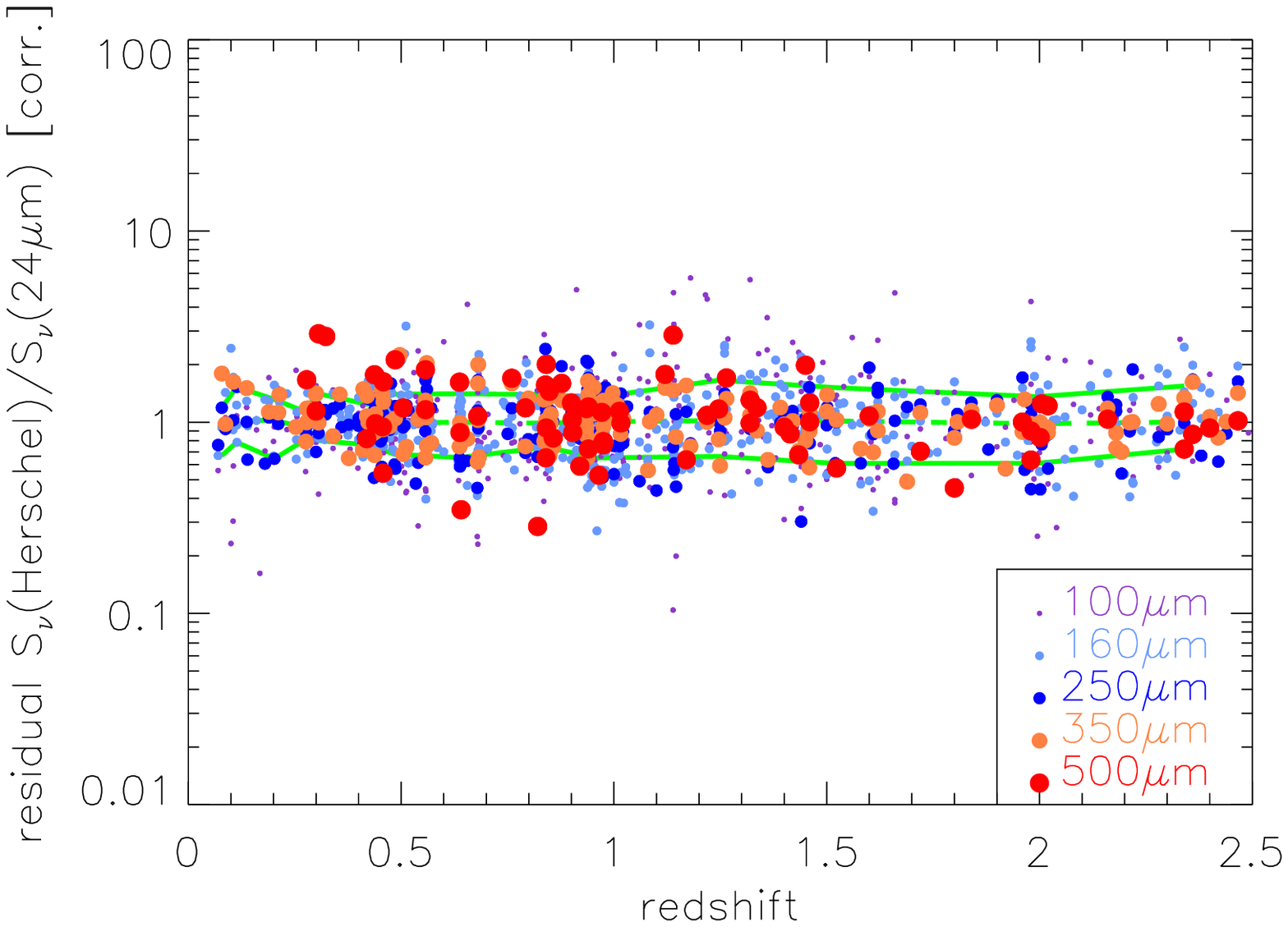}
   	  \caption{Residuals of colour index \herschel\ band minus MIPS-24\,$\mu$m.
   	  with respect to the best-fit relations as a function of redshift.
   	  The \textbf{upper panel} presents the residuals for the observed colour index. 
   	  The 68\,\% dispersion around the median is equal 
   	  to [0.17, 0.16, 0.18, 0.21, 0.31] dex for the five \herschel\ bandpasses when 
   	  accounting for redshift ranges in the best-fit equations, while it would have been 
   	  [0.20, 0.18, 0.20, 0.33, 0.42] without separating galaxies in redshift slices. 
   	 \textbf{Lower panel} shows the dispersion after accounting for the offset 
   	 measured at shorter wavelengths, \eg correcting $S_{160\mu m}$/$S_{24\mu m}$ by the 
   	 distance of $S_{100\mu m}$/$S_{24\mu m}$ to the median, 
   	 the dispersion in colour indices becomes as small as [0.14, 0.12, 0.13, 0.19] 
   	 at [160, 250, 350, 500]\,$\mu$m, respectively (no correction can be 
   	 applied to the 100\,$\mu$m band obviously). 
	 The correction process reduces the dispersion of the colour distribution because 
	 that variation is physical and not due to noise in the measurements.   	 
   	  }
       \label{FIG:rms_deboost}
\end{figure}

\subsection{Universal \textit{Herschel}--24\,$\mu$m colour evolution and dispersion}
The observed dispersion of the colours of galaxies as a function of redshift results from the combination of noise in the measurements and physical differences such as PAH equivalent widths, distributions of grain sizes, nature (silicates, carbon), and temperature. The upper plot in Figure~\ref{FIG:rms_deboost} shows the colour residuals with respect to the best-fit relations given in Sect.~\ref{SEC:colour}. If noise was the dominant cause of this dispersion, we would not expect the offset of a source observed at 100\,\micron\ to be correlated with the one observed at 160\,\micron. Flux boosting, due to a bright neighbour, would artificially induce such correlation but with a different effect as a function of FWHM and, as seen in Figure~\ref{FIG:bestfit}, with a preferential effect of boosting the flux density. However, as we will show in the following sections, offsets to the best-fitting trends are correlated between the \herschel\ bands suggesting that they are due to physical differences. For example, \cite{elbaz11} showed that starbursts tend to exhibit stronger FIR/MIR ratios than normal star-forming galaxies, which result in a boost in all \herschel\ bands. The effect of dust temperature on this boost is a second-order effect.
 
Assuming that differences in the FIR/MIR colours of observed galaxies are physical and not dominantly due to noise, we can use the \eg excess of far-IR emission, as probed by the 100/24 flux density ratio, as an indication that the SPIRE/24\,\micron\ ratio will also exhibit an excess, hence use this information to better predict the larger wavelengths flux densities.  

Since the presence of neighbours results mostly from projection effects of sources that are not physically connected, the selection of clean sources, \ie sources selected to have at most one bright 24\,\micron\ neighbour, is not expected to bias our analysis. Since we allow at most one bright neighbour but reject sources with two or more neighbours, we do not expect our sample to be biased against interacting systems of close pairs.  

 The resulting best-fitting laws that describe 
how colours change as a function of $(1+z)$, 
as shown in Figure~\ref{FIG:bestfit}, are the following:

\begin{equation}
\begin{array}{cll}
      S^{best-fit}_{100\mu\mathrm{m}} &= S_{24\mu\mathrm{m}} \times 30.0 \times (1+z)^{-1.32} &,  z \le 1 \\
                                     &= S_{24\mu\mathrm{m}} \times 1.7 \times (1+z)^{2.80} &,  1 < z < 1.4 \\
                                     &= S_{24\mu\mathrm{m}} \times 103 \times (1+z)^{-1.88} &,  1.4 < z < 2.3 \\
                                     &= S_{24\mu\mathrm{m}} \times 11.0 								  &,  z > 2.3
\end{array}
\label{EQ:bf100}
\end{equation}

 \begin{equation}
  \begin{array}{cll}
       S^{best-fit}_{160\mu\mathrm{m}} &= S_{24\mu\mathrm{m}} \times 41 \times (1+z)^{-0.77} &,  z \le 0.9  \\ 
                                              &= S_{24\mu\mathrm{m}} \times 6.9 \times (1+z)^{2.01} &,  0.9 < z < 1.4 \\ 
                                              &= S_{24\mu\mathrm{m}} \times 146 \times (1+z)^{-1.48} &,   1.4 < z < 2.3 \\ 
                                              &= S_{24\mu\mathrm{m}} \times 25       						      &,  z > 2.3  
  \end{array}
  \label{EQ:bf160}
 \end{equation}

  \begin{equation}
  \begin{array}{cll}
         S^{best-fit}_{250\mu\mathrm{m}} &= S_{24\mu\mathrm{m}} \times 23 \times (1+z)^{1.18} &,  z \le 0.6  \\ 
                                                &= S_{24\mu\mathrm{m}} \times 58 \times (1+z)^{-0.78} &,  0.6 < z < 0.9 \\ 
                                                &= S_{24\mu\mathrm{m}} \times 15 \times (1+z)^{1.30} &,   0.9 < z < 1.5 \\ 
                                                &= S_{24\mu\mathrm{m}} \times 471 \times (1+z)^{-2.45} &,   1.5 < z < 2 \\ 
                                                &= S_{24\mu\mathrm{m}} \times 8.3 \times (1+z)^{1.23} &,    z > 2

  \end{array} 
  \label{EQ:bf250}
    \end{equation}

 \begin{equation}
  \begin{array}{cll}
         S^{best-fit}_{350\mu\mathrm{m}} &= S_{24\mu\mathrm{m}} \times 8 \times (1+z)^{2.09}&,  z \le 1.4   \\
        										  &= S_{24\mu\mathrm{m}} \times 207 \times (1+z)^{-1.62}&,  1.4 < z < 2.1   \\
        										  &= S_{24\mu\mathrm{m}} \times 4 \times (1+z)^{1.85}&,   z  > 2.1  

  \end{array}
  \label{EQ:bf350}
 \end{equation}

 \begin{equation}
  \begin{array}{cll}
      S^{best-fit}_{500\mu\mathrm{m}} &= S_{24\mu\mathrm{m}} \times 4 \times (1+z)^{2.30}&,  z \le 1.4 \\
  										    &= S_{24\mu\mathrm{m}} \times 61 \times (1+z)^{-0.82}&,  1.4 < z < 2 \\     
       										&= S_{24\mu\mathrm{m}} \times 2.7 \times (1+z)^{2}&,   z > 2 \\
  \end{array}
  \label{EQ:bf500}
 \end{equation}

\noindent where $S^{best-fit}_{\lambda}$ 
and $S_{24\mu\mathrm{m}}$ are flux densities
and $S^{best-fit}_{\lambda}$ were measured in the \herschel\
images using 24\,\micron\ prior positions as in \cite{magnelli09} 
using the sample of clean sources.

These best-fitting laws define the reference 
colours from which any excess or lack of flux of the real 
\herschel\ sources can be determined: in the former case, 
a departure from these laws can, in principle, mean that the source 
is gaining flux from a companion due to blending (\textit{boosting}), a sign that 
sources are affected by confusion (alternatively, 
sources can also be experiencing a starburst phase that 
can resemble a boosted source) or, in the latter case,
 the source may be giving up part of its flux to boost others.

Based on these findings, we believe that we can use 
the $S_{Herschel}/S_{24\mu\mathrm{m}}$ best-fit colours  
to extrapolate \herschel\ flux densities based on 24\,\micron\ 
measurements 
for objects below the nominal detection limit of \herschel.

To construct our list of best-fitted fluxes for all prior sources,
we calculated the $S^{best-fit}_{\lambda}$
 for $\sim$ 2700 sources in the ancillary MIPS catalogue, adopting the 
 spectroscopic redshift where available (48\%) and otherwise the 
 photometric redshift (37\%).  For the remaining 15\% lacking a 
 redshift, we assumed that they were faint, distant 
galaxies with colours following the behaviour observed 
in the high redshift domain in Eqs.~\ref{EQ:bf100} to \ref{EQ:bf500}.

\begin{figure*}
\centering
\includegraphics[width=16cm]{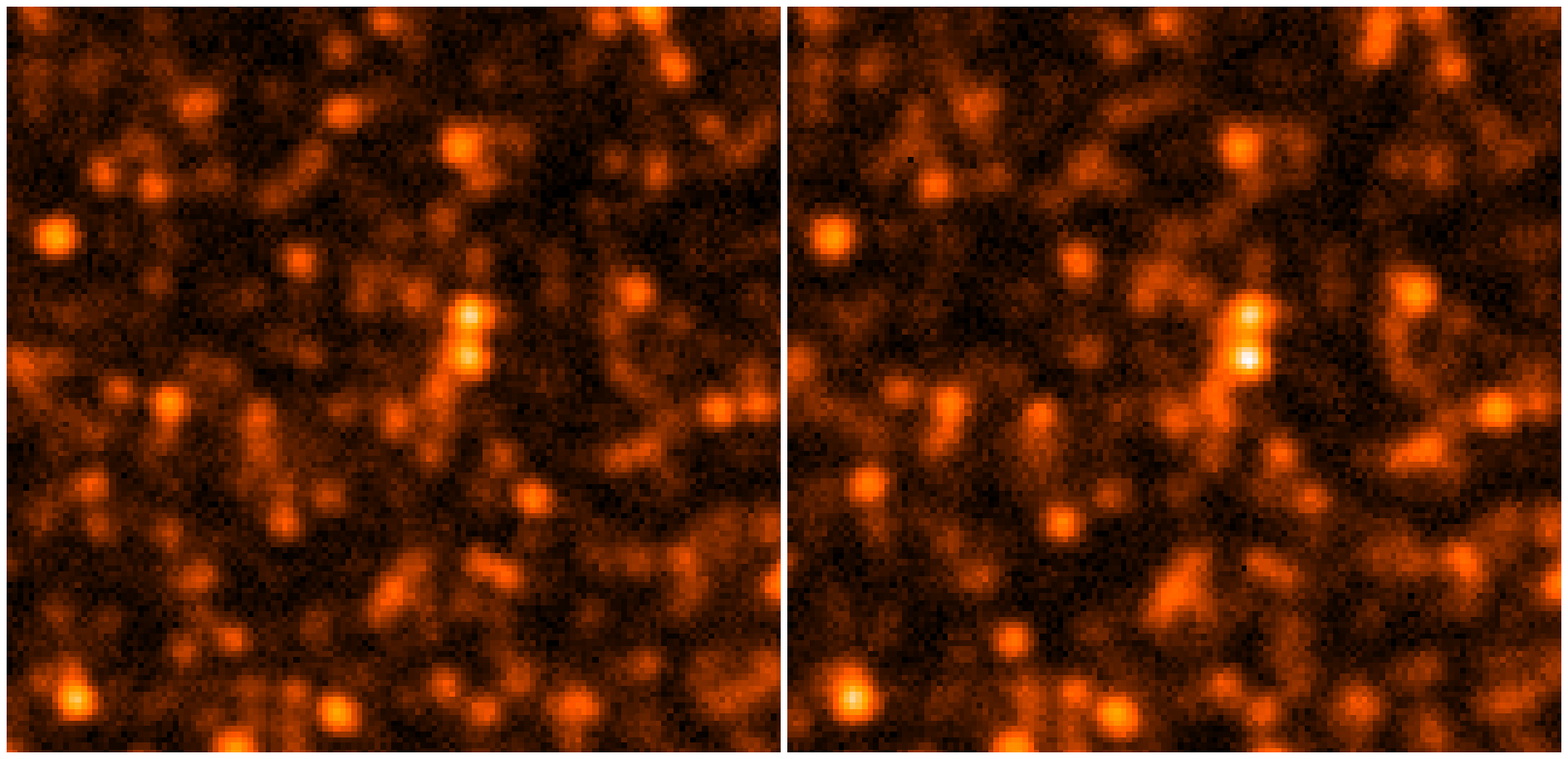}
\caption{Selected area of the real (left) and the simulated maps (right) for GOODS-North field in 250$\mu$m. Each stamp is $\sim$9'$\times$9' (see \S~\ref{SEC:imagemake}).
} 
\label{FIG:stamp}
\end{figure*}

\subsection{Iterative correction}
 \label{SEC:colourcorr}
The second step to produce the final flux 
densities, which will be used in the realistic mock images,
 exploits the advantage of the multi-wavelength association of data
and the observed colour deviations with respect to the best-fitting relations, to 
iteratively correct the colour of clean sources from 
shorter to longer wavelengths. 

We begin by considering the shortest wavelengths that show very 
low or no confusion at all. That is the case for 24\,\micron\
 and 100\,\micron\ sources, at least down to the flux density limit above which 
 sources are detected in the 
160\,\micron\ and SPIRE bands.
Here we will test the hypothesis that the colour dispersion 
for clean sources around the median colour trend, found in 
the previous section, is physical and not artificially produced by noise
and/or confusion. If this hypothesis is correct, then we can estimate
the distance in colour of an individual galaxy to the median colour trend 
at the shortest wavelength, \eg 100\,\micron, 
and use this information to predict the flux density at the largest
wavelengths, \eg 250\,\micron. In other words, we should 
be able to reduce the dispersion around the median in the 
\spire bands by using the colour distance measured
in the shorter wavelength bands. 

For any given source detected at 24\,\micron\ for which we have a redshift determination,
we start by determining its $S^{best fit}_{100\mu\mathrm{m}}$ from Eq.~\ref{EQ:bf100}. Then, this
extrapolated flux density is compared to the observed one and its excess (or deficit) is 
defined by $C_{100}$ as :

    \begin{equation}
      C_{100} = S^{best fit}_{100\mu\mathrm{m}} / S_{100\mu\mathrm{m}}^{\rm observed}
      \label{EQ:c100}
   \end{equation}
 
\noindent where $C_{100}$ contains the colour 
information, from a well sampled set of data,
that will serve to correct any excess (or lack) of flux for that source 
also detected in a longer wavelength band. In the case that 
the source is not detected at 100\,\micron\ the
constant is set to 1, \ie, we adopt the best fitted 
100\,\micron\ flux as the flux density for that 24\,\micron\ prior
in the simulations.

To determine whether offsets at 100 and 160\,\micron\ are correlated,
we define a corrected 160\,\micron\ flux density as :
    \begin{equation}
    \begin{array}{cll}
      S_{160\mu\mathrm{m}}^{corrected} &=& S^{observed}_{160\mu\mathrm{m}} \times C_{100}\\
      \\
                                     &=& S^{observed}_{160\mu\mathrm{m}} \times S^{best fit}_{100\mu\mathrm{m}}/S^{observed}_{100\mu\mathrm{m}}                         
      \end{array}
      \label{EQ:f160cor}
   \end{equation}

Figure~\ref{FIG:bestfit}-right presents the distribution of the corrected flux densities as a function of redshift. No correction was applied at 100\,$\mu$m since there is no shorter \herschel\ band that can be used for a correction, hence the upper-left and upper-right panels are identical. Below, the 160\,\micron\ band shows a reduced dispersion which confirms that the dispersion seen at 100 and 160\,\micron\ are not only due to random noise but are correlated. 
Globally, the 68\,\% dispersion around the median of the $S_{160\mu\mathrm{m}}/S_{24\mu\mathrm{m}}$ ratio is equal to 0.16 dex when galaxies are compared to their median in redshift intervals, as in Figure~\ref{FIG:bestfit}-left. It becomes as low as 0.12 dex in Figure~\ref{FIG:bestfit}-right, when correcting for the offsets observed at 100\,\micron\ using $C_{100}$. This implies that one can extrapolate a value for $S_{160\mu\mathrm{m}}$ with a 0.12 dex accuracy with the knowledge of the redshift of a source and of its flux densities at 24 and 100\,\micron.

To extend the process to the following wavelength, 250\,\micron\, we define $C_{160}$ as the residual offset between the corrected 160\,\micron\ flux density and the median best-fitting trend :
    \begin{equation}
      C_{160} = S^{best fit}_{160\mu\mathrm{m}} / S_{160\mu\mathrm{m}}^{corrected}
      \label{EQ:c160}
   \end{equation}

Using this new correction factor, one can correct the observed 250\,\micron\ flux density by combining both $C_{100}$ and $C_{160}$ as in Eq.~\ref{EQ:f250cor} and then determine the dispersion of resulting flux density distribution around the best-fitting $S_{250\mu\mathrm{m}}/S_{24\mu\mathrm{m}}$ ratio vs redshift trend. The 68\,\% dispersion around the median gets reduced from 0.18 dex to 0.13 dex.
We apply a similar procedure to correct the 350 and 500\,\micron\ flux densities, so that: 

    \begin{equation}
      \begin{array}{cll}
      S_{250\mu\mathrm{m}}^{corrected} &=& S^{observed}_{250\mu\mathrm{m}} \times C_{100} \times C_{160}\\
      \\
      					                                       &=& S^{observed}_{250\mu\mathrm{m}} \times S^{best fit}_{160\mu\mathrm{m}} / S^{observed}_{160\mu\mathrm{m}}
      \end{array} 
      \label{EQ:f250cor}
   \end{equation}

  \begin{equation}
  \begin{array}{cll}
      C_{250} & = & S^{best fit}_{250\mu\mathrm{m}} / S_{250\mu\mathrm{m}}^{corrected} \\
      \\
      C_{350} & = & S^{best fit}_{350\mu\mathrm{m}} / S_{350\mu\mathrm{m}}^{corrected}      
  \end{array} 
  \label{EQ:c250}
    \end{equation}

  \begin{equation}
  \begin{array}{cll}
      S_{350\mu\mathrm{m}}^{corrected} & = & S^{observed}_{350\mu\mathrm{m}} \times C_{100} \times C_{160} \times C_{250}\\
      \\
      															& = & S^{observed}_{350\mu\mathrm{m}} \times S^{best fit}_{250\mu\mathrm{m}} / S^{observed}_{250\mu\mathrm{m}} 
        \end{array} 
        \label{EQ:f350cor}
    \end{equation}

    													   
  \begin{equation}
  \begin{array}{cll}      															   
	S_{500\mu\mathrm{m}}^{corrected} & = & S^{observed}_{500\mu\mathrm{m}} \times C_{100} \times C_{160} \times C_{250}  \times C_{350} \\  
	\\
	  														  & = & S^{observed}_{500\mu\mathrm{m}} \times S^{best fit}_{350\mu\mathrm{m}} / S^{observed}_{350\mu\mathrm{m}}\\
  \end{array} 
  \label{EQ:f500cor}
    \end{equation}
  
When no measurement of the source exists in the band $k$ then
we set the constant as $C_{k}=1$. 

When applying these iterative extrapolations, 
we observe that the original colour dispersion for clean sources 
(left panel of Figure~\ref{FIG:bestfit}) is reduced in all 
\herschel\ bands (right panel in Figure~\ref{FIG:bestfit}).
This means that the colour dispersions are correlated from
one band to the other, hence the 
knowledge of the shorter wavelengths can be used 
to improve the extrapolation to the largest wavelengths. 
We can consider those corrected flux densities 
as good approximations to the real emission of 
the 24\,\micron\ sources in the FIR bands. 
Figure~\ref{FIG:rms_deboost} shows the dispersion
of the observed and corrected colours for all \herschel\ bands. 

We find that the use of this iterative offset correction for the colours at shorter wavelengths (as in Figure~\ref{FIG:bestfit}-right) reduces the MAD of the dispersion of the FIR/24\,\micron\ colour with redshift to 0.15$\pm$0.04 dex, \ie [0.14, 0.12, 0.13, 0.19] dex at [160, 250, 350, 500]\,$\mu$m, respectively (no correction can be applied to the 100\,$\mu$m band obviously), as compared to the initial value of 0.23$\pm$0.07 dex (as in Figure~\ref{FIG:bestfit}-left). Thus, the colour dispersion is typically divided by two with our technique, which implies that the prediction of the flux density in the most extreme band at 500\,$\mu$m is known within 55\% (within 32\% at 250\,$\mu$m), starting from the knowledge of the source redshift and flux density in the shorter bands.

To summarise the previous discussion, our aim here consists in demonstrating that we can robustly predict the flux density of an individual galaxy at wavelength ``$k+$1" (\eg 500\,$\mu$m) on the basis of a best-fitting \textit{Herschel}/24\,\micron\ colour trend that depends on redshift and on a set of colours measured by the coefficients $C_{k}$ (\eg $C_{100}$, $C_{160}$, $C_{250}$, $C_{350}$) that mimic the variety of IR SEDs of individual galaxies. What do not claim to demonstrate that all galaxies share the exact same SED, but instead that the measurement of the $C_{k}$ coefficients can be used to improve the prediction of the flux density of individual galaxies at wavelength $k+$1, where blending is more problematic than in the shorter bands. If our assumption was incorrect, we would not observe a reduction of the dispersion as we do in Figure~\ref{FIG:bestfit}-right and in Figure~\ref{FIG:rms_deboost}-bottom. The residual dispersion combines the uncertain extrapolation from $\lambda$ $\leq$ $k$ to  $\lambda=k$ with noise due to instrumental effects and the source extraction process. In the following section, we will use these colours to inject mock \textit{Herschel} sources in a noise map to produce a fully consistent mock \textit{Herschel} image by accounting for the measured $C_{k}$ coefficients when building band $k+1$.

  \begin{figure}
      \includegraphics[width=8cm]{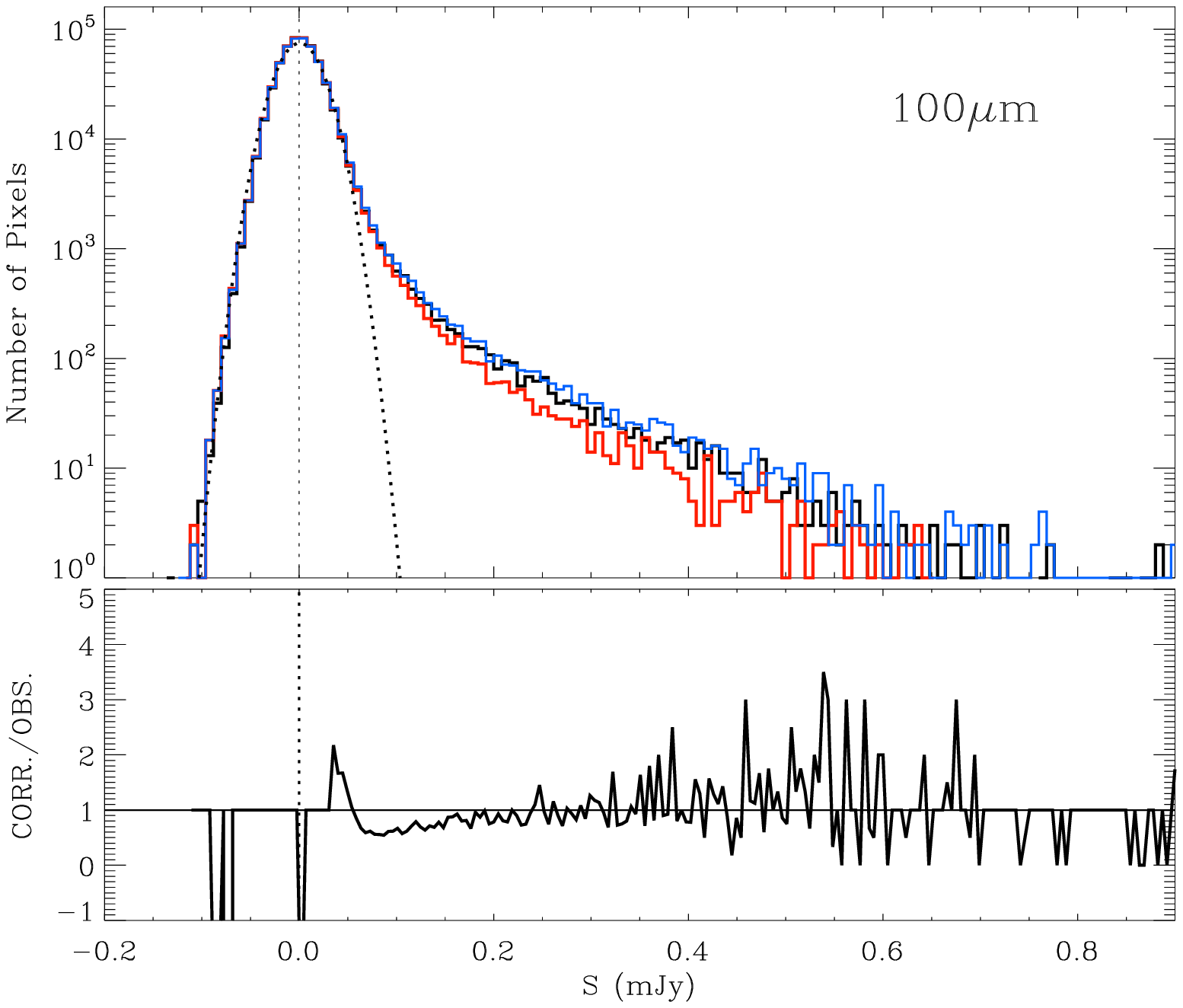}
      \includegraphics[width=8cm]{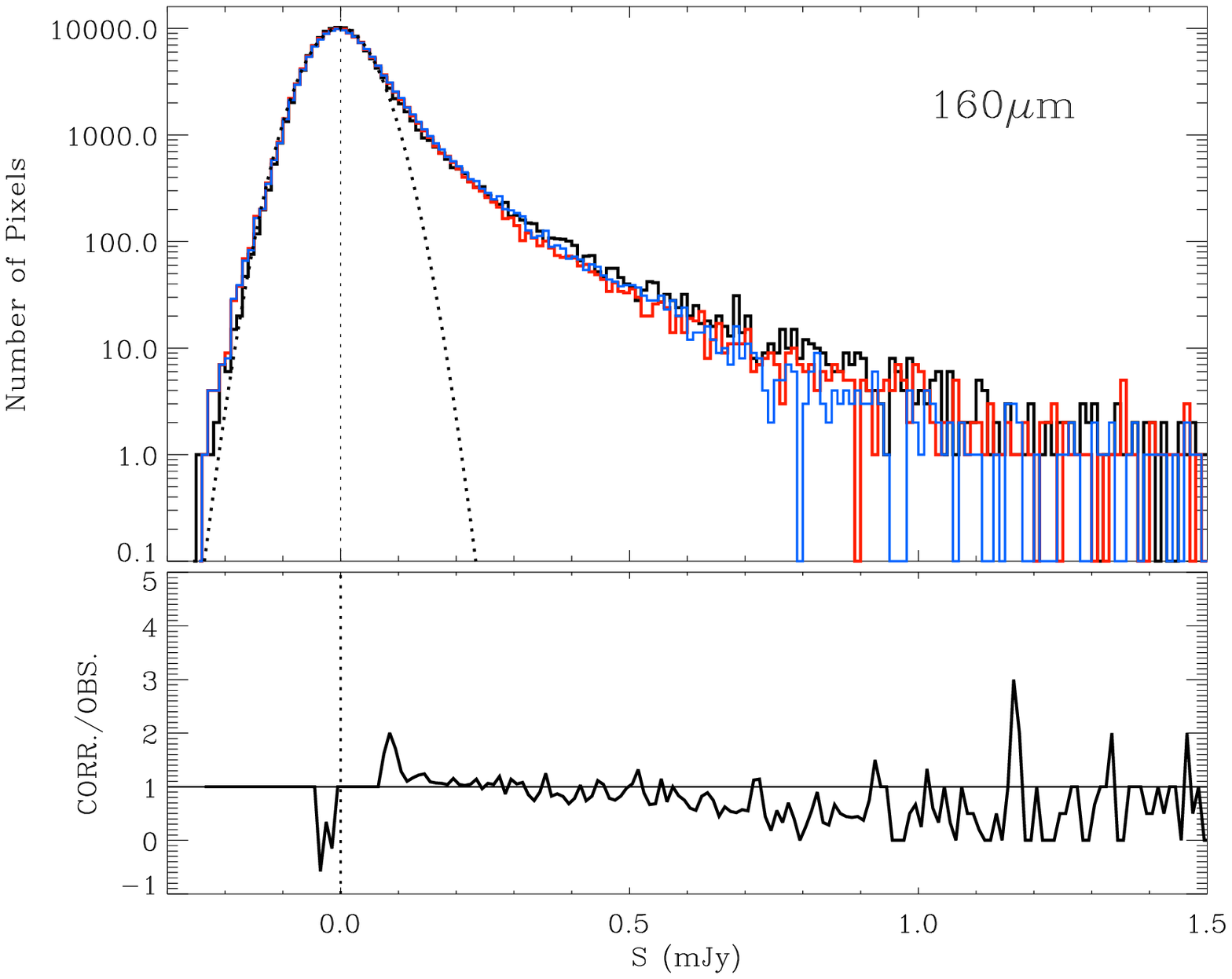}
      \caption{Histogram of the flux distribution of pixels 
      for \herschel\ maps and simulated mock images: observed (black), best-fit (red), and 
      iteratively corrected (blue) fluxes. The corresponding pixel distribution of the noise maps 
      used in the simulations are also presented (dashed lines). 
      The panels below each histogram show residuals coming from ratio of the 
      observed \herschel\ map to the iteratively corrected map after subtracting the instrumental noise
      as a function of the flux density.
      The ratio varies around 1 along the flux bins. The similarity between the observed and mock images 
      is remarkable in particular in the 
      flux bins where most of galaxies contributing to the cosmic infrared background are present (See \S \ref{SEC:cirb})
      } 
       \label{FIG:pofdpacs}  
   \end{figure}      
      
       \begin{figure}
      \includegraphics[width=8cm]{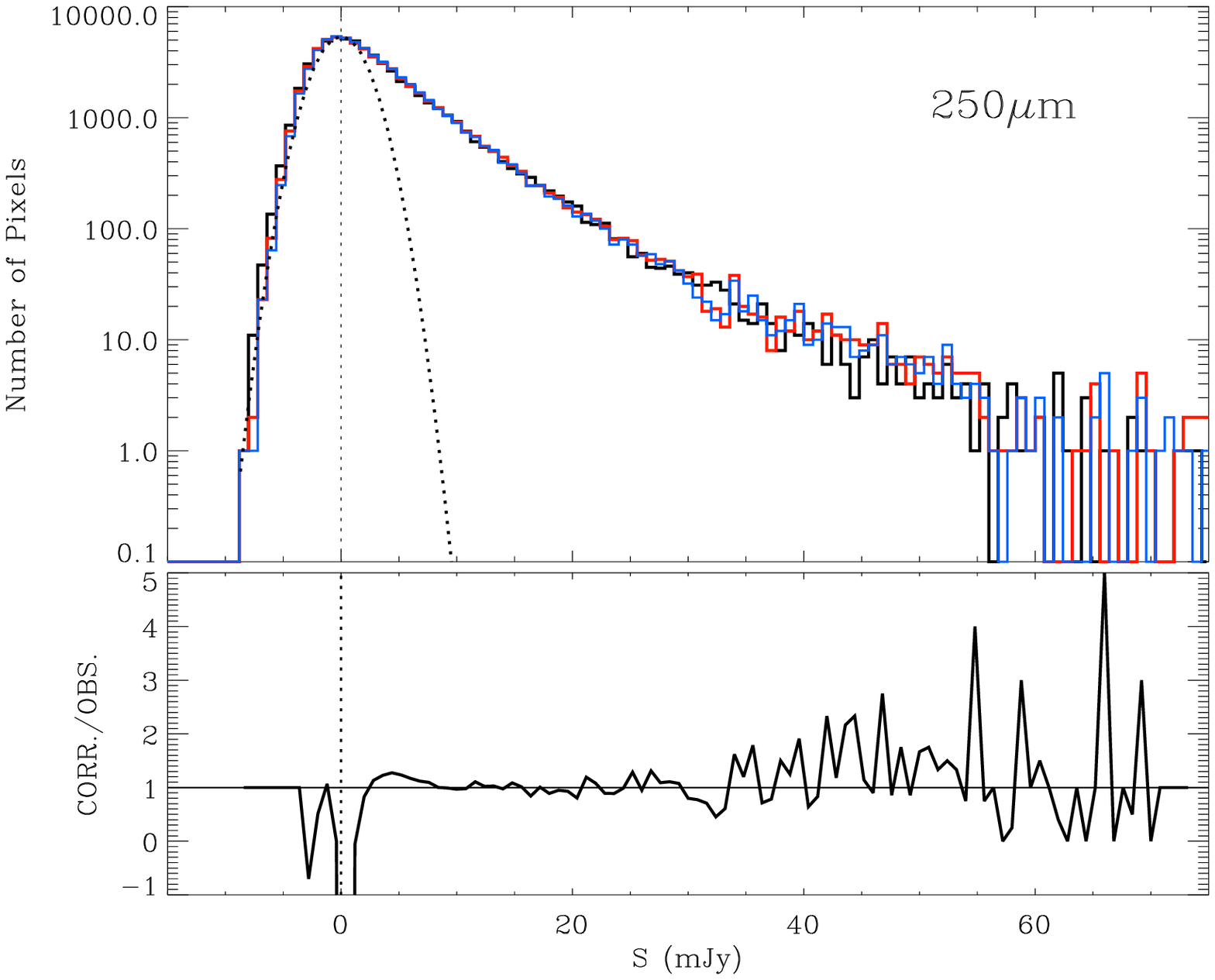}
      \includegraphics[width=8cm]{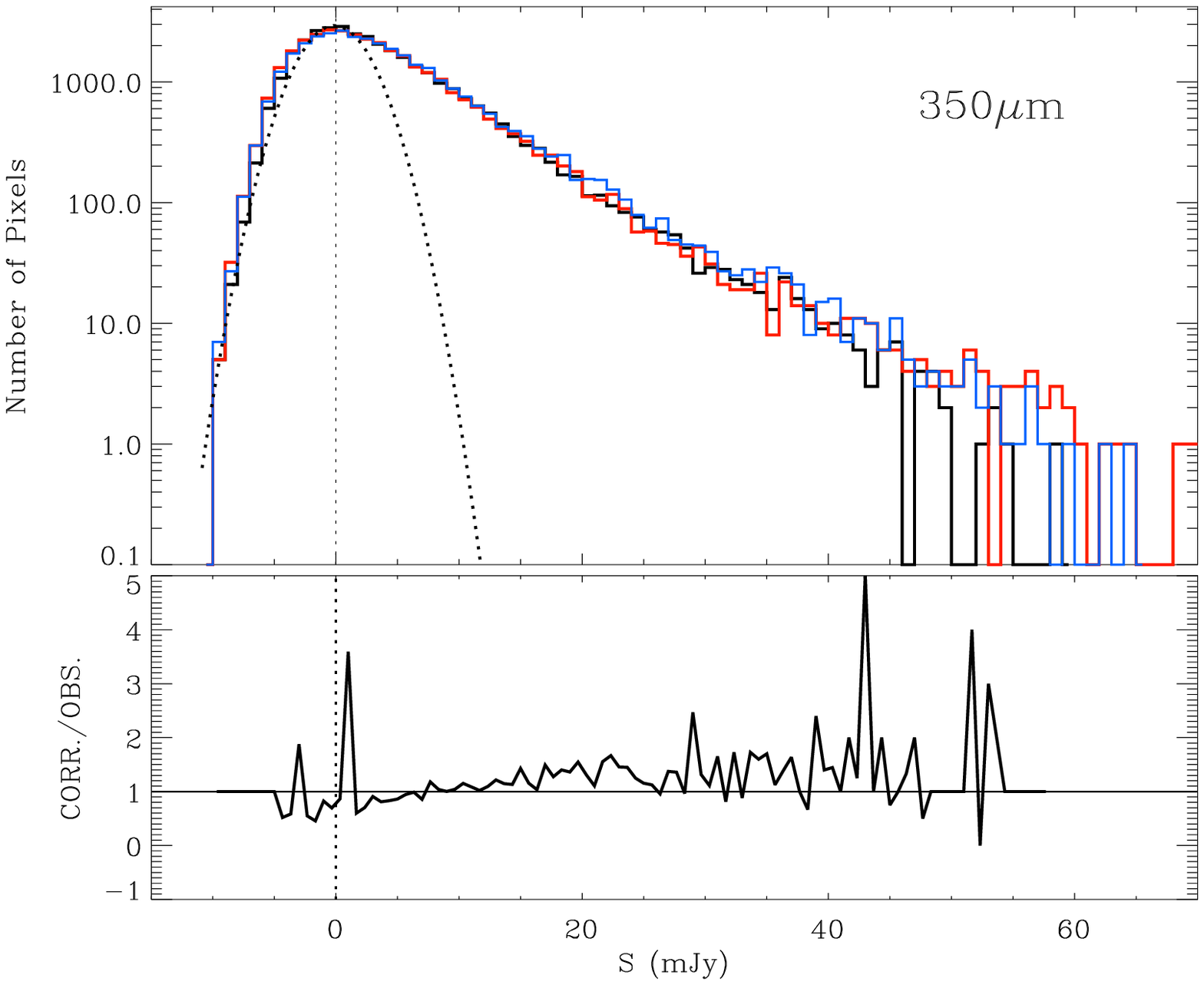}
      \includegraphics[width=8cm]{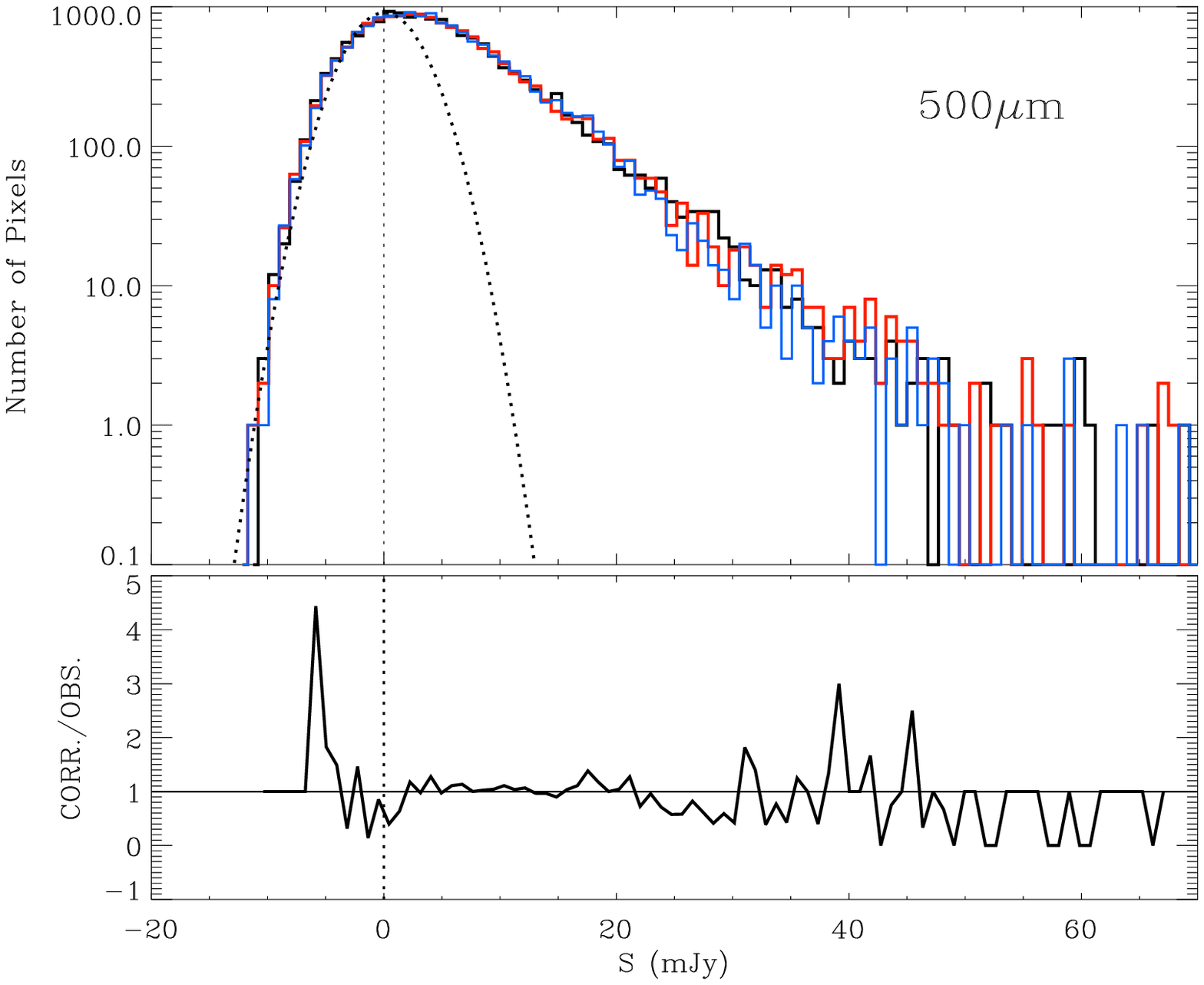}
      \caption{Same as Figure~\ref{FIG:pofdpacs} but for 250\,\micron\ 
	(upper panel), 350\,\micron\ (middle pannel), and 500\,\micron\ (lower panel)
      } 
       \label{FIG:pofdspire}  
   \end{figure}

\section{Production of realistic mock \herschel\ images}
\label{SEC:imagemake}
\subsection{Instrumental noise, PSF, and astrometric uncertainty}
\label{SEC:noiseastrom}
To produce mock images as realistic as possible, 
all characteristics of the actual \goodsh maps need to be reproduced
such as the pixel scale, instrumental noise and point spread function (PSF)
 to convolve the input fluxes.
 
We start with a map only containing instrumental noise, with no source, which we build by dividing the individual scan maps of the real \pacs and \spire observations in two equal sets and subtracting one half to the other. To avoid being affected by different shifts in the positions of the half-maps, we combined separately the odd and even individual scan maps, rather than taking the first half and the second half which would induce a time shift between the two mosaics. The resulting maps are devoid of sources, since those were subtracted in the process being equally present in both mosaics, but contain the instrumental noise. To reach the actual instrumental noise level present in the full mosaics, this noise map needs to be divided by a factor two\footnote{This factor 2 combines two effects: (i) each half-map contains half of the exposure time, hence a noise level that is\,$\sqrt{2}$ larger than the real noise map; (ii) when we subtract the two half-maps, the noise is combined quadratically hence becomes larger by another factor\,$\sqrt{2}$. The two effects combine to make the resulting noise map twice noisier than the real map, consequently the signal in that map needs to be divided by two to mimic the real instrumental noise map.}.
   
Individual sources were then injected by attributing to each one of the 2704 MIPS-24\,$\mu$m priors a flux density in each one of the \textit{Herschel} bands using the method described in Section~\ref{SEC:colour}, \ie the \textit{Herschel} over 24\,$\mu$m colour at the redshift of the sources. Sources with no photometric nor spectroscopic redshift were attributed a universal \textit{Herschel} over 24\,$\mu$m colour as discussed in Section~\ref{SEC:colour}.

The mock images were built using the observed PSF of the asteroid Vesta for \pacs and a Gaussian model for the \spire PSF, including a random astrometric error of $0.35"$ to the source positions. If we had a perfect knowledge of the position of the sources and if there was no instrumental noise, the confusion noise would be highly reduced by the use of priors in the source extraction. This can be tested by generating mock images with these conditions and sources can then be deconvolved with exquisite accuracy. For that reason, even though it is said that the confusion comes from the \textit{Herschel} beam, it is because the PSF is not perfectly sampled, its central position linked to the prior is uncertain, and that there is instrumental noise on top that produce a combined effect that forbids the clear deblending of nearby sources. In order to include a position uncertainty, we have performed a blind source extraction on the real images and compared the centroid of the extracted sources to the one of the prior positions. This comparison gave an offset of $\sim$0.4$"$ rms that we used to randomly offset our mock sources. This uncertainty results from the combination of the \textit{Herschel} relative pointing error (0.2-0.3$"$) with the remaining absolute astrometric uncertainty and with the precision of the determination of the centres of sources in the \textit{Herschel} maps. The absolute pointing accuracy of \textit{Herschel} is $0.8"$ but this was corrected by stacking positions of the brightest 24\,$\mu$m sources on each individual scan maps before projecting them in the final mosaic. An offset was then computed by comparing the centroid of the detection in the stack to the centroid of the postage stamp images, \ie the centre of the actual sources.

In the following two sections, we demonstrate that the mock \textit{Herschel} images designed from 24\,$\mu$m priors do contain the same amount of confusion noise than the real images and that they incorporate more than 95\,\% of the SPIRE 250 to 500\,$\mu$m sources at the depths reached in the present survey.

\subsection{Test of the \textit{realism} of the mock images from their pixel distributions}
\label{SEC:nsources}

A critical issue in the production of \herschel\ mock images is to make sure that mock sources reach faint enough flux limits to reproduce the confusion noise level due to the sea of faint undetected sources. Here we have assumed a priori that the list of 24\,\micron\ priors was large and deep enough to include such faint sources making the confusion noise level. In order to test this, one needs to compare the histogram of pixel distributions in the real and mock \herschel\ images. 

The close similarity between the real and mock images can be seen in Figure~\ref{FIG:stamp}. Quantitatively, this similarity goes beyond significantly detected sources as shown by the comparison of the pixel distributions of the real and mock images.
 The upper panels in Figures~\ref{FIG:pofdpacs} and 
\ref{FIG:pofdspire} show the pixel distribution for the observed 
\herschel\ (in black) and mock (red and blue) images. The mock images (in blue) were generated using the iteratively corrected flux densities as described in the previous section. Mock images using only the best-fitting relations between 24\,\micron\ and \herschel\ flux densities as a function of redshift were also built, and their resulting histograms are shown (in red).
 This implies that the simulations contain enough faint sources to mimick the effect of confusion. 
 Therefore, it is not necessary to include additional 
prior sources from other catalogues in shorter wavelengths (\eg optical or UV)
to reproduce the background of faint sources responsible for the bulk of the confusion noise.
The noise introduced in the simulation is also show as a Gaussian distribution of pixels (dashed lines).
The residual of subtracting the instrumental noise 
to the observed and simulated maps corresponds to the light 
produced by the galaxies in all redshifts in the GOODS-North field and its sum is the integrated galaxy light.
The lower panels in the same figures show the ratio between the noise-subtracted distributions of the
iteratively corrected image and the observed \herschel\ map. The ratio fluctuates close to a value of 1 
in all the flux bins.
The recovery of the this fossil light in the simulated and in the real maps is very similar confirming the
again the accuracy of the simulation making when the correct instrumental noise and sources with realistic flux densities are used.
 
Moreover, these faint sources are also the ones that make the cosmic IR background, hence can be used as a reference to determine the fraction of the background that is actually detected as compared to the one present in the images (see \S \ref{SEC:cirb}). This confirms the results of \cite{dole06}, who found that 
24\,\micron\ selected sources contribute to more than 70\% to the cosmic IR background at 70\,\micron\ and 160\,\micron. 

Once the set of realistic mock images are built, it is possible to extract their source fluxes  and compare them with the input data to quantify the level of confusion in each source.
Since we use the actual position of individual sources, the good quality of the mock images indicates that we can use them to not only to quantify statistically our source extraction method and the level of the confusion noise, but also determine the robustness of flux density measurements on individual positions in the images. 

\begin{figure*}
\centering
\includegraphics[width=14cm]{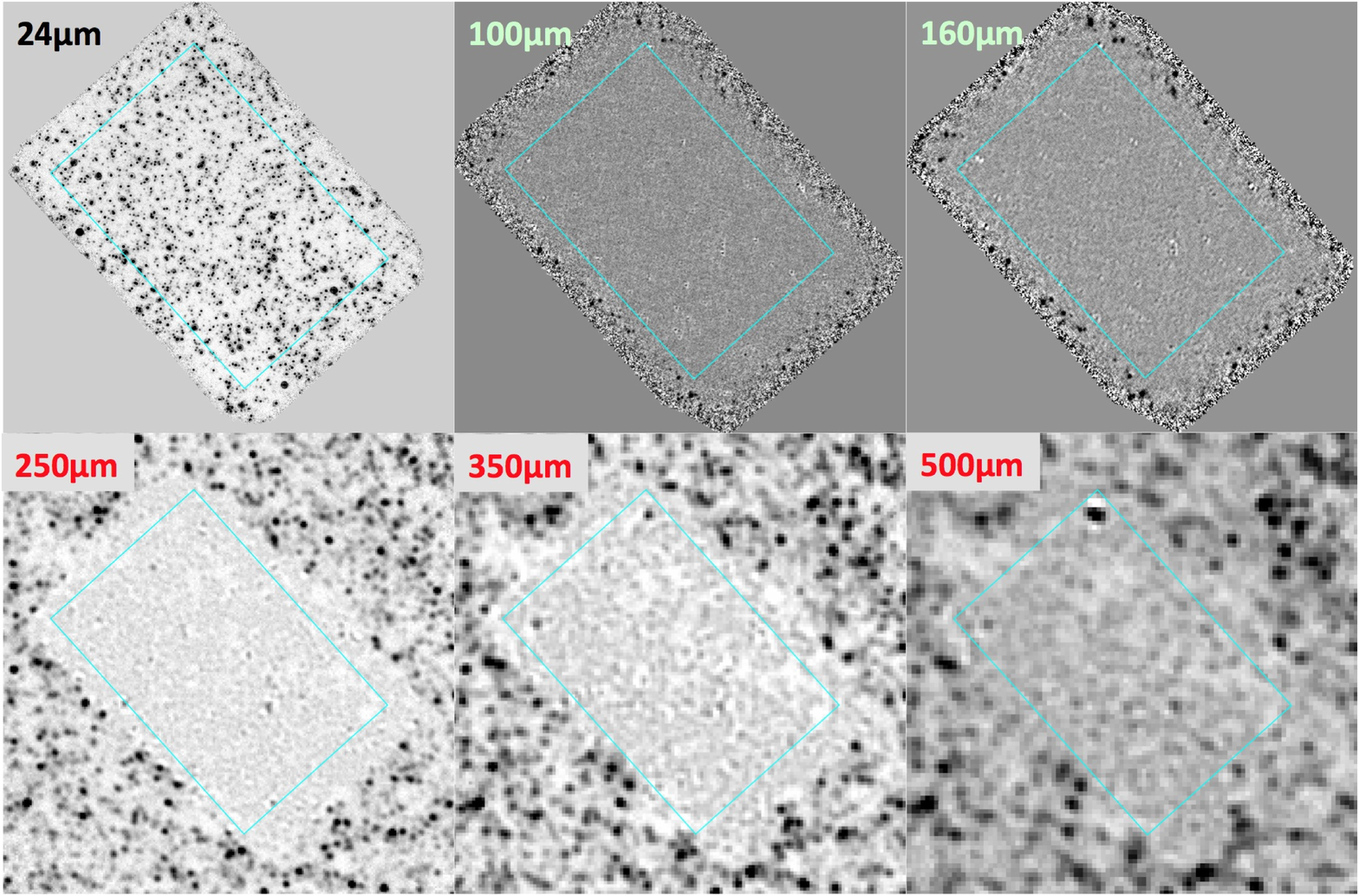}
\caption{Residual images in GOODS-North field after source extraction using 24\,\micron\ priors in each band. The maps show how efficient is the extraction process in the area covered by the prior catalogue (blue rectangle), showing only very few cases where the remnants of bright sources badly subtracted are still present in the SPIRE residual maps (black spots).
}
\label{FIG:resim}
\end{figure*}

\subsection{Are we missing a population of 24\,$\mu$m-dropouts in our 24\,$\mu$m-based mock Herschel images ?}
\label{SEC:nomissedsrc}

A pre-requisite to build the realistic mock \textit{Herschel} images following the process, as described in Section~\ref{SEC:imagemake}, was to start with 24\,$\mu$m seeds. The advantage of this choice is obvious since we start from a wavelength probing dust emission with the best spatial resolution, a resolution that is itself based on the IRAC one, \ie at the sub-arcsec level. However, the number of seeds is limited to 2704 sources detected at 24\,$\mu$m above the threshold of 20\,$\mu$Jy and, although in theory, it was shown that this threshold is enough to detect fainter objects (\ie less star-forming galaxies) than the \textit{Herschel} bands at the depths used in the \goodsh\ survey (see \citealt{elbaz11}), we now wish to confirm that a large population of \textit{Herschel}, and more particularly SPIRE sources, are not absent from our simulated mosaics.

We already discussed in Section~\ref{SEC:dataset}, the existence of $<$1\% 24\,$\mu$m-dropouts that are detected with PACS \citep{magdis11}. We wish here to address the remaining issue of the SPIRE sources. Two possibilities are offered to address this issue. As a first step, one can study the residual maps after having subtracted sources from the SPIRE images at the positions of the 24\,$\mu$m priors. The \textit{Herschel} residual maps are presented in Figure~\ref{FIG:resim}. The area used for the catalogue production where the noise level is homogeneously and deepest in the PACS bands is identified with a blue rectangle in all images. We show for comparison the MIPS-24\,$\mu$m image on the upper-left image. The black spots are sources in all images. Note that the SPIRE images cover a much larger area than the PACS ones, which explains the presence of a large number of sources outside of the blue rectangle. One can see that the PACS images contain no clear residual sources, while there remains 2-3 spots in the SPIRE residual maps. These spots are systematically found close to bright sources in the mosaics making it difficult to know for sure whether they are truly new sources that would be seen with SPIRE but be 24\,$\mu$m-dropouts, or if they are just residual emission from badly subtracted sources due to the combined effects of noise and uncertain position accuracy of the priors. We therefore decided to generate another set of simulations to disentangle these two possibilities, \ie true 24\,$\mu$m-dropouts vs noise artefacts.

We designed a more complete simulation that was made out of optical priors following the recipe described in \cite{schreiber14} (see Appendix B) that we briefly summarise hereafter. A list of 23672 optical priors were used, an order of magnitude more than the 2704 MIPS priors, to generate these simulated mosaics. These sources were selected to have a 5-$\sigma$ detection in the near-IR ($K_{\rm S}<24.5$). Stellar masses were derived from the fit of their optical to near-IR magnitudes, and then converted into a star-formation rate (SFR) using the SFR-M$_{\star}$ scaling law that has been observed from $z$=0 to 3.5 with a 0.3 dex scatter in \cite{schreiber14}. A small percentage of 2\% of the sources were attributed a 0.6 dex higher SFR to account for starbursts. We note that we do not use the optical-NIR SEDs to derive SFR since this would imply having a perfect knowledge of the extinction and star-formation history of galaxies, and that the regions emitting in the optical-NIR are representative of the colours of the ones where star formation is taking place. This is far from clearly demonstrated and in particular the SFR of galaxies, which present an excess with respect to the main sequence one, are systematically underestimated. Consequently, in order to make sure to have both normal star forming galaxies, as well as a population of starbursts that are systematically missing from optical-NIR estimates, we decided to distribute SFR according to the observed distribution of SFR for each stellar mass bin.
Then SFRs were converted into a total IR luminosity, L$_{\rm IR}$(8--1000\,$\mu$m) using the  \cite{kennicutt98} relation SFR=1.72$\times$10$^{-10}$ $\times$L$_{\rm IR}$. The \cite{chary01} IR SEDs that were best-fitting the stacking data were then used to distribute the L$_{\rm IR}$ to each one of the \textit{Herschel bands} in order to generate a catalogue of input fluxes.

\begin{figure*}
      \includegraphics[width=6cm]{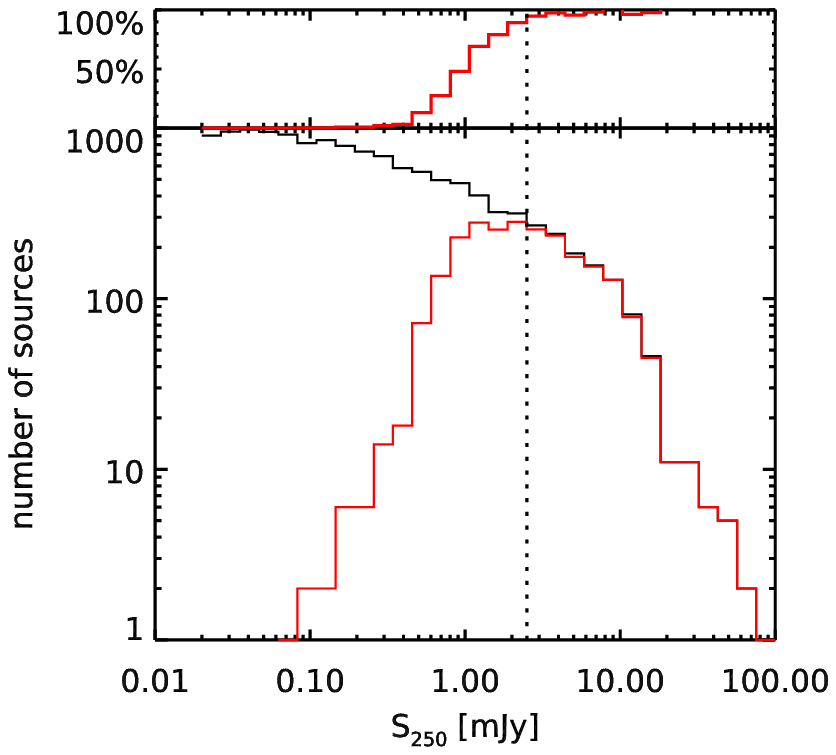}
      \includegraphics[width=6cm]{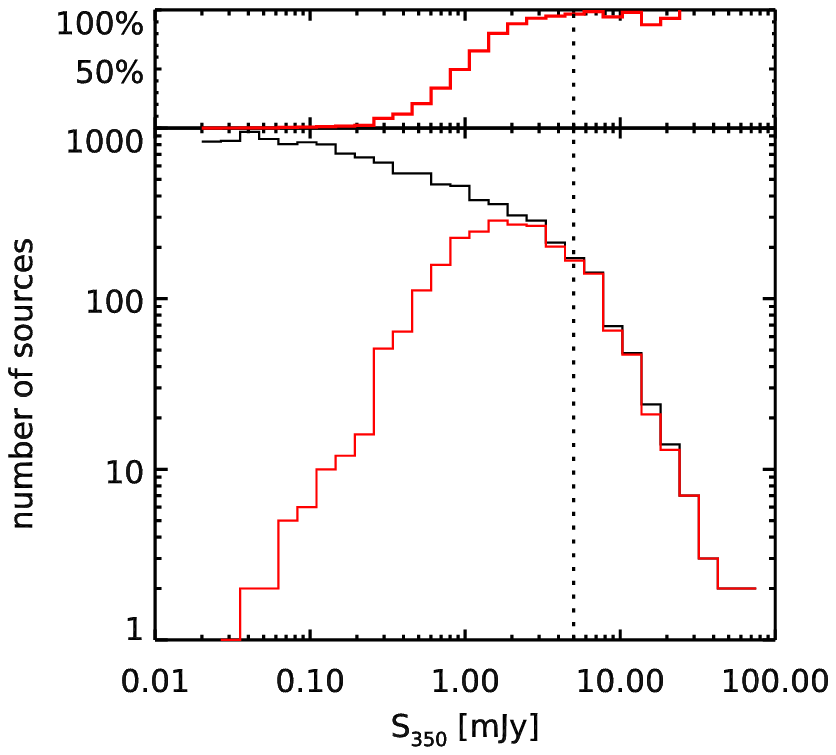}
      \includegraphics[width=6cm]{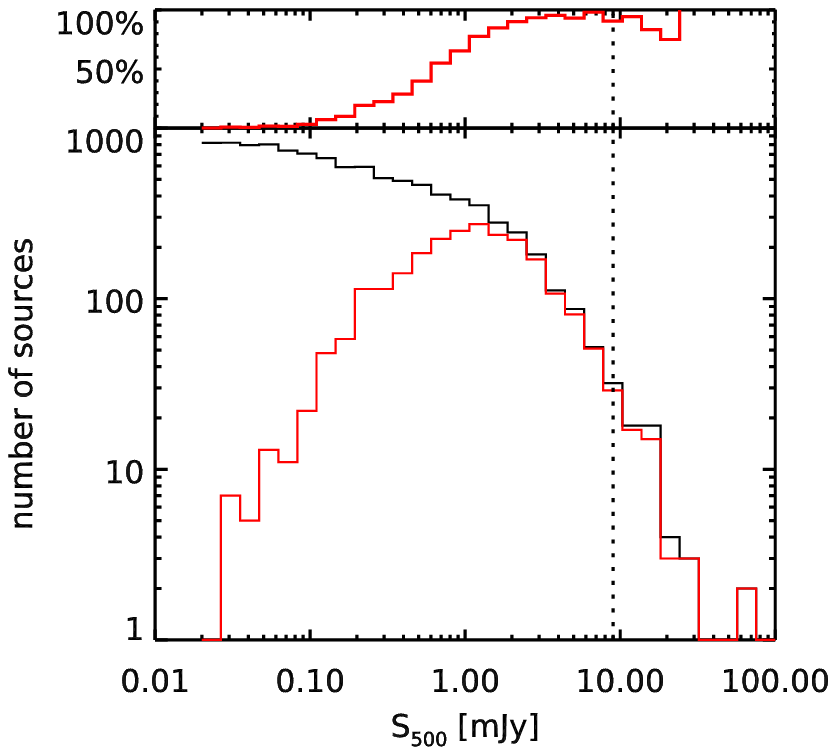}
      \caption{Number (lower panels) and fraction (upper panels) of 
      24\,\micron\ sources having a SPIRE counterpart (red) and those with 
      no associated SPIRE detection as a function of the flux density in 250\,\micron\ 
	(left panel), 350\,\micron\ (central pannel), and 500\,\micron\ bands (right panel).
	Vertical lines show the flux limits reached using the simulations found in the present work
	2.5~mJy for 250\,\micron\, 5.0~mJy for 350\,\micron, and 9.0~mJy for 500\,\micron. 
	(see \S~\ref{SEC:pai} for more details of how these flux limits were obtained).
	 Consequently, the missing 24\,\micron\ population at the flux limits of the simulations
	 is very marginal.
      } 
       \label{FIG:fracmipsSPIRE}  
   \end{figure*}

Note that we validated that the sources produced following that recipe did also follow the colour trends shown in Figure~\ref{FIG:bestfit}. The advantage of this other set of mock \textit{Herschel} sources is that it goes much deeper than the present one. The inconvenient of this method is that it only statistically reproduces the \textit{Herschel} sources, \ie sources are given a SFR that reproduces the dispersion of the SFR-M$_{\star}$ ``main sequence", while here we try to stay as close as possible to the real \textit{Herschel} fluxes. However this deeper simulation can be used to determine the fraction of SPIRE sources that are missed above our detection thresholds when we limit ourselves to 24\,$\mu$m priors. This is illustrated in Figure~\ref{FIG:fracmipsSPIRE}, where the number of SPIRE sources as a function of flux density for each one of the three SPIRE bands are shown (black line), as well as the sources that are also detected at 24\,$\mu$m above the threshold of our prior list (red line). As one can see, above our detection thresholds of 2.5, 5, and 9 mJy at 250, 350, and 500\,$\mu$m the fraction of sources missed when requiring a 24\,$\mu$m prior is at the percent level.

\section{The confusion noise} 
\label{SEC:confusion}

Ironically, there is a non negligible confusion around the definition of \textit{confusion noise}.
It is often defined as the limiting depth below which the number of sources gets so large that there is less than an arbitrary number of beams per sources. Using this definition, \citet{oliver10} obtain a confusion limit of $\sim$20 mJy for the three SPIRE bands defined from the depth corresponding to 40 beams/source in the SPIRE deep surveys and BLAST data predicts confusion limits for SPIRE bands of 22, 22, and 18 mJy at 250, 350, and 500\,\micron\ also equivalent to 40 beams/source \citep{devlin09}.

Alternatively, one can estimate the variance of the sky map due to the presence of \textit{unresolved sources} following the recipe of \citet{condon74} and labelled as the \textit{photometric confusion noise} in \citet{dole03}. \citet{nguyen10} estimate this noise to be as large as 5.8, 6.3, and 6.8 mJy/beam at 250, 350, and 500\,\micron, hence leading to lower limits to the 3-$\sigma$ detection limit of 17.4, 18.9, and 20.4 mJy, respectively (lower limits since instrumental noise is not included here, but discussed in the paper). Interestingly, the 40 beams/source definition computed by \citet{nguyen10} decreases from 250 to 500\,\micron\ going from 19 mJy at 250\,\micron\ to 18mJy at 350\,\micron, and 15 mJy at 500\,\micron. 

One important caveat in the use of number of beams per source is that such systematic definition not only relies on an arbitrary number of 40 instead of, \eg, 20, but more importantly assumes that the impact of confusion is the same independently of the slope of galaxy counts below this detection threshold. It is obvious that if the depth considered is close to the plateau of the integral counts, then few sources will appear at fainter flux densities, while if the number of sources below this depth presents a steep increase with decreasing flux density, the impact on the fluctuations of the sky will be much stronger.

In practice, what one needs is not a systematic definition but rather a quantitative estimate of the photometric accuracy that can be reached on most of the extracted sources. This is particularly true when confusion can be due to very faint undetected sources, as well as relatively bright sources that are located at specific positions in the maps, introducing a spatial variation of confusion. If we consider the standard Gaussian value of an \textit{rms} of 68\,\%, we can set our limit to be the depth above which more than 68\,\% of the sources (at the faintest flux limits) can be extracted with a photometric accuracy better than 32\,\%, meaning a S/N$>$3. This definition was introduced by \citet{chary04} and later used for both \textit{Spitzer} and \textit{Herschel} \citep{magnelli09,magnelli13,elbaz11}. The photometric accuracy on the extraction of \textit{Spitzer} MIPS--24\,\micron\ sources was estimated by injecting mock sources on the real maps in small number, to avoid overcrowding the field, but a large number of times. The limiting condition above, which corresponds to \dfout $= |(S_{out}-S_{in})|/S_{out}$ $\leq$0.32 for $>$68\,\% of the sources, was reached at $\sim$20\,$\mu$Jy. At such limiting depth, the number of beams per source is as low as 7 beams/source, yet leading to reliable photometric measurements for most sources.

A potential drawback of the latter method is coming from the fact that sources are injected at some specific positions in the map and even though they can be randomly disposed, one cannot estimate the photometric uncertainty on positions in the map were bright sources are located. In ultra-deep images, \textit{bright} sources are located all over the image, hence making it difficult to study variations of the confusion noise as a function of position. In order to account for the specific position of bright sources in the \herschel\ images, \citet{hwang10} and \citet{elbaz10} noted that there was a correlation between the presence of bright sources in the neighbourhood of a specific source in the \spitzer\ 24\,\micron\ and \herschel\ maps which led them to select sources using the \textit{clean index} defined in Section~\ref{SEC:clean}. 
Using the clean index to reject such objects in the GOODS fields, \citet{elbaz11} were able to reduce the detection threshold accounting for confusion (corresponding to the 3$\sigma$ limit) by a factor $\sim$3 in the SPIRE bands reaching 5.7, 7.2, and 9 mJy at 250, 350, and 500\,\micron\, respectively.

In the following, we will test this method on our realistic mock \herschel\ mosaics, improve it by studying all positions on the mosaics and quantify the limiting depth reaching the ``3$\sigma$ accuracy for 68\% of the sources" at the actual positions of the \herschel\ sources. We will show that one can reach a detection limit, that we will from now on call the \textit{local confusion limit} (since it accounts for the local noise due to bright sources as determined from shorter wavelengths), 7 times deeper than the \textit{global confusion limit} of $\sim$20 mJy at 250\,\micron\ (4 times at 350\,\micron\ and 2.5 times at 500\,\micron).

  \begin{figure}
    \includegraphics[width=8cm]{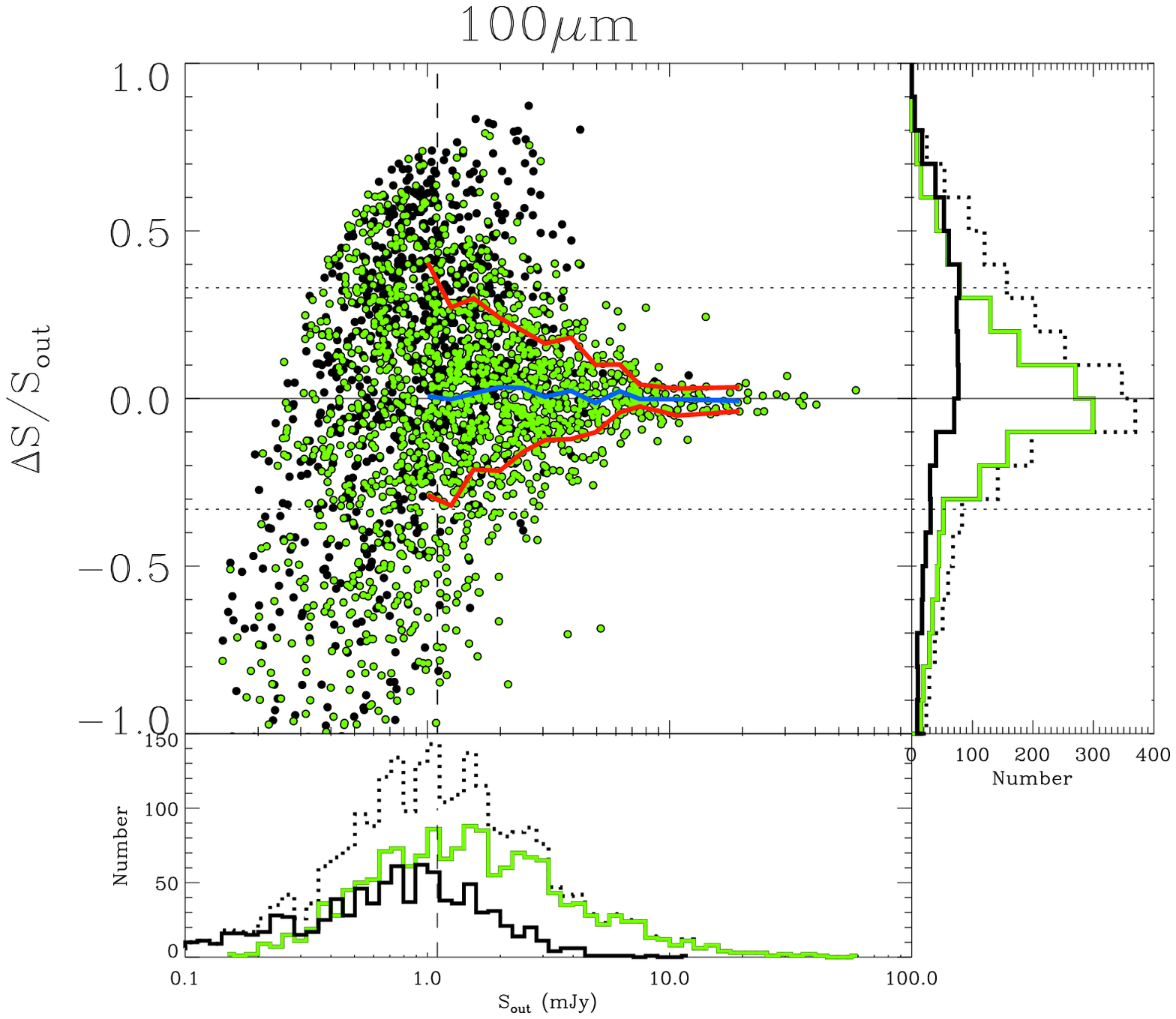}
    \includegraphics[width=8cm]{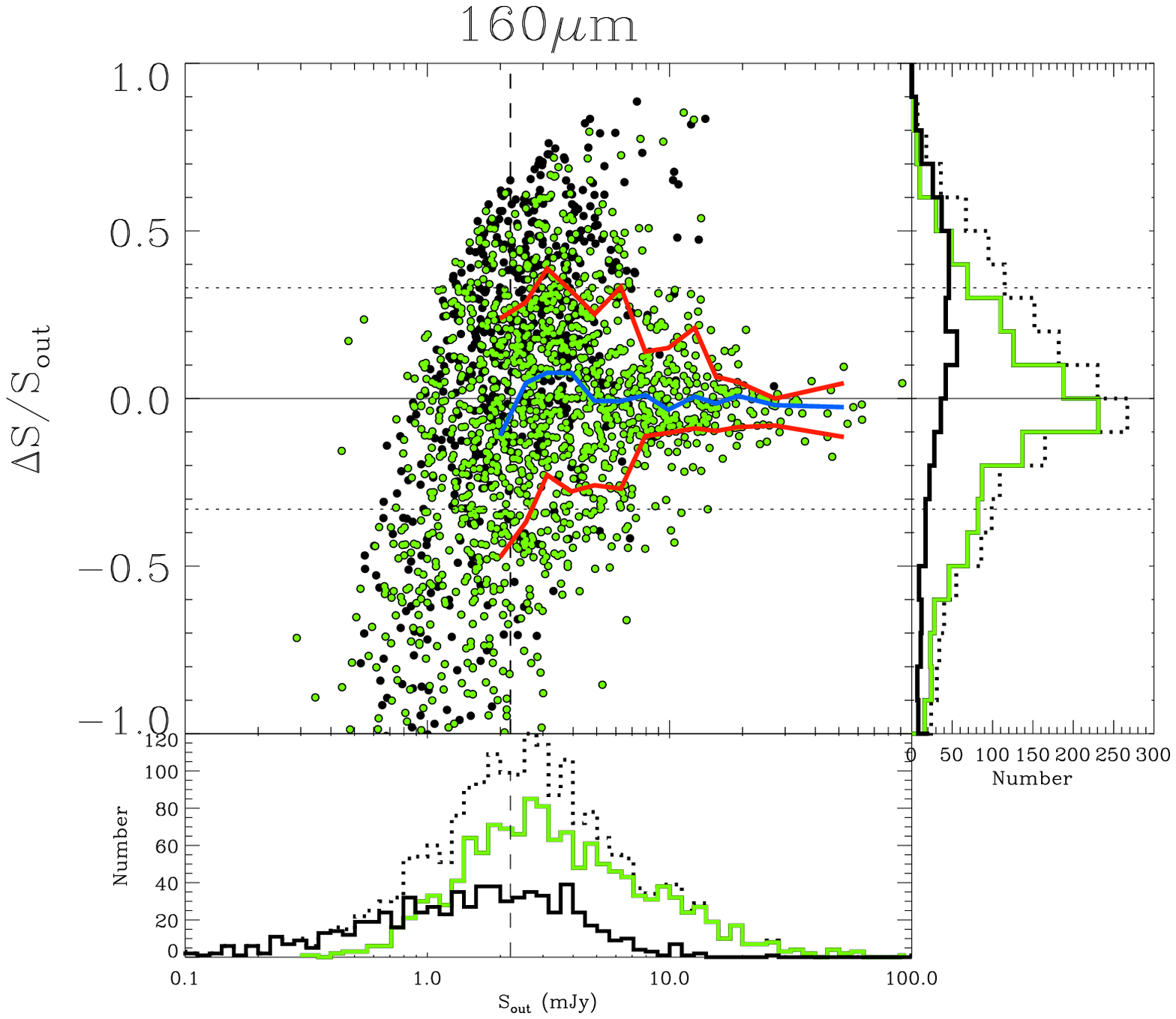}
     \caption{ Photometric Accuracy index  ($PAI$) ,
     $\Delta S/S_{out}$ $= (S_{out}-S_{in})/S_{out}$, as a function of the output flux $S_{out}$.
      The median  trend of \dfout is shown by the blue line and the red envelope 
      lines contains the 68\%  of the clean galaxies, \ie sources 
      associated with a 24\,$\mu$m source having a CI$\le$1 (green dots).
      Dotted horizontal lines at \dfout$=\pm$ 0.32 mark the
      limits that define sources having good photometry.
      The intersection between the 68\% envelope line and the good photometry
      limit defines the depths for \herschel\ data based on the simulations 
      (vertical dashed lines; 1.1 mJy, 2.2 mJy, 2.5 mJy, 5.0 mJy, and
      9.0 mJy for the 100\,\micron, 160\,\micron, 250\,\micron, 350\,\micron, 
      and 500\,\micron\ \herschel\ bands, respectively). The histograms on the bottom and right sides of each panel show 
      the distribution of the whole sample of galaxies (dotted line), the clean galaxies (solid green line) and the complementary non-clean 
      galaxies (solid black line) as a function of the flux density (\textbf{horizontal histogram}),
      and as a function of $\Delta S/S_{out}$ (\textbf{vertical histogram}). The comparison of the solid green and dotted lines shows that the proportion of clean sources remain identical in all flux density bins, \ie, there is no bias against faint or bright sources when dealing with clean sources alone.
      }
     \label{FIG:dfinpacs}
  \end{figure}

  \begin{figure}
    \includegraphics[width=7.6cm]{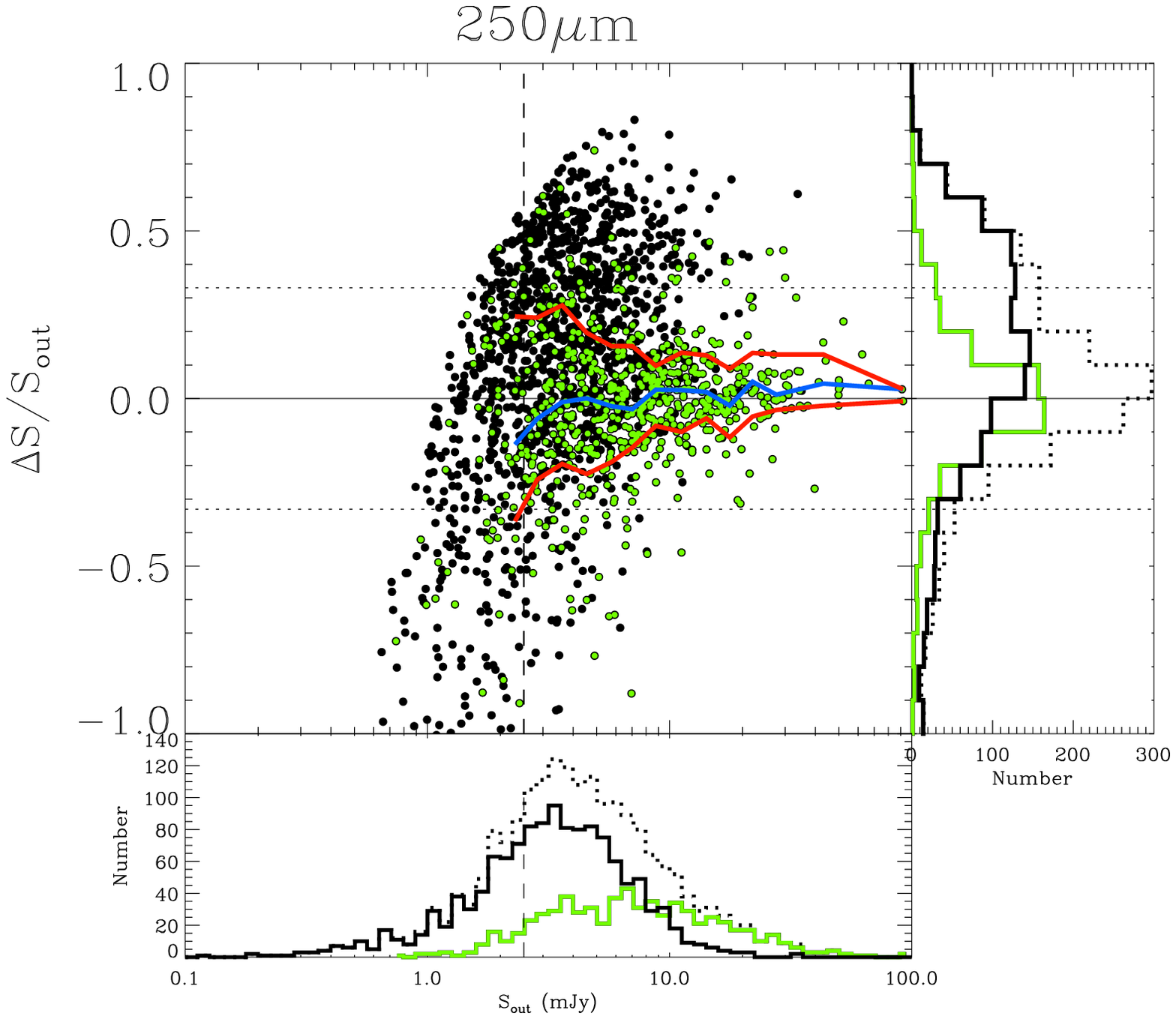}
    \includegraphics[width=7.6cm]{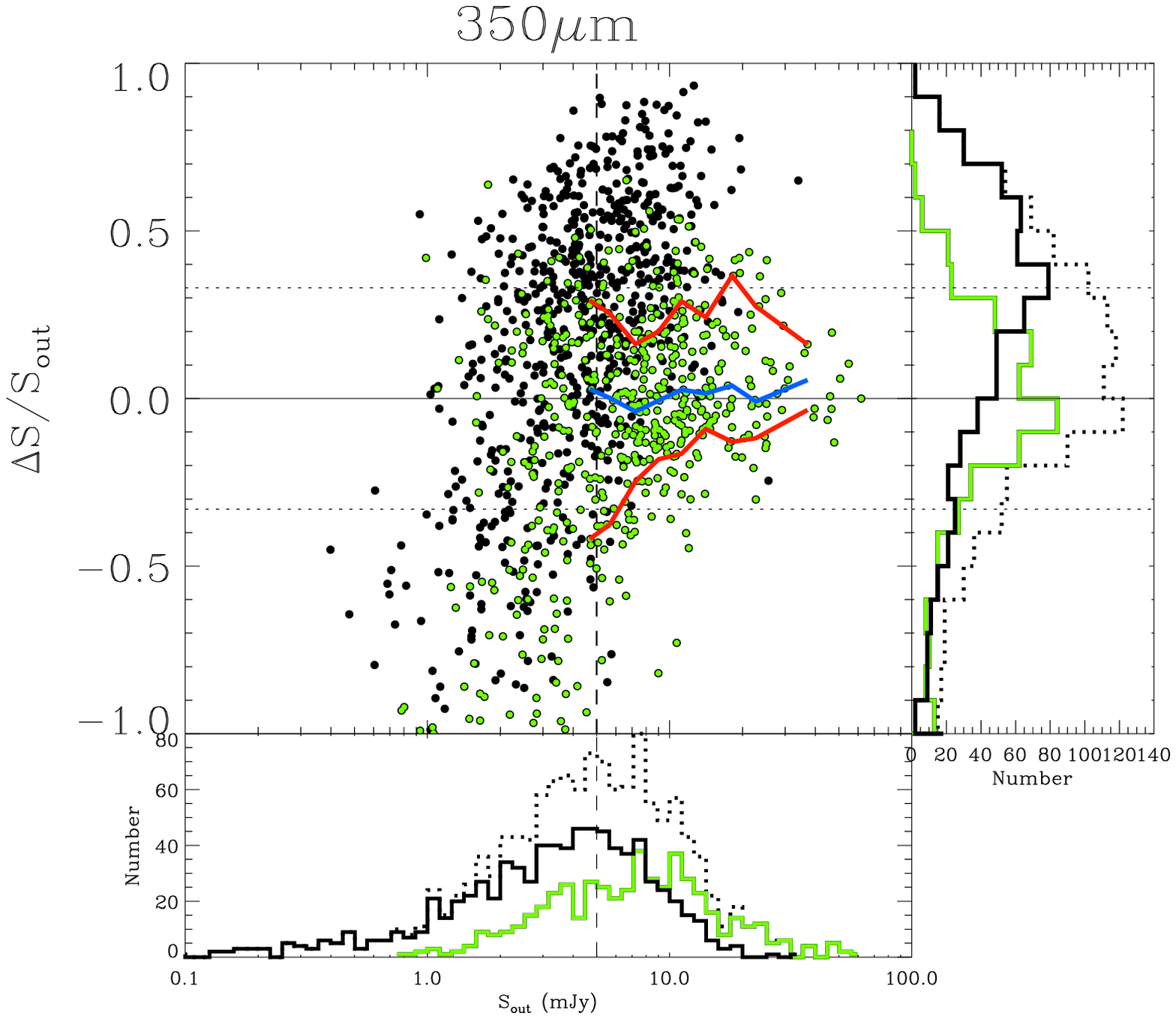}
    \includegraphics[width=7.6cm]{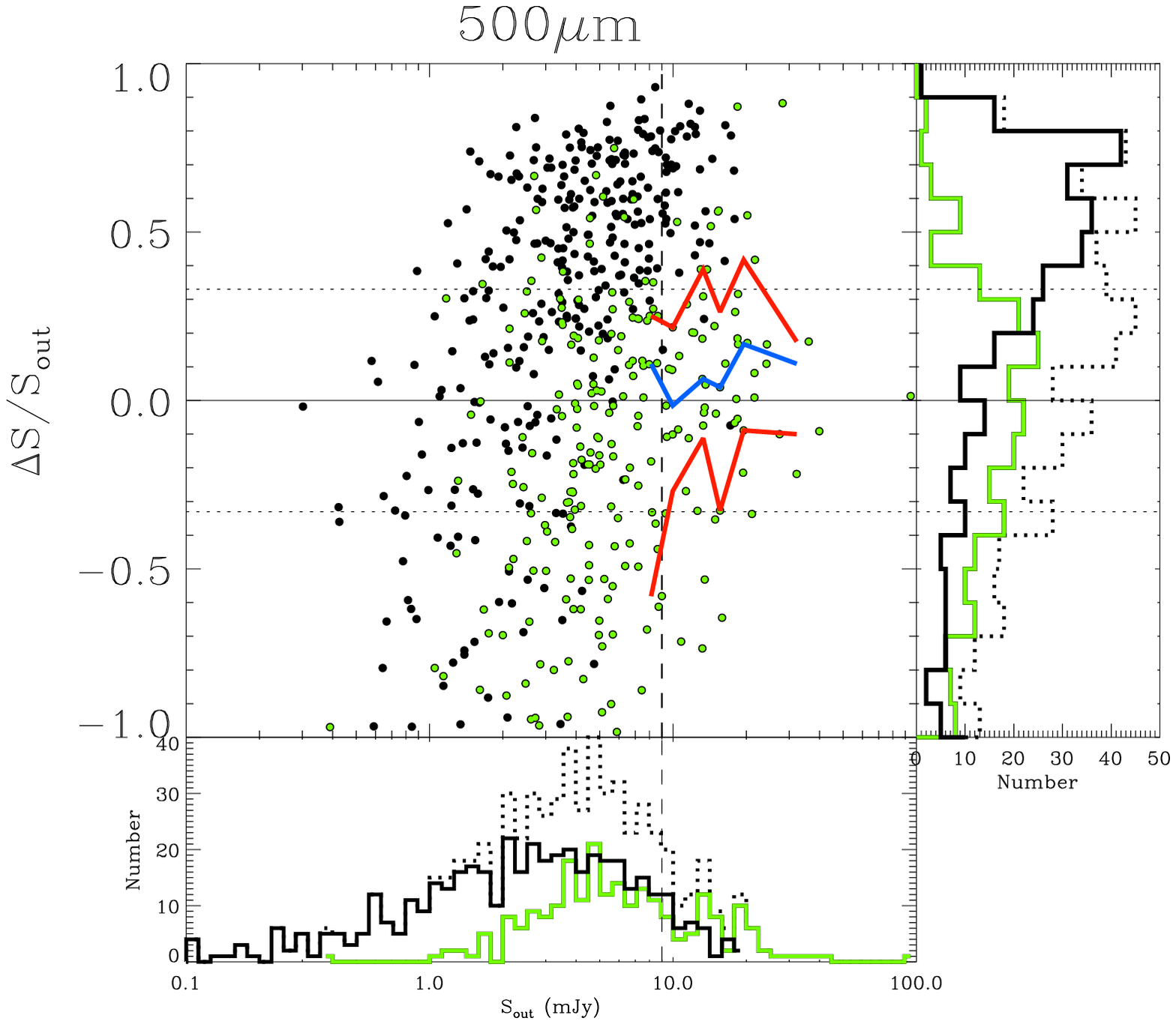}
     \caption{Same as Figure~\ref{FIG:dfinpacs} but for 250\,\micron\
        (upper panel), 350\,\micron\ (middle pannel), and 500\,\micron\ (lower panel)
      }
     \label{FIG:dfinspire}
  \end{figure} 
\subsection{The Photometric Accuracy Index  ($PAI$) }
\label{SEC:pai}
For a valid comparison between the confusion levels in the real and the simulated maps, not only the mock images have to be built as realistically as possible but also the extraction method applied must be identical to the method used to produce the \goodsh catalogues. 

The photometric accuracy of the flux extraction in the mock images will allow us to quantify with high precision in what degree the extracted sources are affected by bright neighbours and by the confusion produced by faint sources. We will use that knowledge to explore deeper levels in the detection limits of the \goodsh images. 

To quantify the confusion noise in the simulations, we define the \textit{Photometric Accuracy Index} ($PAI$) as in \cite{magnelli09}:
    \begin{equation}
         PAI = |\Delta S|/S_{out} = |(S_{out} - S_{in})| / S_{out}.
    \end{equation}
    
In the following, we will consider that sources of a given flux density, $S_{in}$, can be accurately measured if the extracted flux density, $S_{out}$, of at least 68\% of them can be extracted with an accuracy better than 32\%. We wish to introduce here an important distinction between what we will call the ``global confusion limit" and the ``local confusion limit". The global confusion limit is the minimum flux density that fulfills the requirement that $PAI$ $\leq$0.32 for 68\% of the sources. We have seen in the previous sections that we can use the information at shorter wavelengths to improve our knowledge at larger wavelengths. In particular, a source that is far from any 24\,\micron\ bright source has on average a \herschel\ flux density that is less affected by blending. Since our mock \herschel\ sources are located at the same positions as the real ones and with similar flux densities, we can use the simulations to reject a fraction of the sources that are more likely to be affected by the presence of bright neighbours than others. By doing so, we will reduce the effective area of the survey but with the advantage of decreasing the flux limit above which output fluxes are reliable. In the following, we will call such improved flux density limit the ``local confusion limit", local in the sense that we account for what happens at the location of sources. 

Figures~\ref{FIG:dfinpacs} and ~\ref{FIG:dfinspire} show $\Delta S/S_{out}$ as a function of \fout for the simulations of the GOODS-North field. 
The plots show how the photometric uncertainty increases inversely with the flux densities. Positive values of $\Delta S/S_{out}$ above 0.32 indicate that the extracted fluxes in that flux bin are overestimated. There is a characteristic asymmetry in that direction in the SPIRE bands that is typical of what one would expect from the contamination of bright nearby sources. However, the behaviour of the clean sources in these plots will allow us to obtain new depths, presented below, based on the simulations.

\subsection{Validating the ``clean index" (CI) with the ``photometric accuracy index" (PAI)}
In Figures~\ref{FIG:dfinpacs} and ~\ref{FIG:dfinspire}, all sources are shown with a filled black dot, while clean sources with at most one bright neighbour (CI$\leq$1, which is going to be our clean criterion) are represented by open green circles. 
Clean sources, as determined from the 24\,\micron\ clean index (sources with at most one bright neighbour at 24\,\micron\ with 12 and 20\arcsec in the PACS and SPIRE bands, respectively), are shown with light green dots. Clean sources are nearly indistinguishable from the full set of sources in Figure~\ref{FIG:dfinpacs} showing the 100 and 160\,\micron\ PACS results. On the other hand, the clean sample tends to be centred on the \dfout = 0 line in the SPIRE bands (Figure~\ref{FIG:dfinspire}), while the rest of the sources, those which happen to fall close to a bright 24\,\micron\ neighbour, are systematically above the dotted \dfout = 0.32 line.

This shows that the clean index is a robust method for rejecting sources that are heavily affected by confusion. At 250\,\micron\, for example, one would need to consider a limiting depth of $\sim$20 mJy to reach the requirement of 68\% of the sources with  \dfout $\leq$ 0.32, which would correspond to the global confusion limit. Instead, we can see that this requirement is reached at much fainter levels, down to a depth of 2.5 mJy after rejecting non clean sources (vertical dashed line). Using the clean index, the local confusion limit is set by the intersection between the red envelopes that encompass 68\% of the sources around the sliding median of the clean sample (blue line) and the horizontal dotted lines at \dfout = $\pm$ 0.32, which define our 3-$\sigma$ limit.
In these mock \herschel\ images, we can see that one can expect to reach a local confusion limit of 1.1 mJy, 2.2 mJy, 2.5 mJy, 5.0 mJy, and
9.0 mJy for the 100\,\micron, 160\,\micron, 250\,\micron, 350\,\micron, and 500\,\micron\ \herschel\ bands, respectively (intersection of the vertical lines with the \dfout = $\pm$ 0.32 in Fig. ~\ref{FIG:dfinpacs} and ~\ref{FIG:dfinspire}). These limits are slightly deeper than the \cite{elbaz11} ones at 250 and 350\,\micron\ as discussed in the next section.

\subsection{Comparison of the noise level in the real and mock \herschel\ images}
\label{SEC:noise}

The $PAI$ can obviously only be determined in the mock \herschel\ images for which we know both $S_{in}$ and $S_{out}$.
However, we can check the consistency of the noise level present in the real and mock \herschel\ images by comparing
the $PAI$=\dfout to the noise level attributed to individual detections in the real images as estimated from the residual maps (see \cite{magnelli13} for a more detailed description). Once detections are extracted from the real \herschel\ images, one can estimate the local noise at the position of a source by the fluctuations measured in the residual map. The robustness of the noise estimate obtained on residual images has been tested by injecting a limited number of mock sources on the real \herschel\ images, and extracting their flux density to derive their \dfout. As discussed in \cite{elbaz11} and \cite{magnelli13}, while both noise estimators agree for PACS, there is a systematic offset for SPIRE. The noise obtained from the residual maps in the SPIRE bands is lower than the actual uncertainty in the measurements of the \dfout. Typically, the 5-$\sigma$ detection threshold estimated from the residual maps corresponds to an effective threshold of 3-$\sigma$. In order to derive a local noise estimate, one needs to use the residual maps and correct the \textit{rms} by this systematic offset.

\begin{figure}
    \includegraphics[width=9cm]{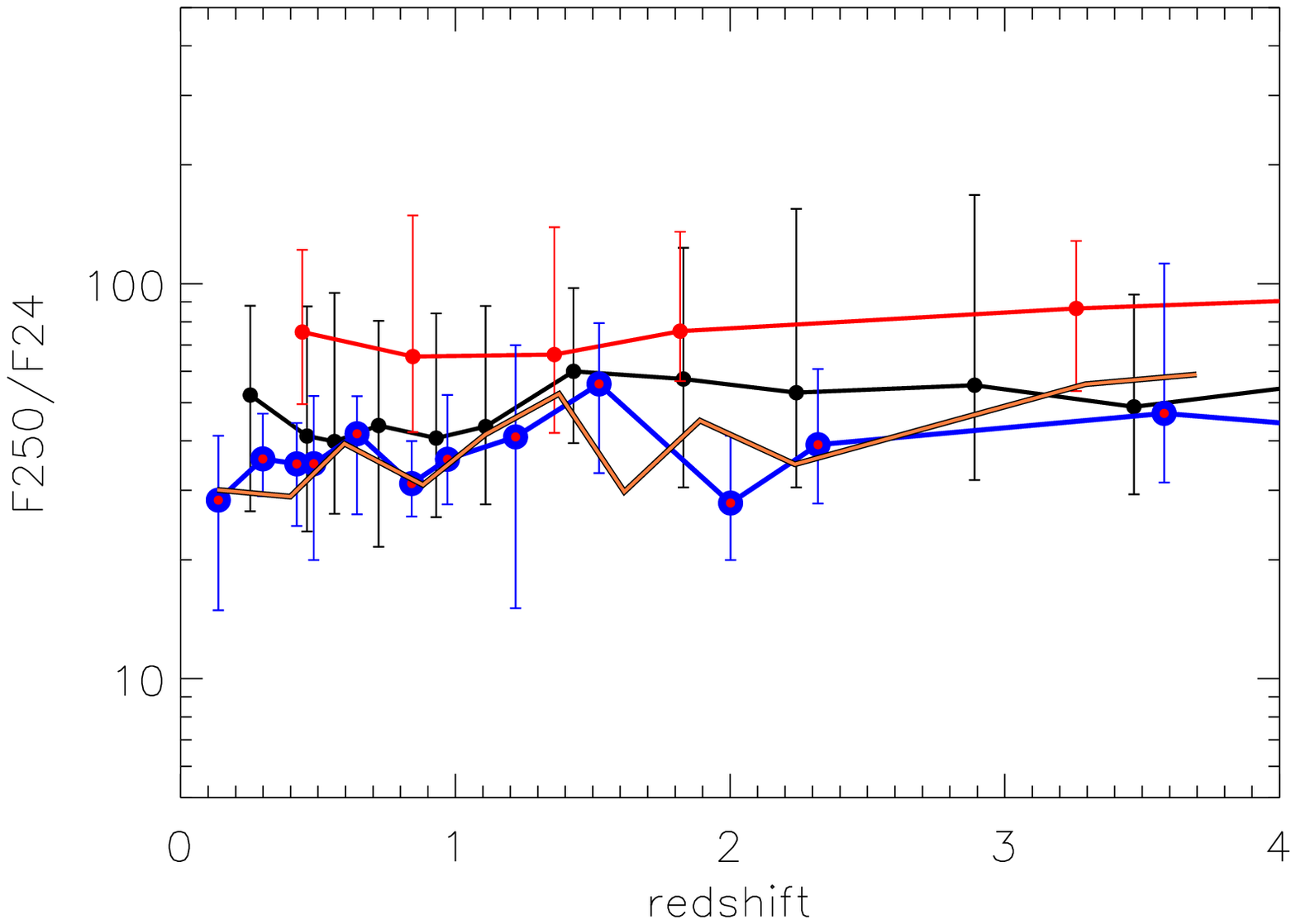}
   	  \caption{Comparison of observed $S_{250}/S_{24}$ colours for
   	  sources with flux densities between  3 $<$ $S_{250}$ $<$ 5.7~mJy
   	  selected by their associated photometry accuracy from the simulations. 
   	  The median colours of sources with a good photometric 
   	  accuracy $|\Delta S|/S_{out}$ $\le$ 0.32  (black line) follow the colour trend,
   	  of clean-selected sources CI $\le$ 1  (blue line) and the stacked sources (orange line) 
   	  while the median colours of sources having poor photometric accuracy 
   	  $|\Delta S|/S_{out}$ $>$ 0.32 (red line) are 
   	  systematically displaced to brighter 250$\mu$m fluxes, an indication of
   	  blending. 
   	  }
       \label{FIG:valid_ds}
\end{figure}
\begin{figure}
	\centering
    \includegraphics[width=7cm]{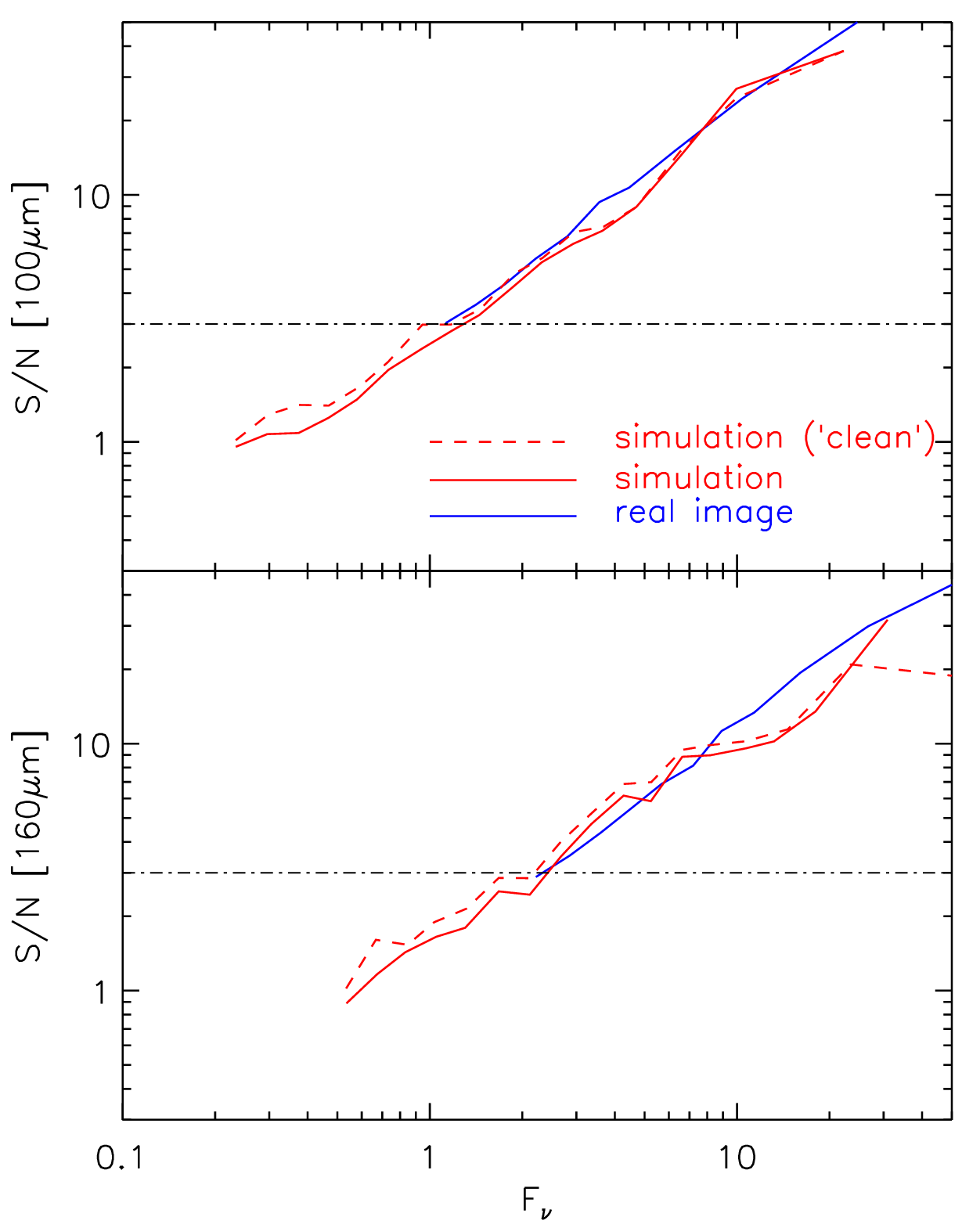}
   	  \caption{Signal-to-noise ratio as a function of flux for extracted sources in the \herschel\-PACS maps (blue), 
   	  the simulated images (solid red) and for the subset of clean sources in the simulation (dashed red line).  
   	  The horizontal lines marks the 3-$\sigma$ level.  
   	  The similarity in the trends demonstrates that the simulations reproduce well the sources of noise in 
   	  the real image, \ie the faint population of 24\,\micron\ prior sources, as well as the instrumental noise. 
   	  Flux limits can also be defined for the \herschel\ data by the 3-$\sigma$ limits reached by clean 
   	  sources (intersection between the dashed red line and the horizontal line).
   	  }
       \label{FIG:snrcompPACS}
	 \end{figure}
\begin{figure}
	\centering
    \includegraphics[width=7cm]{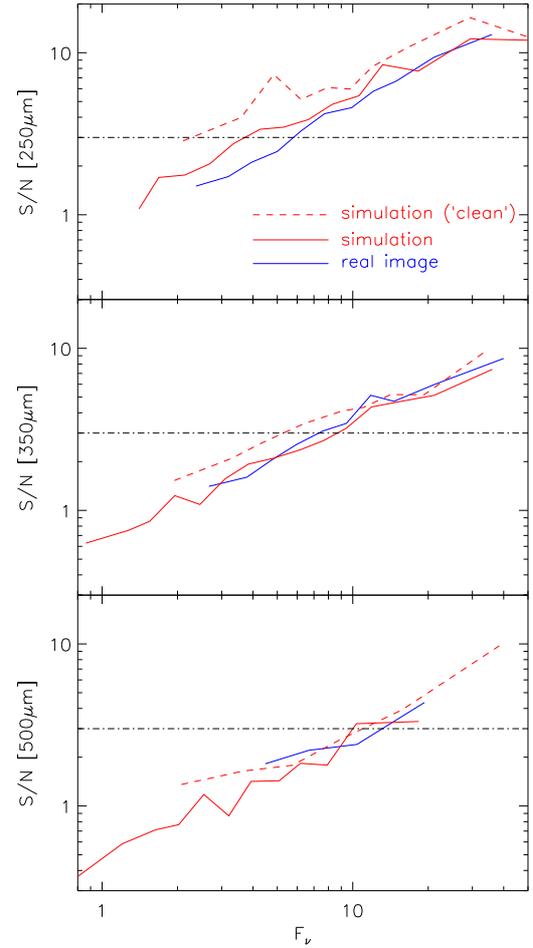}           
   	  \caption{Same as Fig.~\ref{FIG:snrcompPACS} but for the SPIRE bands. 
   	  }
       \label{FIG:snrcompSPIRE}
	 \end{figure}

After accounting for this systematic offset, we can perform a comparison of the noise estimated by the $PAI$ in the mock images (solid red line in Figures~\ref{FIG:snrcompPACS},\ref{FIG:snrcompSPIRE}) with the noise estimated on the real \herschel\ images from the residual maps (solid blue line). We have here directly applied the strategy described above for the real images and, without adding any correction factor, one can see that the red and blue solid lines perfectly match, confirming again that the mock \herschel\ images are indeed close to the real ones. We reach very similar detection limits. In \cite{elbaz11}, the detection limits were based on the blue solid line. The red dashed line shows the S/N obtained for the clean sources only in the mock \herschel\ images, we can see that at 250 and 350\,\micron\ the 3-$\sigma$ limits are slightly deeper ($\sim$2.2 and 5~mJy).

\begin{figure}
    \center
    \includegraphics[width=8cm,height=7cm]{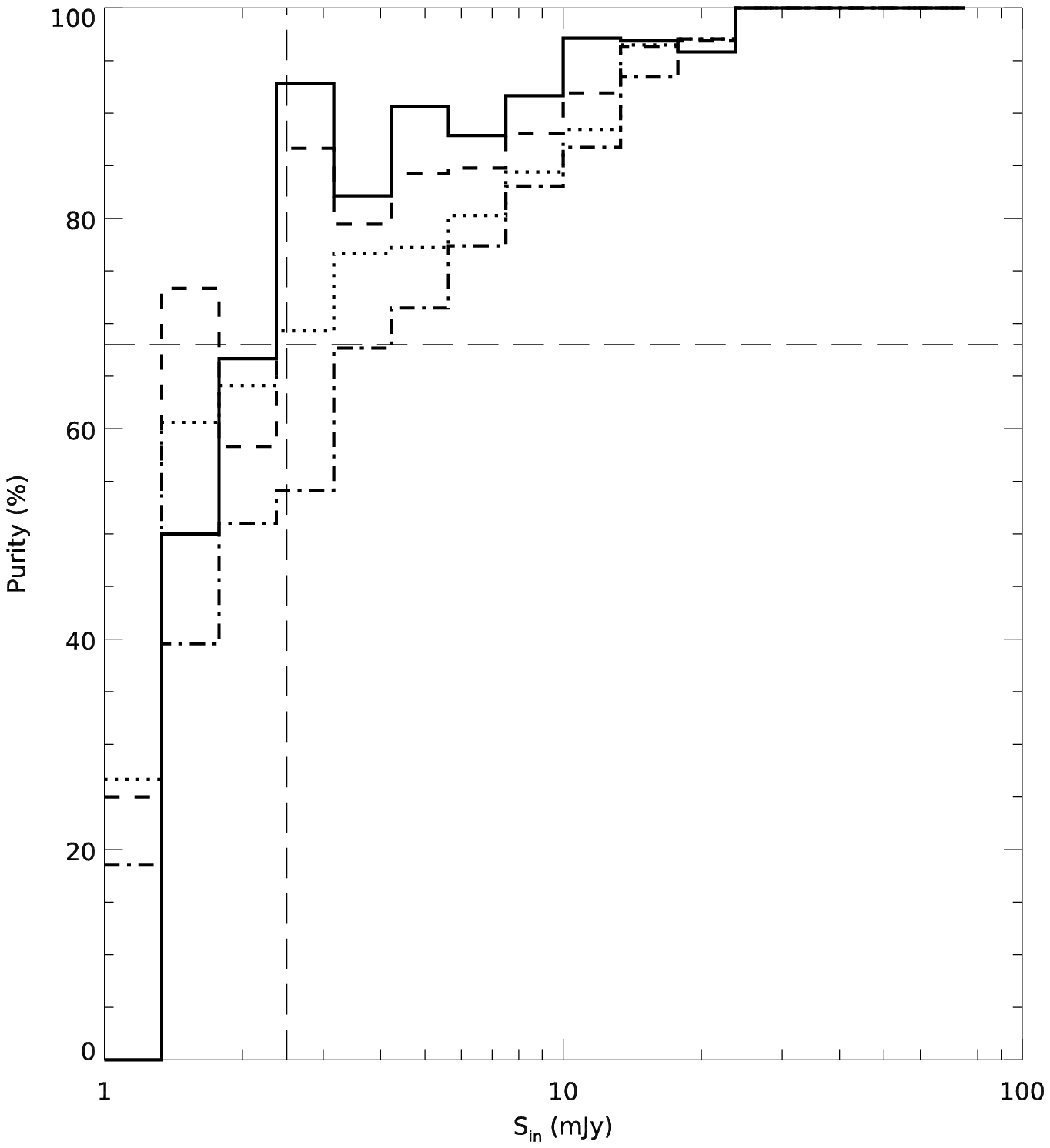}
    \includegraphics[width=8cm,height=7cm]{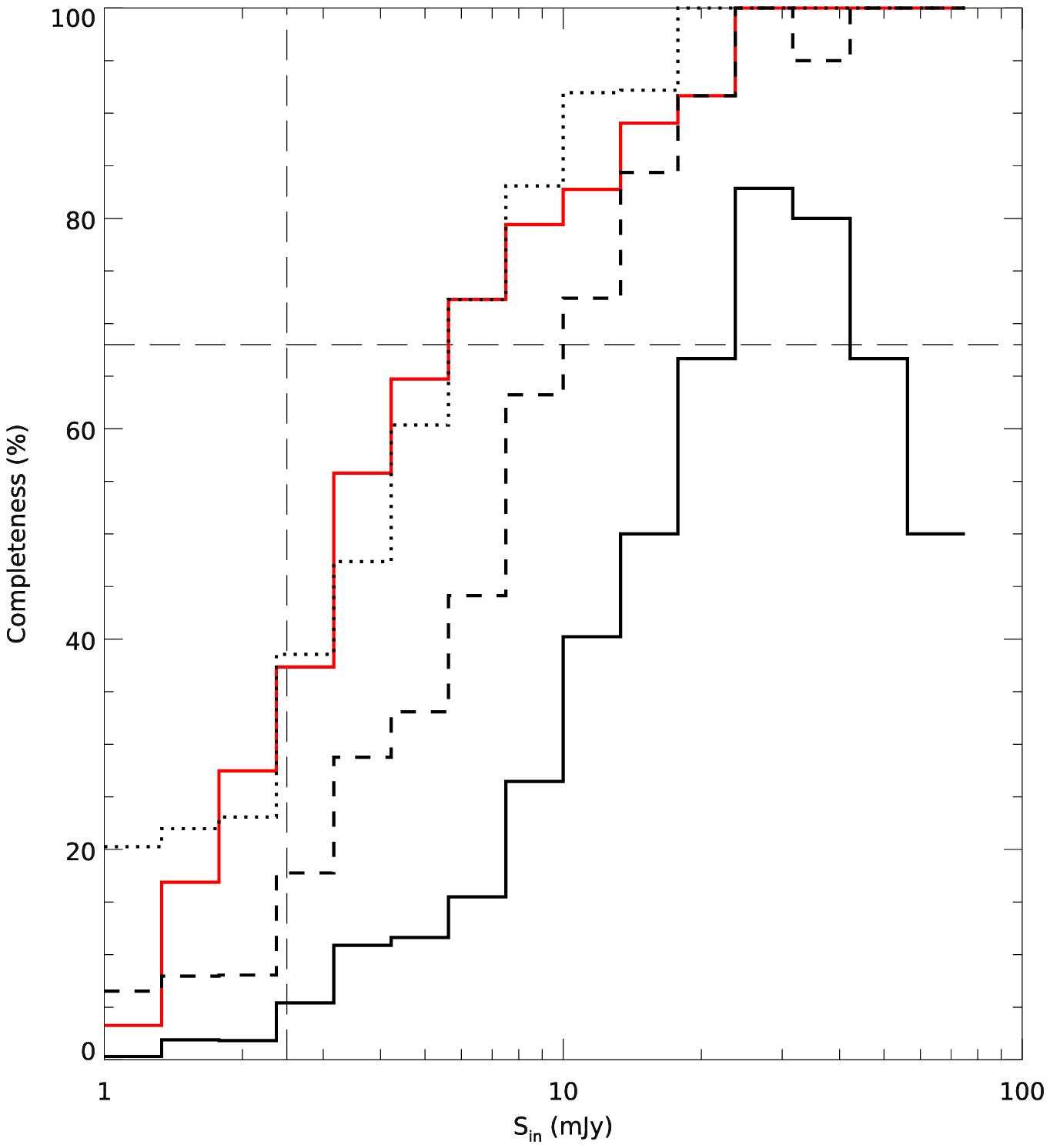}
\caption{Plots presenting the quality of the source extraction on the mock \textit{Herschel} images from a ``purity" (top) and ``completeness" (bottom) criterion. We present here the example of the 250\,$\mu$m band. \textbf{\textit{top:}} purity of the sources extracted in the mock images defined as the fraction of sources with accurate photometry, \ie $|\Delta S|/S_{out}$ $\le$ 0.32, among the sources selected using the clean index, CI. 
The different lines present the sources with no bright 24\,\micron\ neighbour, CI=0 (solid line), at most one 24\,\micron\ neighbour, CI=1 (dashed line), at most 2 neighbours, CI=2 (dotted line), and the full sample of sources without any filtering (dash-dotted line). \textbf{\textit{bottom:}} completeness of the samples selected using the clean index (same definition of the lines), defined as the ratio in percentage of the number of sources filtered by the clean index over the total number of sources in each flux density bin (differential completeness). The red solid line shows the completeness of the sources with $PAI$=\dfout $\leq$ 0.32, which are by definition associated with a 100\% purity.
}
       \label{FIG:comp250}
\end{figure}

\subsection{Representativity of ``clean" sources}
\label{SEC:lci}
We have demonstrated that the clean index (CI) offered a reliable way to pre-select prior positions for which \herschel\ flux density measurements are obtained with a photometric accuracy better than 32\%. Imposing on the list of 24\,\micron\ priors the condition that they have at most one bright neighbour in that band is an efficient method to reject galaxies with poor photometry. However, one may wonder whether this result is conditioned by the fact that we used \herschel\ images based on 24\,\micron\ sources only. We showed that the histograms of pixel values of the mock and real images were comparable (Figures~\ref{FIG:dfinpacs},~\ref{FIG:dfinspire}), and that the S/N ratio as a function of flux density were also comparable (Figures~\ref{FIG:snrcompPACS},\ref{FIG:snrcompSPIRE}).
In this section, we wish to present a complementary test that reinforces already existing evidence that selecting sources by the clean index does not bias their properties, in terms of FIR over MIR colour, and to compute the completeness of such selection.

The following test has been carried out for all \herschel\ bands but for brevity is shown here for the 250\,\micron\ band alone. We reproduce in Figure~\ref{FIG:valid_ds} the 250/24 colour as a function of redshift as in Figure~\ref{FIG:bestfit}. The measure of the average 250\,\micron\ flux density obtained by mean stacking images at the positions of all 24\,\micron\ sources is the same as in Figure~\ref{FIG:bestfit}, and shown with an orange line. It provides a reference for the typical trend of this colour. We note that the 250/24 colour presents nearly no variation with redshift because, while the redshift increases, the 250\,\micron\ band goes across the $\sim$100\,\micron\ bump due to the thermalised big dust grains, and in parallel the 24\,\micron\ band goes across the $\sim$8\,\micron\ PAH bump. If we select the 24\,\micron\ prior positions for which we get a $PAI$=\dfout $>$ 0.32 in the mock \herschel\ image and plot in Figure~\ref{FIG:valid_ds} the 250/24 colour measured on the real \herschel\ image on those positions, we get a 250/24 ratio, which is typically twice as large than the typical one (red solid line in Figure~\ref{FIG:valid_ds}). For this test, we have selected the faintest flux density bin, where the photometry is more likely to be uncertain, \ie 3~mJy $<$ $S_{250}$ $<$ 5.7~mJy. Even at such faint flux density levels, we find that if we instead select sources that exhibit a $PAI$=\dfout $\leq$ 0.32 in the mock \herschel\ image the colour of these sources in the real \herschel\ image is very close to the one obtained from stacking as shown by the solid black line in Figure~\ref{FIG:valid_ds}. This shows that the simulations do provide an efficient way to identify sources with reliable photometry among the sources that are detected in the real images. The $PAI$=\dfout is a solid reference indicator for the local confusion noise but it requires to produce realistic \herschel\ images and one would like to test if a systematic criterion can provide a reliable selection mechanism as well. Clean sources, identified with the clean index, are shown with a solid blue line with red dots in Figure~\ref{FIG:valid_ds}. They follow with a remarkable accuracy the trend seen for the $PAI$=\dfout $\leq$ 0.32 and stacked sources. Therefore, selecting sources with the clean index does not induce a systematic bias in their IR SED.

The quality of the source extraction done in the mock \textit{Herschel} images is described in the following using two criteria: completeness and purity. The purity defines the photometric reliability of the catalogue of sources. It is measured with the PAI, \ie the relative difference between the input flux density and the measured one, and sources are considered as ``pure" if their photometric accuracy is better than 32\,\%. The dash-dotted line in Figure~\ref{FIG:comp250}-top shows the purity of the whole sample of sources, \ie using as priors all the 24\,$\mu$m positions, in the case of the 250\,$\mu$m band. At flux densities lower than 4 mJy, less than 68\,\% of the sample is pure. If instead we require that prior positions be limited to only 24\,$\mu$m sources with at most 2 neighbours, then this limit is reached at 3 mJy. Sources with CI$\leq$1 (\ie, having at most one bright neighbour), are pure for 68\,\% of them at 2.5 mJy. Asking CI=0, does not improve this limit on the purity of the sample.

The second criteria, completeness, quantifies the fraction of sources detected in comparison with the total number of sources that have been injected in the mock \textit{Herschel} images. The dotted line Figure~\ref{FIG:comp250}-bottom, which presents the CI$\leq$2 condition, is obviously the one with the largest completeness, but at the expense as discussed above of a worse purity. In other words, adding sources by allowing to go from CI=1 (dashed line) to CI=2 (dotted line) enhances the completeness by adding sources with a photometric accuracy worse than 32\,\%, the limit that we have here defined as that of ``pure galaxies". The solid red line in Figure~\ref{FIG:comp250}-bottom shows the fraction of sources that are detected with a PAI$\leq$0.32, \ie, the fraction of pure sources among all measurements at the positions of the priors. If the mock \textit{Herschel} images were exactly reproducing the amount of noise that each source suffers from then we could use the positions of these pure sources to obtain the best completeness in the catalogue of sources built out of the real \textit{Herschel} image. However, we can also see that by requiring systematically CI$\leq$1 (dashed line), we get a completeness that is not so bad in comparison and which is much better than the CI=0 request.

In conclusion, the best compromise to optimise the best purity for the largest completeness appears to be the requirement of CI$\leq$1.
We note, however, that the completeness correction factors estimated from the simulations is only a lower limit, in the case of \spire particularly, since the mock \herschel\ images were built from a list of 24\,\micron\ priors hence miss a potential population of sources below the 24\,\micron\ detection limit. 
%

  \begin{figure*}
    \centering
    \includegraphics[width=14cm]{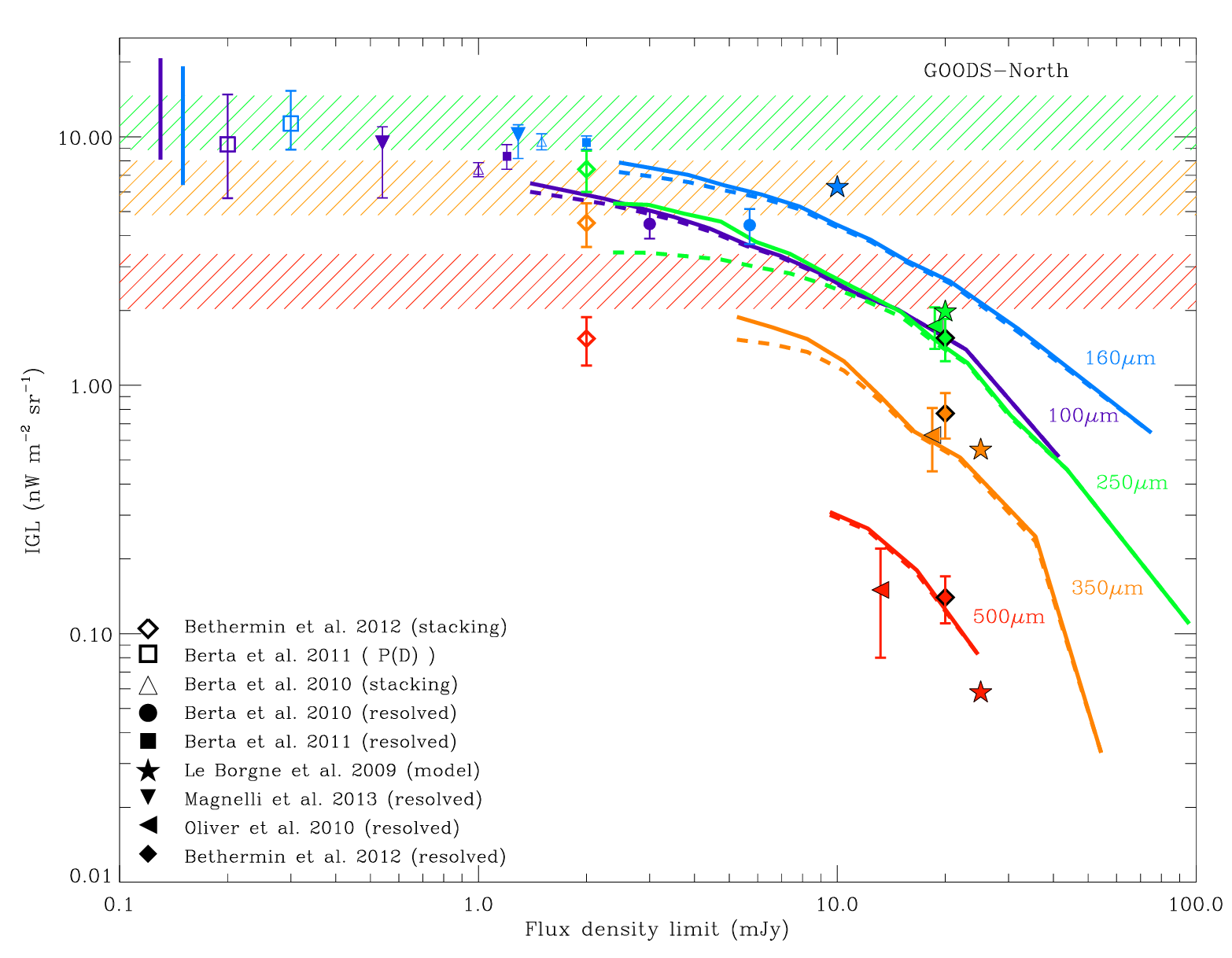} 
     \caption{Cumulative contribution to the CIRB of the IGL as 
     function of the flux density limit in 100\,\micron\ (violet), 
     160\,\micron\ (blue), 250\,\micron\ (green), 
     350\,\micron\ (orange), and 500\,\micron\ (red) in the \goods north field.
     \textit{Dashed lines} show the cumulative flux density directly from the 
     observed \goodsh sources and the \textit{solid lines} are the same but corrected
     by completeness from the simulations in this work.  
     Resolved \citep{berta10,oliver10,berta11,magnelli13},  
     stacked \citep{berta10, bethermin12} and $P$($D$) results \citep{berta11}
      at different adopted $S_{lim}$ are also shown for comparison, 
     as well as IGL predictions for \herschel\ based on
     a model of multi-wavelength galaxy counts \citep{leborgne09}.
     Horizontal dashed bands show the total extrapolated CIRB in the SPIRE bands adopted from \cite{bethermin12}
     using the same colour coding than for the curves and their widths represent
     the limits of the errors.
     In the case of PACS total values, vertical thick bars mark 
     the error in the extrapolations of the CIRB calculated by \cite{berta11}.
     }
     \label{FIG:cirb}
  \end{figure*}

\section{Resolving the cosmic infrared background with \goodsh}
\label{SEC:cirb}

The realistic mock \herschel\ images have helped us calibrate the robustness of the source extraction above a limiting depth, where more than 68\% of the sources are detected with a photometric accuracy better than 32\%. They also provide a way to estimate the completeness of the catalogues extracted above this depth as a function of flux density and thus to correct for incompleteness the contribution of \herschel\ sources to the cosmic IR background (CIRB) down to the local confusion limit ($\sim$2 mJy), which is nearly ten times deeper than the global confusion limit ($\sim$20 mJy).

While statistical studies provide estimates of the amount of EBL due to galaxies \citep{patanchon09,dole06, marsden09, bethermin10a, glenn10, bethermin11, berta11, bethermin12}, they do not allow the study of the properties of the galaxies making this contribution. Using the method of the $PAI$ and/or CI described above to identify sources at the faintest levels in the \herschel\ maps and correcting their contribution to the EBL from their incompleteness as estimated from simulations (as a lower limit), we can study a greater population of galaxies making a stronger contribution to the EBL than extraction methods relying on a single wavelength and at the same time have the possibility to study the nature of these individual galaxies.

  \begin{table*}
    \centering
    \begin{tabular}{c c c c  c  c}
     \hline\hline
      wavelength  & $S>S_{lim}$ & resolved IGL                     & fraction of CIRB & Comments                                    & reference CIRB$^{*}$ \\    
      ($\mu$m)    & (mJy)            &  (nWm$^{-2}$sr$^{-1}$)  &                          &                                                     & (nWm$^{-2}$sr$^{-1}$)\\
       \hline																
       \textbf{100}& \textbf{1.1} &   \textbf{6.9}                    & \textbf{54\%}    & \textbf{after completeness corr.}  &\textbf{12.61 [-1.74;+8.31]$^{a}$}  \\     
       \textit{100} & \textit{1.1}  &   \textit{6.1}                   & \textit{48\%}     &  \textit{before completeness corr.}&\\     
       100             &              1.2 &              8.35                    & 66\%                 &  Berta et al. (2011)                        &\\     
       100             &            0.54 &              9.45                    & 75\%                 &  Magnelli et al. (2013)                   &\\     
       \hline																
       \textbf{160}& \textbf{2.2} &   \textbf{8.2}                  & \textbf{60\%}    & \textbf{after completeness corr.}  & \textbf{13.63 [-0.85;+3.53]$^{a}$}\\  
       \textit{160} & \textit{2.2}  &   \textit{7.2}                   & \textit{53\%}     &  \textit{before completeness corr.}&  \\    
       160             &            2.0   &              9.49                    & 70\%                 &  Berta et al. (2011)                        &\\    
       160             &            1.29 &            10.27                    & 75\%                 &  Magnelli et al. (2013)                   &\\  
       \hline																
       \textbf{250}& \textbf{2.5} &    \textbf{6.0}                   & \textbf{55\%}    & \textbf{after completeness corr.}   &\textbf{10.13 [-2.33;+2.60]$^{b}$}  \\       
       \textit{250} & \textit{2.5}  &    \textit{3.4}                  & \textit{34\%}     & \textit{before completeness corr.} &   \\
       250             &            20    &              1.73                    &            17\%      &  Oliver et al. (2010)                       & [resolved sources] \\    
       250             &            20    &              1.55                    &            15\%      &  B\'{e}thermin et al. (2012)            & [resolved sources] \\    
       250             &            2.0   &              7.4                      &             73\%     & B\'{e}thermin et al. (2012)             & [using stacking]\\  
       \hline																
       \textbf{350}& \textbf{5.0} &    \textbf{2.1}                   & \textbf{33\%}    & \textbf{after completeness corr.}   &\textbf{6.46 [-1.57;+1.74]$^{b}$}   \\       
       \textit{350} &  \textit{5.0} &    \textit{1.5}                  & \textit{24\%}      & \textit{before completeness corr.}&   \\
       350             &            20    &              0.63                    & 10\%      &  Oliver et al. (2010)                       & [resolved sources] \\    
       350             &            20    &              0.77                    & 12\%                 &  B\'{e}thermin et al. (2012)            & [resolved sources] \\     
       350             &           2.0    &              4.50                    & 70\%                 & B\'{e}thermin et al. (2012)             & [using stacking]\\  
       \hline																
       \textbf{500}& \textbf{9.0} &    \textbf{0.37}                   & \textbf{13\%}     & \textbf{after completeness corr.} &\textbf{2.80 [-0.81;+0.93]$^{b}$}   \\      
       \textit{500} & \textit{9.0}  &    \textit{0.30}                  & \textit{11\%}      & \textit{before completeness corr.}&   \\
       500             &            20    &              0.15                    & 5\%      &  Oliver et al. (2010)                       & [resolved sources] \\    
       500             &           20     &               0.14                   &              5\%       &  B\'{e}thermin et al. (2012)           & [resolved sources] \\     
       500             &         2.0      &               1.54                   &            55\%       & B\'{e}thermin et al. (2012)            & [using stacking]\\  
     \hline  
     \end{tabular}
\caption{Summary of the resolved IGL, its contribution to the CIRB down to specific flux density limit $S_{lim}$ in each SPIRE band for the GOODS-North field. 
The percentages correspond to the fractions of the IGL with respect to the total extrapolated estimations of the CIRB reported by ($^a$) \cite{berta11} and by ($^b$) \cite{bethermin12}. The new flux limits $S_{lim}$ in the PACS bands are similar to \cite{elbaz11} but the SPIRE depths are deeper.}
  \label{TABLE:cirb}
  \end{table*}

  \begin{table*}
    \centering
    \begin{tabular}{c c c c c c c}
     \hline\hline
      wavelength  & $S_{lim}$  & $A_{beam}$ & $N_{compl}$ & $IGL_{compl}$ & $N_{real}$ & beams$/$source\\    
      ($\mu$m) & (mJy)  & (arcmin$^{2}$)  &  (\%) & (\%) &  & ($N_{beams}$/$N_{real}$)\\
       (1) & (2)  & (3) & (4) & (5) & (6) & (7)\\    
       \hline   
100  &   1.1  &   0.021 &     80 &           89 &           568 &       12 \\
160  &   2.2  &   0.057 &     78 &           88 &           589 &         4\\
250  &   2.5  &   0.154 &     41 &           61 &           342 &         3\\
350  &   5.0  &   0.292 &     58 &           72 &           185 &         3\\
500  &   9.0  &   0.632 &     78 &           81 &            43  &         5\\
     \hline  
     \end{tabular}
       \caption{Summary of the number of sources and beams involved 
       in the simulations, as well as in the \herschel\ maps.
        Col.(2): Flux density above which sources are detected in the simulation, \ie confusion limit defined as the depth above which 68\,\% of the sources get measured with a photometric accuracy better than 32\,\%.
        Col.(3): $A_{beam}$ is the beam area at the central wavelength listed in Col.(1) defined as $\pi$ $\times$ (0.73$\times$FWHM)$^2$ (\citealp{dole03,oliver10}).
        Col.(4): $N_{compl}$ is the completeness in \% in number of clean sources  ($neib24$ $\le$1), \ie the \% of sources injected with $S^{in}$$\geq$$S_{lim}$ that are detected in the simulation as clean and with $S^{out}$$\geq$$S_{lim}$.
        Col.(5): $IGL_{compl}$ is the completeness in \% in contribution to the IGL.
        Col.(6): $N_{real}$ is the number of clean sources detected in the real $Herschel$ maps above $S_{lim}$. 
        Col.(7): beams$/$source is the number of beams per source, where $N_{beams}$= 150$'^2$$/$$A_{beam}$.
        }
  \label{TABLE:beam}
  \end{table*}

  \begin{table*}
    \centering
    \begin{tabular}{c c c c c c}
     \hline\hline
      wavelength  & $S>S_{lim}$ & $\langle z \rangle$ &  $\langle z \rangle_{f}$ & $\langle M_\star \rangle$ & $\langle M_\star \rangle_{f}$ \\    
      ($\mu$m) & (mJy) &  & & \multicolumn{2}{c}{(10$^{11}$M$_{\odot}$)}\\
       \hline   
       100 & 1.1 &   0.94 & 0.86 & 0.8 & 1.4 \\     
       160 & 2.2 &   0.96 & 0.98 & 0.8 & 1.6 \\    
       250 & 2.5 &   0.96 & 1.10 & 0.9 & 2.0 \\      
       350 & 5.0 &   0.94 & 1.17 & 1.2 & 2.4 \\      
       500 & 9.0 &   1.26 & 1.54 & 2.3 & 3.6 \\                                                                                                                              
     \hline  
     \end{tabular}
       \caption{Summary of the median $\langle z \rangle$ and 
        flux-weighted average redshift $\langle z \rangle_f$, as well as the median
        $\langle M_\star \rangle$ and the flux-weighted stellar mass 
        $\langle M_\star\rangle_f$ of the galaxy population that produce the resolved IGL  
        down to specific flux density limit $S_{lim}$ in each SPIRE band found based on the simulations of the GOODS-North field. }
  \label{TABLE:zm}
  \end{table*}

  \begin{figure*}
    \centering
    \includegraphics[width=8cm]{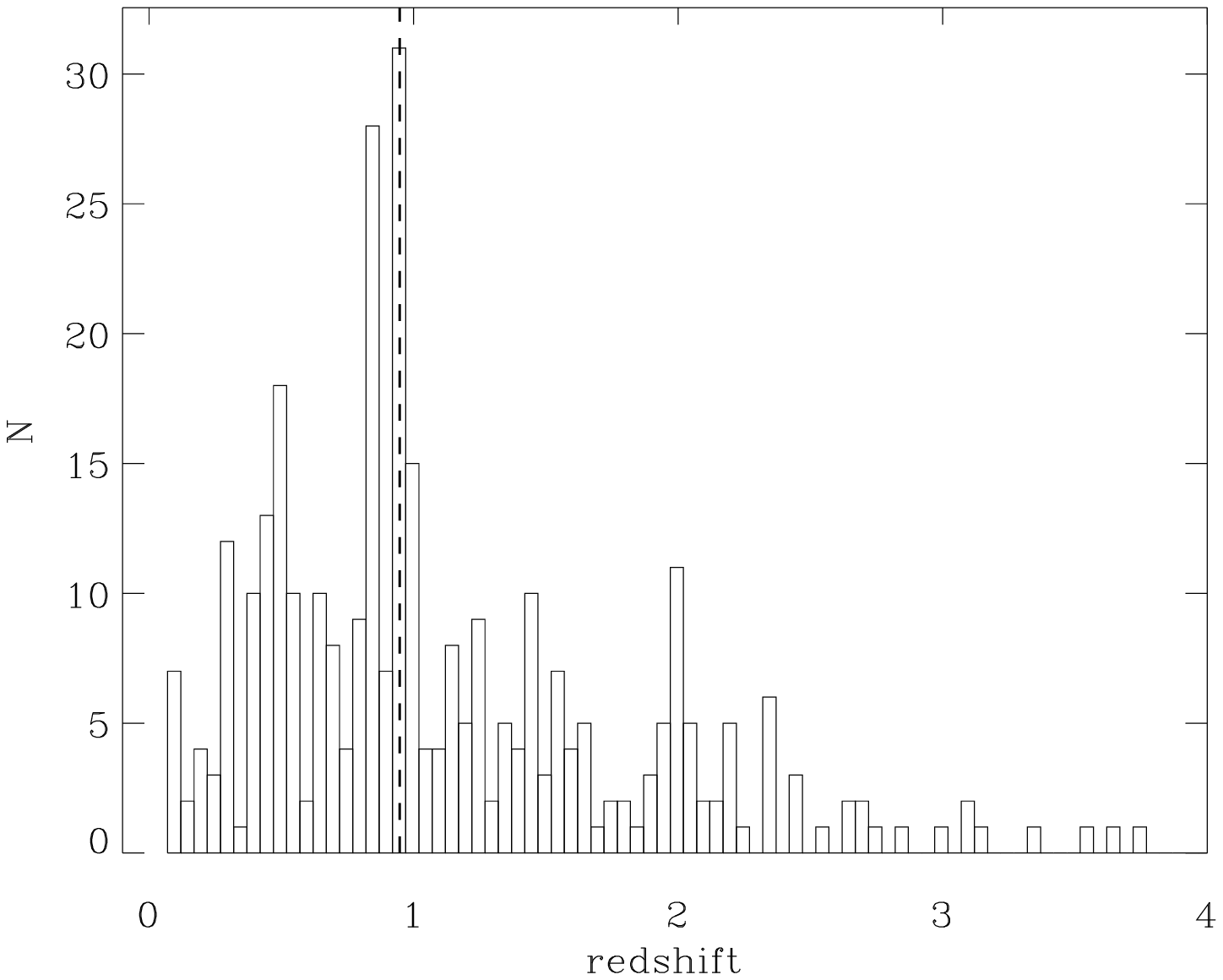}
    \includegraphics[width=8cm]{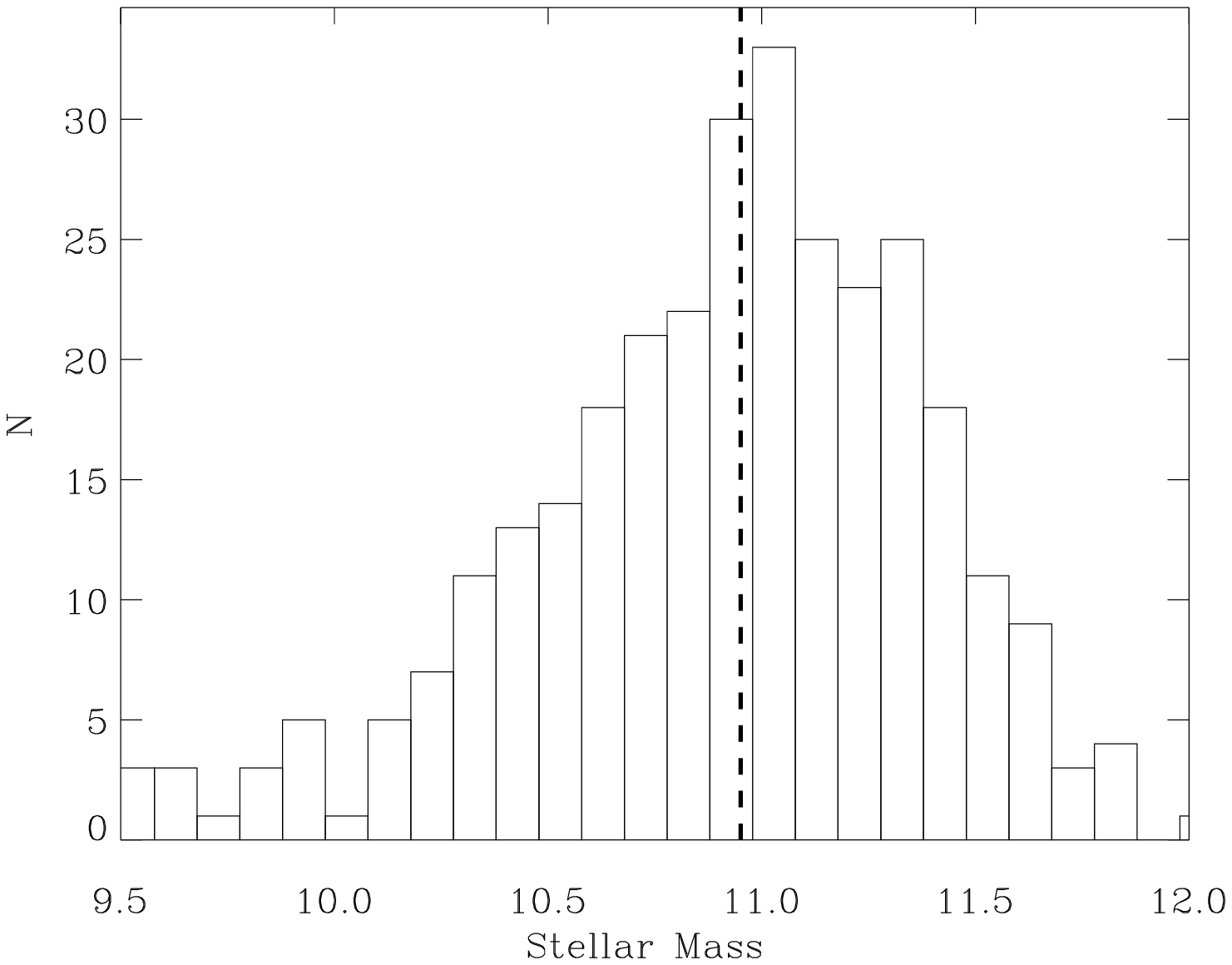}
     \caption{Distributions in redshift (left) and stellar mass (right) of the sources responsible for 55\,\% of the CIRB at 250\,$\mu$m. 
     Vertical dashed lines show the median of the distributions.
     }
     \label{FIG:hisz}
  \end{figure*}

\subsection{Integrated galaxy light : fraction of the cosmic IR background resolved in individual detections}
The combined contribution of individual sources to the CIRB is the \textit{integrated galaxy light} (IGL), which can be expressed in its differential form as:

 \begin{equation}
     \frac{d IGL}{dS} = \frac{d N}{d S} \times \frac{S_{band}}{10^{20}} \times \nu_{band}
     \label{EQ:igl}
  \end{equation}

\noindent where d$N$(sr$^{-1}$) is the surface density of sources with 
flux density $S_{band}$ in mJy over a flux bin d$S$(mJy) and $\nu_{band}$ 
in Hz is the frequency in a given FIR bandpass. 
The constant 10$^{20}$ appears after the 
conversion of 1~mJy = 10$^{-20}$nWm$^{-2}$Hz$^{-1}$.
Integrating Eq.~\ref{EQ:igl} down to the $S_{lim}$, we obtain the 
IGL in units of nWm$^{-2}$sr$^{-1}$.
Our IGL measurements correspond to high fractions 
of the CIRB down to very deep flux limits compared with 
other estimations.

We have compared the estimates of the amount of extragalactic background light produced by individual galaxies resolved by \textit{Herschel} coming from the PAI and CI techniques and found that they were consistent within less than 5\,\%. We present here the values obtained using the CI alone since it does not rely on the assumptions on redshifts and SEDs used to produce the simulations.

We first compute the cumulative contribution of all ``clean" sources individually detected above the \textit{Herschel} flux limits -- numbers in \textit{italics} on the second line at each wavelength in Table~\ref{TABLE:cirb}. Then, we correct these numbers for completeness using the correction factors derived from our realistic mock \textit{Herschel} images and obtain the values listed in bold letters in the Table~\ref{TABLE:cirb}. For example, over a total of 971 sources injected in the mock $Herschel$ image at 250\,$\mu$m above a flux limit of $S_{lim}^{in}=$ 2.5 mJy, the algorithm detects 398 sources above $S_{lim}^{out}=$2.5 mJy that are classified as clean (\ie have at most one bright neighbour at 24\,$\mu$m). Therefore, the completeness of our clean list is $N_{compl}=$41\,\% in number of sources and $IGL_{compl}=$61\,\% in the contribution to the IGL (higher than $N_{compl}$ since detect the brightest sources). 
 
In comparison with the commonly used definition of the confusion limit (\eg 40 beams/source in \citealp{nguyen10}), we have also computed the equivalent of our confusion limits in numbers of beams/source. Various papers have used different definitions of the beam size, here we use a definition of the beam area matching the ones used in \cite{dole03} and \cite{oliver10} for consistency in the comparison: $A_{beam}$ = $\pi$ $\times$ (0.73$\times$FWHM)$^2$. The beam size at 250\,$\mu$m is 0.154 arcmin$^2$ and the number density of clean detections is equivalent to 3 beams/source, knowing that this level was reachable only by the use of the shorter bands to allow the identification of robustly determined photometric measurements. We list in Table~\ref{TABLE:beam} these numbers for all 5 $Herschel$ bands.

The resulting background estimated in nW m$^{-2}$ sr$^{-1}$ are then compared to estimates of the \textit{extragalactic background light} (EBL) at these wavelengths, in order to determine the resolved fraction of the CIRB. Two options exist for this comparison that consists in using as reference values either the COBE measurements or statistical extrapolations of \textit{Herschel} counts using stacking and $P$($D$) analysis. Error bars on the COBE values are larger, especially at 100\,$\mu$m, where the correction for zodiacal light is highly uncertain, hence we use here extrapolated values from ($^a$) \cite{berta11} and by ($^b$) \cite{bethermin12} (listed in the last column of Table~\ref{TABLE:cirb}) as a reference to compute fraction of the resolved CIRB.

In the PACS 100 and 160\,$\mu$m bands, sources brighter than 1.1 and 2.2 mJy make around 50-60\,\% of the CIRB at these wavelengths as compared to the $\sim$75\,\% obtained by \cite{magnelli13} counting sources down to 0.6 and 1.3 mJy in GOODS-South. However, thanks to our \textit{Herschel} mock image we are able to correct these values obtained for clean sources for incompleteness. After having applied this correction, we get fractions of the resolved CIRB of 54\,\% and 60\,\%, which are comparable to the ones obtained using stacking and $P$($D$) analysis as shown in Figure~\ref{FIG:cirb}. This is also true for the SPIRE bands, since in both cases, PACS and SPIRE, our estimates shown with solid coloured lines (after completeness correction) can easily be extended by eye to the symbols showing the locus of statistical estimates using stacking and $P$($D$) analysis.

The most remarkable improvement of our method can be seen in the SPIRE bands, where we resolve 55, 33, and 13\,\% of the background light at 250, 350, and 500\,$\mu$m with individually detected sources thanks to the use of our estimate of the ``local confusion noise" as compared to 15, 12, and 5\,\% as measured by \cite{oliver10} and \cite{bethermin12} with the global confusion limit of $\sim$20 mJy. Our approach has allowed us to resolve in individual galaxies 3.5 times more extragalactic background light at 250\,$\mu$m and more than $\sim$3 times at 350 and 500\,$\mu$m.

We present in Fig.~\ref{FIG:cirb} our estimates of the IGL both before (dashed lines) and after (solid lines) correcting for completeness.
Direct measurements from \cite{berta10}, \cite{berta11}, \citep{magnelli13} for PACS and \citep{oliver10}, \cite{bethermin12} for SPIRE are show for comparison with filled symbols, while extrapolations with stacking and $P$($D$) analysis are shown with open symbols. We also present expectations of the IGL fractions predicted for \herschel\ based on a model of multi-wavelenth galaxy counts by \cite{leborgne09}.

We compared our resolved IGL fractions to the ones predicted by the backward evolution model of \cite{bethermin11b}, which reproduces galaxy counts in all \textit{Herschel} bands. In this paper, the authors calculate the confusion limit expected for different telescope diameters in the FIR using the criterion defined in \cite{dole03} by requiring sources to be brighter than 5-$\sigma^{\rm photometric}$, where $\sigma^{\rm photometric}$ is the noise due to the fluctuations induced by the sources below this detection limit, and such that less than 10\,\% of the sources have a neighbour closer than 0.8$\times$FWHM above that detection limit. The standard global confusion limit fits the predictions of the model for a 3.5m class telescope, but our improved depths in the 250\,$\mu$m band correspond to the standard confusion limit of a telescope with an aperture close to $\sim$10m (see Figure 14 and Tab. 8 in \citealp{bethermin11b}).

\subsection{Nature of the sources making the CIRB}
We present here the characteristic redshift and stellar mass of the galaxies that make the resolved fraction of the CIRB described in the previous section. We provide two methods to estimate these characteristic values, the median -- $\langle z \rangle$ and $\langle M_\star \rangle$ --  and flux weighted average -- $\langle z \rangle_f$ and $\langle M_\star \rangle_f$ computed following Eqs.~\ref{EQ:redshift},\ref{EQ:mass}.

  \begin{equation}
     \langle z \rangle_f = \frac{\sum_{i} S_{i}(\lambda) \times z_{i}}{\sum_{i} S_{i}(\lambda)},     
  \label{EQ:redshift}  
  \end{equation} 

    \begin{equation}
     \langle M_\star \rangle_f = \frac{\sum_{i} S_{i}(\lambda) \times M_{\star i}}{\sum_{i} S_{i}(\lambda)},     
  \label{EQ:mass}  
  \end{equation} 

where $S_i$ are the flux densities of the sources contributing to the CIRB for a \herschel\ band centred at $\lambda$. 

The IGL in the \pacs and \spire bands to the CIRB, as well 
as the typical redshifts and stellar mass of the galaxies 
contributing to the CIRB are summarised in  
Table~\ref{TABLE:cirb}. 
We see that the typical redshift of the galaxies responsible for a dominant 
contribution to the infrared background light increases with increasing 
wavelength. This is naturally expected as a result of $k$-correction, 
since the peak emission of the IR spectral energy distribution of individual galaxies,  
like \eg the closest starburst galaxy M82, lies 
around 80\,$\mu$m (with a moderate $T_{dust}$ variation). 
Thus, the peak of the IR background around 140-160\,$\mu$m is expectedly 
produced by galaxies around z$\sim$0.9. Increasing the wavelength, sources at 
larger redshifts contribute a larger contribution with respect to this redshift,
hence we observe the increasing redshift to z$\sim$1.3 in the higher wavelength bands. 

Our median redshifts in 100 and 160\,\micron\ are consistent 
with the results of \cite{magnelli13} using the PEP/\goodsh\ data
who obtained $z$ = 0.85$^{+0.41}_{-0.33}$ for 
sources $S_{100\mu m}$ $>$ 1.5~mJy, and 
$z$=0.94$^{+0.52}_{-0.38}$ for sources $S_{160\mu m}$ $>$ 2.5~mJy (see also \citealt{berta10,berta11}).
In the case of the SPIRE bands (see Figure~\ref{FIG:hisz}-left), \cite{bethermin12} found that
half of the CIRB is emitted at the typical redshifts of 1.04, 1.20, and 1.25
for 250, 350, and 500\,\micron, respectively, by staking data
in COSMOS and GOODS-north as part of the 
HerMES. 
This  increasing trend in redshift in the higher wavelengths band is also consistent
with our results in the SPIRE bands, while we find larger median
redshift in 500\,\micron.

Thanks to our method, it is now possible to study the individual galaxies responsible for the peak of the CIRB around 250\,$\mu$m. 
The major difference and strength of our approach with respect to other methods (\eg stacking) is the fact that we are identifying the contribution of each individual galaxy to the background instead of studying the average contribution of populations of galaxies. This is important for studies based on large galaxy samples. For example, \cite{daddi07a,daddi07b} reported that, based on stacking, $z$=2 sources with a ``mid-IR excess'' were powered by AGNs in the mid-IR. However, this excess was coming only from a population of about 30\% of the sources that were indeed powered by AGNs and numerous enough to make a strong signal in the stacking. It was impossible from the stacking alone to know if the characteristics found could be applied to 30\% or 100\% of the sources. 

We present in Figure~\ref{FIG:hisz} their redshift and stellar mass distributions. Both distributions are well peaked, and it is quite interesting to see that the stellar mass distribution peaks at a value of M$_{\star}$ $\sim$9$\times$ $10^{10}$ M$_{\odot}$, close to the mass of the Milky Way with a narrow dispersion of only 0.3 dex in log$_{10}$. Interestingly the same happens in the other bands, which present a similar characteristic mass. The redshift peak is at $z$ $\sim$0.95.

In consequence, most of the cosmic infrared background is due to the radiation of siblings of the Milky Way around 7.5 billion years ago. This finding is in agreement with the fact that the stellar mass function of star-forming galaxies does not seem to vary strongly with redshift up to $z\sim$2 \citep{bell03,bell07,pozzetti10,ilbert10} and with the statistical analysis done using stacking of galaxy populations by \cite{viero13}.

\section{Conclusions}
\label{SEC:conclusions}

The main conclusions of this work are summarised below:

- We have developed a method to robustly predict FIR fluxes based on 
the knowledge of their redshift and emission at shorter wavelengths 
starting from 24\,\micron.%

- Using those predicted FIR fluxes, we were able to build realistic simulated \herschel\ images of the GOODS-North that we showed to robustly reproduce the confusion noise present in the real \herschel\ maps. 

- We showed that the extrapolation of the FIR emission of 24\,\micron\ sources detected down to 20~$\mu$Jy make up enough signal in the mock images to sharply reproduce the pixel distribution of the observed \herschel\ maps, suggesting that the bulk of the confusion noise, even fainter than the detection limits, is included in the mock images.

- We used the difference between the extracted photometry of sources 
in the simulated images and the input fluxes used to make the 
mock images -- the Photometric Accuracy Index ($PAI$) -- to determine the degree 
of local confusion in the \goodsh maps, 
\ie the photometric uncertainty at each specific position on the mosaics.

- We validate the Clean Index (CI) approach \citep{hwang10,elbaz10} that identifies clean and confusion dominated \herschel\ sources from the local environment of 24\,\micron\ sources clean sources in the \goodsh data based on the quantification of the number of contaminating close neighbours to a given source. We confirm that the optimal selection rule for clean sources is when CI$\leq$ 1. While the CI identifies sources that respect the $PAI$ criterion (\dfout$\leq$0.32), the mock \herschel\ images combined with the $PAI$ allow one to identify reliable sources with a stronger completeness and to set deeper flux density limits for the data in the \spire bands.

- After correcting for incompleteness, the emission of sources detected with our approach in the GOODS-North field, we find that galaxies individually detected by \textit{Herschel} down to 1.1 mJy at 100\,\micron\ and 2.2 mJy at 160\,\micron\ make up as much as 54\% and 60\% of the extragalactic background light, 
 whereas down to 2.5, 5, and 9 mJy sources produce 55, 33, and 13.0\% of the EBL at 250\,\micron, 350\,\micron, and 500\,\micron, respectively.

- Interestingly we find that the dominant contributors to the CIRB in all \herschel\ bands appear to be galaxies analogous to the Milky Way in stellar mass with M$_{\star}$ $\sim$9$\times$10$^{10}$ M$_{\odot}$ reinforcing the idea of a dominant dark matter halo mass in which stars formed over the Hubble time \citep{conroy09,bethermin13}. 

\begin{acknowledgements}
The authors would like to thank the referee for their valuable 
comments, which helped to improve the manuscript.
RL acknowledges the financial support from FONDECYT through grant 3130558
and CEA-Saclay. 
This research was supported by the French Agence Nationale de la Recherche (ANR) project ANR-09-BLAN-0224. DE acknowledges the contribution of the FP7
SPACE project ASTRODEEP (Ref.No: 312725), supported by the European
Commission.
\end{acknowledgements}

\bibliographystyle{aa} 
\bibliography{ref_rleitonmasterbibtex_revised_1412.bib} 

\end{document}